# Mesoscopic phenomena in multiple light scattering

Academisch Proefschrift
ter verkrijging van de graad van doctor aan
de Universiteit van Amsterdam, op gezag van de
Rector Magnificus Prof. Dr. P. W. M. de Meijer
in het openbaar te verdedigen in de Aula der
Universiteit op donderdag 11 mei 1995 te 10.30 uur

door

Marcus Cornelis Wilhelmus van Rossum
geboren te Utrecht

# Dankwoord

Het boekwerk dat voor u ligt, is niet het resultaat van mijn inspanning alleen, maar was slechts mogelijk door de hulp van anderen. Allereerst wil ik daarvoor mijn begeleider Theo Nieuwenhuizen bedanken. Bijna al het gepresenteerde werk is het resultaat van onze plezierige samenwerking. Zijn snelle inzicht in de fysica en de rest van het leven is opmerkelijk. Zijn betrokkenheid bij mijn promotie en verdere loopbaan waardeer ik bijzonder. Mijn promotor Prof. Aad Pruisken wil ik bedanken voor zijn inzet en behulpzaamheid. Leerzaam was het nauwe contact met de experimentele groep van Prof. Ad Lagendijk. Van hem en de andere groepsleden heb ik veel opgestoken.

Mijn ouders en mijn broer ben ik erg dankbaar voor hun niet aflatende steun en voortdurende interesse in de voortgang van dit werk.

Liefste Karen, ik zou je kunnen bedanken voor het proeflezen van een groot gedeelte van de tekst, maar behalve dat misschien een verkeerde indruk zou wekken, staat het in geen enkele verhouding tot hetgeen je werkelijk voor me hebt betekent in het afgelopen jaar.

Ook gaat mijn dank uit naar Mevr. Wouters-Hellebrekers voor een collectie prachtige, oude natuurkundeboeken, die behalve historisch interessant, ook nog bruikbaar bleken.

The hospitality of Prof. Ilya Polishchuk and Dr. Alexander Burin during my visit to Moscow are acknowledged with warm feelings.

Ich danke Michael Schreiber und Etienne Hofstetter herzlich für die numerische Daten unserer gemeinsamen Arbeit.

Nee, ik ben jullie niet vergeten: Meint van Albada, Johannes de Boer, Hanna Brands, Marco Brugmans, David van Coevorden, Hilde Fleurent, Jan de Gier, Floor Goedemondt, Rogier Groeneveld, Jan Groenewold, Tom Hijmans, Bert Holsbeek, Jiri Hoogland, Wim Koops, Rik Kop, Mark en Ron Kroon, Ad Lagendijk, Jom Luiten, Alec Maassen van den Brink, Peter Molenaar, Allard Mosk, Nathalie Muller, Boris Nieuwenhuis, Peter den Outer, Pepijn Pinkse, Irwan Setija, Tijn Smit, Rudolf Sprik, Ruud Vlaming, Willem Vos, Jook Walraven, Gerard Wegdam en Diederik Wiersma. Door jullie waren de afgelopen vier jaren zo levendig, leerzaam en prettig.

Een heel bijzondere plaats nemen de AIO's en OIO's van de Spectroscopie der Verdichte Materie in. Betere collega's zijn niet voorstelbaar, noch bij menig mensamaal en cappootje, noch bij de inmiddels legendarische AIO-nights.

De geslaagde omslag en titelblad van dit boek zijn van de hand van grafisch ontwerper Karel van Laar, waarvoor ik hem heel dankbaar ben.

Tot slot wil ik de geïnteresseerde lezer graag verwijzen ik naar de Nederlandse samenvatting achterin.

Mark van Rossum

Amsterdam, maart 1995

# List of publications

Works on which this thesis is based:

- Chapter 4: M. C. W. van Rossum and Th. M. Nieuwenhuizen, *Influence of skin layers on speckle correlations of light transmitted through disordered media*, Physics Letters A **177**, 452-458, (1993)

- Section 2.6 and section 3.2: Th. M. Nieuwenhuizen and M. C. W. van Rossum, *Role of a single scatterer in a multiple scattering medium*, Physics Letters A **177**, 102-106, (1993)

- Section 2.6: P. N. den Outer, M. C. W. van Rossum, Th. M. Nieuwenhuizen and A. Lagendijk, *Locating objects with diffuse light* in 'OSA proceedings on Advances in optical imaging and photon migration', R. R. Alfano ed., 297-302 (SPIE, Bellingham, 1994)

- Chapter 6: J. F. de Boer, M. C. W. van Rossum, M. P. van Albada, Th. M. Nieuwenhuizen and A. Lagendijk, *Probability distribution of multiple scattered light measured in total transmission*, Physical Review Letters **73**, 2567-2571, (1994)

- Chapter 7: Th. M. Nieuwenhuizen and M. C. W. van Rossum, *Intensity distribution of waves transmitted through a multiple scattering medium*, Physical Review Letters **74**, 2674-2678, (1995)

- Chapter 5: M. C. W. van Rossum, Th. M. Nieuwenhuizen and R. Vlaming, *Optical conductance fluctuations: diagrammatic analysis and non-universal effects*, to be published in Physical Review E (1995)

- Chapter 6: M. C. W. van Rossum, J. F. de Boer and Th. M. Nieuwenhuizen, *Third cumulant of the total transmission of diffuse waves*, to be published in Physical Review E (1995)

Other works on disordered systems:

- Th. M. Nieuwenhuizen and M. C. W. van Rossum, *Universal fluctuations in a simple disordered system*, Physics Letters A **160**, 461-464, (1991)

- M. C. W. van Rossum, Th. M. Nieuwenhuizen, E. Hofstetter and M. Schreiber, *Band tails in a disordered system* in 'Photonic band gaps and localization', C. M. Soukoulis ed., 509-513, (Plenum Press, New York, 1993)

- M. C. W. van Rossum, Th. M. Nieuwenhuizen, E. Hofstetter and M. Schreiber, *Density of states of disordered systems*, Physical Review B **49**, 13377-13382, (1994)

# Contents

*Il y aurait trop d'obscurité,*
*si la vérité n'avait pas des marques visibles.*

Blaise Pascal, Pensées (Br. 857)

# 1

# Introduction

## 1.1 Diffuse light

Light scattering phenomena are seen at many places in nature. It is often the case that light from a source undergoes no or only a few scatterings before it reaches us. One of the many examples of such a situation is the sun we see on a clear day. The presence of only a few scattering events is also called the "ballistic" regime.

In this work, however, we will deal with light that has been scattered many times, also called diffuse light. This typically happens if the light penetrates a medium with many randomly placed scatterers. The intensity then diffuses through the medium. An important length scale is the *mean free path*. It gives the length-scale on which a plane wave is converted into diffuse light. Whereas in the ballistic regime the mean free path exceeds the system thickness, in the diffuse regime the mean free path is much shorter than the sample thickness. As an everyday life example of diffuse light, one can think of light in the fog or in a glass of milk, or one can try to look through this very piece of paper.

Although on earth diffuse light is common, astrophysicists initiated the first detailed studies in this field. With the starlight seen on earth, they try to reconstruct the spectrum and intensity of stars. Therefore, they need to know how the light from the stars penetrates the interstellar clouds. Doppler shifts and absorption bands make this a complicated problem. Chandrasekhar [1] and Van de Hulst [2] did important work in this field.

Closer to earth there is by now a vast range of applications of diffuse light scattering: With remote sensing, one extracts from satellite data the health of vegetation, or one can investigate the diffuse reflection of ice and oceans. The reflectance of laser light in the sky can be used to study the composition of the air (this technique is called "lidar"). There are also medical applications: using diffusely scattered light one tries to locate cancer cells or one measures the oxygen content of blood. The harmlessness of this technique is of course of great importance, yet extracting the information from the scattered light is a complicated problem.

The typical laboratory systems used to study diffuse light are small slabs. These slabs are made of a substance similar to white paint, and are some hundred microns thick. The slabs are illuminated by a laser source and the transmission or reflection is measured. Very small mean free paths can be obtained in these samples, which is advantageous for the study of corrections to the diffusion theory. We will apply our theory to slab geometries and compare it to the experimental data from such slabs.

Very precise measurements are possible for these systems.

## 1.2    Light and other waves

Many principles of diffuse transport of light also apply to other diffuse wave phenomena. Diffuse transport as well as the corrections on diffusion are also studied for neutrons, sound waves [3, 4], microwaves and electrons [5]. Scattering of sound or seismic waves is studied, for instance, to understand propagation of shock waves caused by earth quakes. Oil companies use sound when trying to locate oil. On smaller scales one uses ultrasound to detect cracks or other faults in, for instance, metal products. Finally, also phonons in solid state fall in the category of diffuse wave scattering.

Closely related to light scattering is microwave scattering. In that case the wavelength is a few centimeters, and a typical sample can consist of a "box" of say, one meter in size, filled with metal spheres sized a few centimeters. These length scales are of course convenient for experimenting. Presently, the research on microwaves is mainly carried out by the group of Genack at Queens College in New York.

Electronic measurements are typically performed on samples sized a few microns or less, so called micro or nanostuctures. Using etching techniques samples with a wide range of shapes and properties can be manufactured. One calls the diffuse regime for electrons also the metallic or Ohmic regime.

All systems have their particular advantages and disadvantages. The electronic systems, studied very intensively, have the advantage that localization (see below) can be easily achieved. An applied magnetic field enters conveniently as an extra tuneable parameter. Disadvantages in electron systems are scattering mechanisms that destroy the phase coherence. One such mechanism is phonon scattering, which can be reduced, however, by cooling the sample to, say, $0.2\,K$. Also the electron-electron interaction can obscure the interference effects we are after. Nevertheless, their effect on localization is quite interesting, see, for instance, the book edited by Efros and Pollak [6].

Electromagnetic waves, notably microwaves and light, have the advantage that both photon-photon interaction and photon-phonon interaction are nearly absent. They allow for "clean" experiments on "dirty" samples. Both angular resolved and angular integrated measurements are possible for these systems, resulting in three quite different transmission quantities. This stands in contrast with the electronic systems where usually only the total conductance is measurable. A disadvantage of electromagnetic waves is that they are not easily localized, as the scattering is often weak.

Most of the phenomena we are interested in, occur in all systems mentioned. The reason for this is that, up to some variations, the same equation, namely the scalar wave equation with a random potential, describes the systems. Therefore the study of analogies between different fields has received some popularity in recent years. For the analogies between optics and electronics see, for instance, Van Haeringen and Lenstra [7, 8]. The study of the analogy between the scattering of light and sound dates back somewhat more years already [9].

The wave character, which is present in all these systems, makes this transport different from particle diffuson. The (self-)diffusion of a particle is a subject on its own, and is ruled by exclusion, interaction and entropy effects. The diffusion of waves is different as such effects are mostly absent. Instead, interference is the most important process. It leads to the diffusion, yet also to corrections on the diffusion theory.

## 1.3 Anderson localization

Anderson pointed out in 1958 that in multiple scattering media the interference processes may become so strong that normal diffusion vanishes [10]. It is not simply that the diffusion constant tends linearly to zero as scattering increases. In that case localization could only be reached in the limit of infinitely strong scattering. Rather, the return probability of the intensity becomes higher and higher, reducing the diffusion constant in a stronger fashion. The diffusion constant can become zero, which means that the wave cannot escape anymore from its original region in space. The intensity then typically decays exponentially over one localization length. Yet, in the diffuse regime the wave function extends up to infinity. There are thus two different phases, the diffuse, metallic regime and the localized, insulating regime. In three dimensions a phase transition from the extended to the localized state can occur. In one and two dimensions the states are almost always localized. Yet for a finite sample the localization length can be much larger than the system size, in that case the states appear to be extended and the conductance does not vanish. The localized state is solely the result of the interference of the waves scattered by the disorder.[1]

The precise behavior near the transition is of particular interest and still not fully understood. The standard diagrammatic technique, which we will use, works well for the description of diffusion and the first corrections. It is, however, not suited for the study of the transition. Therefore, various techniques were developed to study the behavior near the transition and the phase transition itself. We mention the most important ones.

An important step was the scaling theory of localization put forward in 1979 by Abrahams *et al.* [12]. It states that near the localization the only parameter of importance is the dimensionless conductance $g$ (which is the conductance measured in units of $e^2/h$). The scaling of $g$ as a function of sample size was studied earlier by Thouless [13, 14] and Wegner [15]. Abrahams *et al.* extended those ideas and derived the renormalization group equations. They concluded that in one and two dimensions there is no real phase transition; the states are always localized. Only in three dimensions a phase transition can occur. The states far from the band center are localized, whereas the states near the band center with high energy are extended. A change in the Fermi energy, for instance by doping, can thus cause the transition from a metal to an insulator.

It is well known that classical diffusive transport is described by the *ladder diagrams*. The so called *most crossed diagrams* are the next most important diagrams in the diffuse regime, they describe the leading interference terms. By self-consistently summing the ladder diagrams and the most crossed ones in the diffuse regime, Vollhardt and Wölfle found a transition to an insulating phase [16, 17, 18]. The most crossed diagrams represent loops of the intensity (as we will discuss below), and the approach can be looked upon as a self-consistent one loop resummation. Although a self-consistent approach will certainly not include all diagrams of higher order, the method works fine even near the transition, because the first few higher order correction terms vanish. Kroha preformed numerical calculations for the transition on a lattice using the self-consistent technique [19, 20]. Good agreement between the self-consistent theory and numerical simulations is found, confirming that the theory contains the most important diagrams, even relatively close to the transition.

---

[1]It would be wrong to think that this is the only scenario for a metal-insulator transition in electron systems. In fact, electron systems allow for a whole range of possible transitions between the conducting and insulating regimes. Due to the band structure, a maximally filled band can cause insulation, while all electron states remain extended. Another mechanism in electron systems is the Mott-Hubbard transition, where the electron-electron repulsion hinders the electron hopping, see Mott [11].

It is also possible to perform an exact average over the disorder within a field theoretic approach. The resulting non-linear action is not very useful. Yet one can subsequently integrate over the fast fluctuations and consider only the "slow" variables of the system, i.e. variables with small spatial variation. This yields the so called *non-linear sigma model* [21, 22, 23, 24], or even a super-string model [25]. With this action it is possible to generate systematically all corrections to the diffusion process (in $2 + \epsilon$ dimensions), thus allowing for an explicit foundation of the scaling theory. The same approach can also be used to generate all scattering diagrams contributing to a certain physical process. However, for more difficult interference processes also this theory becomes quite elaborate, as proving the cancellation of divergent terms has to be done at the level of the standard diagrammatic approach.

A rather recent approach is the *random matrix theory*. The basic assumption here is that the total scattering matrix of the system, as it is very complicated, can be described just by putting random elements respecting the symmetries of the system. Surprisingly this method works well, and just from the assumption of the random ensemble, it predicts many features of the systems correctly and in a simple way. Its is, unfortunately, only applicable to quasi one-dimensional situations, and it also misses some delicate effects (backscatter cone, the naive value of the universal conductance fluctuations, ...) which are present in both experiment and a diagrammatic approach.

Of course all the theoretical results need to be checked. Yet, specially the study of the critical behavior is difficult in experiments: either the system is clean, but the critical region is far away (in optical measurements) or the critical regime is easily reached, but unwanted interactions spoil the situation (in electron systems). Therefore, one also uses numerical data, provided by computer simulations to check the theories, see e.g. relevant articles in Ref. [26].

Although some five years ago there was hope that the Anderson transition might be reached for light scattered on disordered samples, this turned out be based on a misconception. It was observed that in time-resolved transmission measurements, the average transmission time was very long. This was interpreted as an indication that the mean free path was very small. This could mean that the light was close to localization. It turned out, however, that the long transmission times arise since the light spends much time inside the scatterers. This discovery, made by Van Aldaba, Van Tiggelen, Lagendijk and Tip [27], showed that Anderson localization was still out of reach.

A recent development is the construction of so called photonic bandgap materials. In analogy with electron systems, where the lattice parameter is of the order of the Fermi-wavelength, one makes a structure with a lattice parameter comparable to the wavelength of the light ( $\sim 0.5\mu$m). Like in the electronic case, this introduces a band structure. The band-structure for electromagnetic waves is more complicated, as the vector nature comes into play. Still there are some structures (although less numerous than for scalar waves) that exhibit a bandgap [28]. A small distortion of the gap by an impurity could introduce a localized state. This would then be a nice playing ground for studying atoms, which can no longer decay at a certain frequency. It probably also allows for technical applications such as very efficient light emitting diodes (LEDs)[29].

## 1.4  Not quite Anderson localization

As the localized state is hard to reach for scattered light, we will concentrate on pre-cursors of localization. As stated, the simplest leading process is given by the so called ladder diagram or diffuson. It brings the diffuse transport of intensity. In the well understood "classical" or "Boltzmann" regime, one only considers these processes. The situation becomes more interesting if scattering is stronger, and less probable processes start to play a role. In the case of very strong scattering, these interference processes will finally lead to localization, yet already in the diffuse regime they can be detected. Such effects are also called mesoscopic effects. One can roughly distinguish two types:
1) The interference of the intensity with itself, thus creating loops. This leads to cor-rection effects such as negative magneto-resistance in electronic systems [30, 31], and the corrections to the diffusion constant. This is also known as *weak localization*. In the optical case such effects are often too small to be seen: their relative contribution is small and, in contrast to the electronic case, the magnetic field couples only very weakly. Nevertheless, in reflection measurements these processes can be seen as the enhanced backscatter cone.
2) Another possibility is interference of one intensity or light beam with another one, as occurs, for instance, in the correlation of two transmitted beams. In this work we concentrate on such processes. They are responsible for the observed correlations and the non-trivial distribution functions.

There are various books on localization and mesoscopics that, though mainly con-centrating on electron systems, are also of interest here. We mention the proceedings Refs. [5, 26, 32, 33, 34]. Classical waves are discussed particularly in Refs. [35, 29]

## 1.5  Outline of this thesis

In this work we focus on interference effects that occur in the diffuse transport of light. In particular, we study the correlation of two or more light beams traversing the sam-ple. They can have a slightly different frequency and/or different incoming angle. The intensities will turn out to be correlated much stronger than one would expect from simple classical diffusion arguments. As a result, the intensities do not propagate fully independently, but more collectively, causing large fluctuations and yielding interesting distribution functions. These effects become of order unity near the localization tran-sition. Although our system is far from localization and most effects are quite weak, the experimental detection turns out to be very well possible. We do not focus on the transition to the localized state itself or the associated higher order loop effects, instead we perform a diagrammatic expansion in the diffuse regime only including the leading processes.

In the following chapter we study the diffuse transport of light. The basic results presented there are long known. We, however, include the precise treatment of internal reflection and absorption, important in experiments. We also investigate the possibility of locating objects hidden in diffusely scattering media. This is relevant for medical applications in order to detect, for instance, tumors.

In chapter 3 we go beyond the diffusion approach. We introduce the so called Hikami vertex, which describes how two diffuse intensities can interfere with each other. Whereas diffuse transport corresponds to long-range processes, this interference happens on short distance scales, typically one mean free path. The number of interference vertices will act as an expansion parameter in the theory. The effects rapidly become smaller if the number of interference vertices increases. In the following three chapters we do explicit diagrammatic calculations with maximally two of these vertices. In chapters 4 and 5 we study the correlation between two diffuse beams. As mentioned, due to the interfence effects two diffuse beams do not propagate independently through a diffusely scattering medium, but have some correlation. Which correlation process is leading, depends on the transmission quantity considered. We distinguish three correlation functions that can be observed in three different measurements: 1) correlations in the speckle pattern, 2) correlations in the total transmission, and 3) the correlations in the conductance of the sample. The last are for electron systems also known as the *universal conductance fluctuations*. In chapter 6 we extend our calculation of the total transmission correlation to the correlation between three beams, or, equivalently, to the third cumulant of the total transmission. Although for optical systems this is a weak effect, it has been measured recently. In chapter 7 we extend our analysis to arbitrarily many beams. This allows us to reconstruct the distribution of intenstities on the outgoing side for both the angular resolved transmission and the total transmission.

Many effects calculated and discussed in this thesis were first observed in experiments. Often we find remarkable agreement with the experiments. On the one hand, this is due to the relatively "clean" experimental situation, and the high quality of the experimental data. On the other hand, the close collaboration with the experimentalists at the University of Amsterdam has been very fruitful, as it enabled us to focus on experimentally relevant situations and to incorporate the dominant corrections.

# 2

# Diffuse Light

In this chapter we introduce a general setting for basic aspects of diffuse transport of light. We introduce the amplitude Green's function, the averaging over disorder, the diffuse intensity and the Schwarzschild-Milne equation or radiative transport equation for multiple scattered light. The content of this section is known from diverse works in the literature. This is also the chapter where we define our system and introduce most approximations.

## 2.1 Mesoscopics

We consider a slab of thickness $L$, and width $W$. The slab boundaries are at $z = 0$ and $z = L$ (this choice of coordinates reflects the astrophysical history of the problem). Most of the time we deal with a three dimensional slab, such that the cross-sectional area is $A = W \times W$. Yet we can also consider a quasi one or quasi two-dimensional slab (then the width is much smaller than the thickness, yet much larger than the mean free path). The slab contains randomly placed scatterers. The scatterers in the slab are often static, as is the case for dry white paint samples. It is also possible to suspend the scatterers in a liquid. For the short time that the light needs to pass through the sample, the scatterers can be considered as fixed at random positions. However, on the time scale of, say, milliseconds or seconds, they change their position through motion in the fluid. Integrating the signals over a longer time thus probes different sample configurations, and an ensemble average is easily obtained.

The scatterers used in the experiments are often $TiO_2$ particles or latex spheres (both are a main ingredient of white paint). For simplicity we model the individual scatterers by point scatterers, that scatter isotropically, i.e. they scatter equally strong in all directions. We consider a low density of these point scatterers and assume that their positions are uncorrelated. These are very crude approximations indeed. Firstly, the density of scatterers is not low in the experiments, as in order to reduce the mean free path, the packing fraction in the experiments is often around 40%. The isotropic scattering approximation originates from studies on electronic systems, where scattering is dominated by $s$-wave scattering. However, isotropic scatterers do not exist for vector waves and small scatterers are Rayleigh scatterers. Yet this is also not appropriate for the experiment, as the diameter of the scatterers is somewhere about the wavelength of the light. This choice of diameter again reduces the mean free path, as the scatterers are now resonant and scatter very strongly. Moreover, the shape of the $TiO_2$ particles is irregular (potato-like). Clearly the scattering is not isotropic and hard to model at all. Nevertheless, for many aspects the precise individual scattering process is not so important. In practice, the effects of realistic scattering are usually coded in an experimentally determined transport mean free path that replaces the scattering mean free path in the theoretical results for isotropic scatterers. Other theoretical parameters are left unchanged. Therefore, we do not distinguish between the transport mean free path and the scattering mean free path. Indeed, experimental comparison under this assumption, when available, is good. Only recently this assumption has also been founded theoretically [36].

The diffuse character of the light is guaranteed if the thickness is much larger than the mean free path $\ell$, so that on the average many scattering events occur if the light traverses the slab. Such a situation is termed *optically thick*. If the slab is (optically) thin, the system is no longer in the diffuse regime but in the *ballistic* regime. An expansion in the number of scattering events becomes then more appropriate than the diffusion approximation that we use here. A hazy air or a piece of smoked glass are good examples of systems in the ballistic regime.

The mean free path in its turn is supposed to be much longer than the wavelength of the light. This makes it possible to speak of a light ray between the scatterers. If, instead, the mean free path and the wavelength become of comparable scale, localization is expected to set in. This is the Ioffe-Regel criterion, which states that localization occurs if [37, 38]

$$\ell \leq \frac{\lambda}{2\pi}, \tag{2.1}$$

where $\lambda$ is the wavelength of the light. However, in this work we are far from localization and the parameter $1/(k\ell)$ is small, where we introduced the wave-vector $k = 2\pi/\lambda$. In the experiments we typically have $1/(k\ell) \sim 0.01$.

We assume that the spectrum of the light, usually a narrow laser band, is not changed by the scattering. (In principle, there is a Doppler shift if the scatterers are moving, but this is in our case negligible.) Furthermore, we allow for weak absorption, the decay length due to absorption, $L_{abs}$, is assumed to be much larger than the mean free path. We assume that absorbed light really has disappeared from our observation. It can, for instance, be emitted at a wavelength that is not detected by the detector or be converted into heat.

The *mesoscopic, diffuse* regime is in summarized in the inequalities

$$\lambda \ll \ell \ll L, L_{abs}. \tag{2.2}$$

For electron systems there is an extra condition that $L \ll L_{inc}$, where $L_{inc}$ is the incoherence length. This incoherence length is somewhat comparable to the absorption length. The phase coherence is lost beyond this length scale, but in contrast to absorbed photons, the de-phased electrons still contribute to the transport. If $L \gg L_{inc}$, the interesting interference effects will average out. Basically we are now dealing with particle diffusion, see Kaveh [39], who also discusses some other interesting variations of the inequalities (2.2).

Furthermore, in practice there is often a different index of refraction inside the sample and the outside. This causes diffuse intensity propagating outwards to be reflected at the surface, changing the intensity as a function of depth. We do incorporate a refractive index mismatch in our approach.

## 2.2 Amplitude properties

In this section we introduce the wave equation and its Green's functions, which we will need for the description of intensity transport. The Green's function also forms the basis for our subsequent diagrammatic expansion in the next chapters. The techniques employed are well known and are, for instance, treated in the books of Economou [40], Mahan [41] and Abrikosov, Gorkov, and Dzyaloshinski [42].

### 2.2.1 Wave equations

The Maxwell equations inside a medium are generally given by [43]

$$\nabla \cdot \mathbf{D} = 4\pi \rho_f, \quad \nabla \times \mathbf{H} = \frac{1}{c}\frac{\partial \mathbf{D}}{\partial t} + \frac{4\pi}{c}\mathbf{j}_f,$$

$$\nabla \cdot \mathbf{B} = 0, \quad \nabla \times \mathbf{E} = -\frac{1}{c}\frac{\partial \mathbf{B}}{\partial t}.$$

As the medium is not conducting, the free charge $\rho_f$ and free current $\mathbf{j}_f$ vanish. Since we only deal with nonmagnetic materials, we take $\mu = 1$ and we find

$$-\nabla^2 \mathbf{E} = -\frac{\epsilon}{c^2}\frac{\partial^2 \mathbf{E}}{\partial^2 t}. \tag{2.3}$$

It looks as if the components of $\mathbf{E}$ are now decoupled, which would allow for studying them independently. Yet they are still coupled, as we still have $\nabla \cdot \mathbf{D} = 0$. Nevertheless we make a scalar approximation for the electromagnetic field of the light. For bulk quantities this is justified since the initial polarization is scrambled after a few scattering events. The precise consequence of this approximation on the possibility of localization [44] and on skin-layer effects is still unknown. For other applications, such as acoustic waves and spinless electrons (that applies to electron scattering for which the spin direction is conserved), the scalar property is immediate.

The time-independent scalar wave equation at a frequency $\omega$ reads

$$\nabla^2 \psi(\mathbf{r}) + \frac{\omega^2}{c^2}\epsilon(\mathbf{r})\psi(\mathbf{r}) = 0, \tag{2.4}$$

where $c$ is the vacuum speed of light. We shall consider a slab with dielectric constant $\epsilon_0$ and density $n$ of small spheres with dielectric constant $\epsilon_2$ and radius $a_0$, located at random positions $\mathbf{r}_i$. In the limit of point scatterers we obtain

$$\epsilon(\mathbf{r}) = \epsilon_0 + \frac{4}{3}\pi a_0^3(\epsilon_2 - \epsilon_0)\sum_i \delta(\mathbf{r} - \mathbf{r}_i), \quad (0 < z < L), \tag{2.5}$$

$$= \epsilon_1, \quad (z < 0; \ z > L), \tag{2.6}$$

where $\epsilon_1$ is the dielectric constant in the surrounding medium, for instance the glass container or air. The vacuum value of the dielectric constant is unity.[1] The wave

---

[1] The average index of refraction is caused by atoms acting like small dipoles. They change the dielectric constant but cause no scattering.

numbers in the surrounding medium and in the random medium are, respectively,

$$k_1 = \frac{\omega}{c}\sqrt{\epsilon_1}, \qquad k = mk_1 \equiv \sqrt{\frac{\epsilon_0}{\epsilon_1}}k_1, \tag{2.7}$$

respectively. The relative index of refraction $m$ relates to the dielectric constants as $m = \sqrt{\epsilon_0/\epsilon_1}$.

Note that the wave equation resembles the Schrödinger equation for electrons in a random potential

$$-\frac{\hbar^2}{2m}\nabla^2\Psi(\mathbf{r}) + [V(\mathbf{r}) - E]\Psi(\mathbf{r}) = 0. \tag{2.8}$$

Note that in Eq. (2.4) the frequency enters into the potential term. If one considers time independent problems the equations and their solutions are similar. This similarity is the reason for the analogy between light scattering and electron systems. However, differences show up in the time evolutions of both systems.

### 2.2.2   Propagators

In our diagrammatic approach we first need the bare Green's function $G_0$. It is defined as the "inverse" of the wave operator in a medium without scatterers

$$(-\nabla^2 - k^2)G_0(\mathbf{r}, \mathbf{r}') = \delta(\mathbf{r} - \mathbf{r}'). \tag{2.9}$$

For later use in the bulk we have neglected boundary effects, or, equivalently, we set $\epsilon_0 = \epsilon_1$. As then the scatterer-free problem is translationally invariant, the Green's function is easily found in the momentum representation as

$$G_0(\mathbf{p}) = \left[\mathbf{p}^2 - k^2 - i0\right]^{-1}. \tag{2.10}$$

We call this the retarded Green's function; likewise, $G_0^*$ is called the advanced Green's function. The bare Green's function describes the propagation of an amplitude through a non-scattering medium. It has a pole at $p = k + i0$. The real-space function only depends on $|\mathbf{r} - \mathbf{r}'|$, by Fourier transformation[2] we find

$$G_0(\mathbf{r}, \mathbf{r}') = \int \frac{d^3\mathbf{p}}{(2\pi)^3}\frac{e^{i\mathbf{p}\cdot(\mathbf{r}-\mathbf{r}')}}{\mathbf{p}^2 - k^2 - i0} = \frac{e^{ik|\mathbf{r}-\mathbf{r}'|}}{4\pi|\mathbf{r}-\mathbf{r}'|}, \tag{2.11}$$

which indeed obeys Eq. (2.9). The $1/r$ decay is just a geometrical factor describing spreading in the 3d space. Apart from this there is no decay.

### 2.2.3   The $t-$matrix

We now include scattering by introducing first one single scatterer. The propagation between two points in space can now occur with an arbitrary number of scattering events from this single scatterer. This we depicted in Fig. 2.1. The sum of this series is called

---

[2]In this work we define the Fourier transform as $f(\mathbf{r}) = \int \frac{d^3\mathbf{p}}{(2\pi)^3}f(\mathbf{p})\exp(i\mathbf{p}\cdot\mathbf{r})$, $f(\mathbf{p}) = \int d^3\mathbf{p}f(\mathbf{r})\exp(-i\mathbf{p}\cdot\mathbf{r})$.

Figure 2.1: The propagation in the presence of one scatterer. The dressed Green's function $G$ contains an arbitrary number of scatterings from this scatterer.

the dressed Green's function $G$. In real-space this diagram corresponds to

$$
\begin{aligned}
G(\mathbf{r}, \mathbf{r}') &= G_0(\mathbf{r}, \mathbf{r}') + \int d\mathbf{r}_1 G_0(\mathbf{r}, \mathbf{r}_1) V(\mathbf{r}_1) G_0(\mathbf{r}_1, \mathbf{r}') \\
&+ \int d\mathbf{r}_1 \int d\mathbf{r}_2 G_0(\mathbf{r}, \mathbf{r}_1) V(\mathbf{r}_1) G_0(\mathbf{r}_1, \mathbf{r}_2) V(\mathbf{r}_2) G_0(\mathbf{r}_2, \mathbf{r}') + \dots \quad (2.12)
\end{aligned}
$$

Already in 1908 Mie has derived the exact solution for scattering of vector waves from spherical particles of arbitrary size [45], but these results are too involved for our calculations. We rather model the scatterers with the isotropic point scatterers, thus we have the much simpler form $V(\mathbf{r}) = \delta(\mathbf{r} - \mathbf{r}_0) V$. For our problem the bare scattering strength reads

$$
V = 4\pi k^2 a_0^3 (\epsilon_2/\epsilon_0 - 1)/3. \quad (2.13)
$$

Thus

$$
\begin{aligned}
G(\mathbf{r}, \mathbf{r}') &= G_0(\mathbf{r}, \mathbf{r}') + G_0(\mathbf{r}, \mathbf{r}_0) V G_0(\mathbf{r}_0, \mathbf{r}') + G_0(\mathbf{r}, \mathbf{r}_0) V G_0(\mathbf{r}_0, \mathbf{r}_0) V G_0(\mathbf{r}_0, \mathbf{r}') + \dots \\
&\equiv G_0(\mathbf{r}, \mathbf{r}') + G_0(\mathbf{r}, \mathbf{r}_0)\, t\, G_0(\mathbf{r}_0, \mathbf{r}'),
\end{aligned}
$$

where the $t-$matrix is defined as the sum of all possible scatterings from a single scatterer. Hence [3]

$$
t = V + V G_0 V + V G_0 V G_0 V + \dots. \quad (2.14)
$$

This series is the *Born series*. For disordered electron systems one often deals with weak scattering and the diagrammatic expansion can be carried out in second order Born approximation, i.e. $t = V + V G_0 V$. The real part of the return Green's function diverges in three dimensions, since

$$
G_0(\mathbf{r}, \mathbf{r}') \approx \frac{1}{4\pi |\mathbf{r} - \mathbf{r}'|} + \frac{ik}{4\pi} + O(|\mathbf{r} - \mathbf{r}'|). \quad (2.15)
$$

As a result, the point scatterer approximation is unphysical. A cut-off length could be introduced to solve this problem. As in the second order Born approximation the renormalized expression is multiplied by the small factor $V^2$, it then leads to a result that is much smaller than the linear term in $V$, so the real part of $G$ can just as well simply be left out. Yet, the imaginary part should be kept as it describes the scattering, resulting in

$$
t \overset{2^e\,\text{Born}}{=} V + i\frac{V^2 k}{4\pi}. \quad (2.16)
$$

---

[3] In quantum mechanics textbooks $t_{\mathbf{pp}'}$ denotes the scattering amplitude for scattering of a plane wave with momentum $\mathbf{p}$ into one with momentum $\mathbf{p}'$. Here we can work in coordinate space representation. The effective (re-summed) scattering matrix $\delta(r)\delta(r') t$ is then diagonal, and yields $t_{\mathbf{pp}'} = t$. Though, strictly speaking, $t$ is only the value of the matrix elements, it is commonly again called "t-matrix".

In experiments on optical systems, however, efficient scattering is achieved by taking resonant scatterers. This means that higher order terms in the Born series cannot be neglected and one has to calculate the full Born series. By summing Eq. (2.14) we find

$$t = \frac{V}{1 - V G_0(\mathbf{r}_0, \mathbf{r}_0)}, \tag{2.17}$$

for which we could use the same cut-off procedure, i.e. omitting the divergent real part of $G_0$. However, physically this is not a well motivated step. Moreover, it would be nice to keep the resonant properties. This can be done by taking the point scatterer limit, while keeping the total scattering strength constant. Here we follow Nieuwenhuizen *et al.* [46]. We introduce a cut-off, corresponding physically to the inverse scatterer diameter. The regulated return Green's function is denoted $\tilde{G}_0$ and reads

$$\tilde{G}_0(\mathbf{r}, \mathbf{r}) = \int_{|\mathbf{p}| < 1/a_0} \frac{\mathrm{d}^3 \mathbf{p}}{(2\pi)^3} \frac{\mathrm{e}^{i\mathbf{p} \cdot \mathbf{r}}}{\mathbf{p}^2 - k^2 - i0} = \frac{1}{2\pi^2 a_0} + \frac{ik}{4\pi}. \tag{2.18}$$

This yields for the $t$-matrix

$$t = \frac{V}{1 - \frac{V}{2\pi^2 a_0} - i\frac{Vk}{4\pi}}. \tag{2.19}$$

Due to the $k-$dependence of $V$ this $t-$matrix still contains one resonance, occurring if the real part in the denominator of Eq. (2.19) vanishes, or $k^2 = 3\pi/[2a_0^2(\epsilon_2/\epsilon_0 - 1)]$. Note that in electronic systems $V$ is $k$-independent, and the $t-$ matrix has no resonance.

### 2.2.4   The self-energy

Let us now consider the presence of many identical scatterers placed at random positions. The potential now becomes a sum of point scatterers. Connecting scattering events on the same scatterer with a dashed line we obtain Fig. 2.2. This brings for the dressed Green's function

$$\begin{aligned} G &= G_0 + G_0 \Sigma G_0 + G_0 \Sigma G_0 \Sigma G_0 + \dots \\ &= \left( G_0^{-1} - \Sigma \right)^{-1} = \left( -\nabla^2 - k^2 - \Sigma \right)^{-1}. \end{aligned} \tag{2.20}$$

Before averaging over the locations $\mathbf{r}_i$ of the $N$ scatterers we have, to leading order in their density, $\Sigma(\mathbf{r}) = \sum_{i=1}^{N} \delta(\mathbf{r} - \mathbf{r}_i)t$. The average over the scatterer positions $\mathbf{r}_i$ then yields the space-independent result,

$$\Sigma = nt, \tag{2.21}$$

where $n = N/V = N/(AL)$ is the scatterer density. We will use this approximation for the self-energy, also known as the *independent scatterer* approximation, throughout this work. We consider no recurrent scattering. In recurrent scattering the amplitude returns to a previous scatterer after scattering from another one, such as the rightmost self-energy diagram in Fig. 2.2, see for a treatment of such processes Van Tiggelen *et al.* [47]. This process indeed seems improbable in the diffuse regime. Another improvement on our approximation would be the self-consistent calculation of the $t$−matrix. This leads to a rich behavior in which band-gaps in the density of states can appear [48][4].

Figure 2.2: First five lines depict the average Green function (thick line) in a medium with many scatterers. The thin lines are the bare Green's functions $G_0$. We depict the scattering potentials by crosses. The dashed lines indicates that the scattering occurs from the same scatterer. Next two lines define the self energy. The circles represent the $t-$matrices of the different scatterers. The self-energy $\Sigma$ contains irreducible diagrams only. In our approximation the self-energy contains only one $t-$matrix; the rightmost diagram for the self-energy, corresponding to recurrent scattering, is thus neglected. The bottom line gives a general expression for $G$ in terms of $\Sigma$.

For our purpose such refinements would complicate the problems too much. The problem is that in order to respect conservation laws a more detailed approach on this amplitude level also has to be made on intensity level, i.e. for two connected amplitude propagators, and also on vertex level (four or more propagators). This would be horrible because the number of diagrams then increases dramatically. Thus we stick to our simple approximation of the self-energy. Later we will see some examples of errors that occur if different approximations are made on different levels. These errors are a manifestation of the breaking of conservation laws.

In the bulk (i.e. the part not under the influence of the boundaries) the average or dressed retarded Green's function thus reads

$$
\begin{aligned}
G(\mathbf{p}, \mathbf{p}') &= G(\mathbf{p})\delta_{\mathbf{p},\mathbf{p}'} \\
G(\mathbf{p}) &= \left(\mathbf{p}^2 - k^2 - nt\right)^{-1}.
\end{aligned}
\tag{2.22}
$$

Compared to the bare Green's function, the pole of this dressed Green's function, shifts to

$$
p = \sqrt{k^2 + nt} \approx [k + n\mathrm{Re}(t)/2k] + i[n\mathrm{Im}(t)/2k].
\tag{2.23}
$$

The real part of the pole shifts a little but this can be neglected, or can be absorbed in the definition of $k$. More important is the small but finite imaginary part. It causes an exponential decay. (This essentially is the difference in the optical properties of milk versus water.) Indeed, the unscattered (or coherent) intensity is given by the absolute value squared of the averaged Green's function. It decays as

$$
G^*(\mathbf{r}, \mathbf{r}')G(\mathbf{r}, \mathbf{r}') = \frac{\mathrm{e}^{-|\mathbf{r}-\mathbf{r}'|n\mathrm{Im}(t)/k}}{16\pi^2|\mathbf{r}-\mathbf{r}'|^2} = \frac{\mathrm{e}^{-|\mathbf{r}-\mathbf{r}'|/\ell_{\mathrm{ex}}}}{16\pi^2|\mathbf{r}-\mathbf{r}'|^2}.
\tag{2.24}
$$

We defined the extinction mean free path as

$$
\ell_{\mathrm{ex}} = \frac{k}{n\mathrm{Im}(t)}.
\tag{2.25}
$$

The extinction mean free path describes the decay of coherent intensity either through scattering or absorption. Indeed, our formalism is capable to describe absorption as well: If the dielectric constant $\epsilon_2$ of the scatterers is complex, the potential $V$ in Eq.(2.13) has an imaginary part, and there will be an extra imaginary contribution to, e.g., Eq. (2.16). The leading absorption comes from the linear term in $V$. The scattering part comes from second and higher order terms in the Born series. Eq. (2.24) is a manifestation of the Lambert-Beer law.

## 2.3 Bulk intensity transport: the diffusion equation

---

[4]The local density of states $\rho(\mathbf{r})$ relates to the dressed amplitude Green's function as $\rho(\mathbf{r}) = \mathrm{Im}G(\mathbf{r}, \mathbf{r})/\pi$. See Refs. [49, 50, 51] and references therein for detailed calculations of $\rho(\mathbf{r})$.

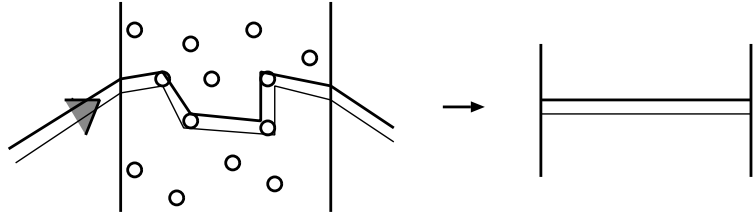

Figure 2.3: Left: an example of an actual scattering process; a retarded (thick line) and an advanced amplitude (thin line) come from the left and share the same scatterers and the same path through the sample. Right: schematic representation of the average process, the diffuson. We will use this representation in the following chapters. The scattering events are not drawn for simplicity.

The intensity of a wave is the square of the absolute value of the amplitude. Squaring the average propagator, one finds the unscattered intensity that decays exponentially over one mean free path, as we saw in Eq. (2.24). In the diffuse regime the mean free path is small as compared to the system sizes and the unscattered beam gives thus only a small contribution to the transport. Examples of the coherent part in diffuse scattering in daily life are the weak sun to be distinguished in the fog, or the contours of the filament inside an opaque bulb. In the rest of this work we will neglect the small coherent contribution to the transmission.

A different process is responsible for diffuse transport. Intensity transport in transmission is dominated by ladder diagrams or diffusons, which can only be obtained by averaging *after* taking the product of $G$ with $G^*$. A diffuson is made up by pairing one retarded and one advanced propagator that visit the same scatterers in the same chronological order. In other words, the diffusion of intensity corresponds to an amplitude and its complex conjugate following the same path along the scatterers. We will see that in the idealized case no phase difference between the propagators builds up, such that averaging does not suppress this process. [5] The diffuson diagram is depicted in Fig. 2.3 and Fig. 2.4. The diffuson is built of the *ladder kernels* tied after each other. The kernel contains $G$, $G^*$, and one common scatterer. We will define the diffuson below to start and end on a scatterer.

The consistent choice of the intensity vertex follows from the intensity conservation. Conservation laws are in a field-theoretic approach given by the Ward identities. They are the field-theoretic extensions of the Noether currents in classical mechanics. The Ward identity for the intensity conservation, as derived by Vollhardt and Wölfle in this context [17], says that the irreducible intensity vertices are generated by cutting the propagators in the self-energy diagrams and then flipping one part down. This is the way we constructed the upper line of Fig. 2.4. It is then clear that taking only a *first* order Born approximation in the self-energy ($\Sigma = nV$), would not yield diffusons at all (there would be no propagator to cut). In the second order Born approximation the scatterers are described by "bare potentials", leading to a vertex factor $V\bar{V}$. For the full Born series the vertex factor consists of two $t-$matrices, i.e., the product $t\bar{t}$, as illustrated in Fig. 2.4.

---

[5]Similarity to meson physics, where quark and anti-quark are paired together on long distances, was noted by Prof. Dr. J. Smit.

Although the scattering mean free path is incorporated in the diffuson equation, it is useful to define it on forehand. In second order Born approximation it equals $\ell = 4\pi/(n|V|^2)$. For the full Born series we define it as

$$\ell = \frac{4\pi}{n t \bar{t}}. \tag{2.26}$$

The scattering mean free path $\ell$, in contrast to the extinction mean free path $\ell_{ex}$, comes solely from scattering and not from absorption. The two relate as $\ell_{ex} = a\ell$, where $a$ is the albedo of the single scatterer. With the definition of the $t-$matrix Eq. (2.19) we see that

$$a = \left[1 - \frac{4\pi}{k}\mathrm{Im}(1/V)\right]^{-1}. \tag{2.27}$$

If the scatterer-potential is real, there is no absorption, corresponding to $a = 1$, this is also called *conservative* scattering for obvious reasons. The case $a = 0$ means that the scatterer absorbs all light that hits it. Equivalently we can derive the relation

$$\frac{k t \bar{t}}{4\pi} = a\mathrm{Im}(t). \tag{2.28}$$

For $a = 1$, which corresponds to no absorption, this is also known as the *optical theorem*. It says that intensity lost in the decay of the coherent beam is seen back in the scattered intensity. If $a < 1$ it is partly absorbed.

The diffuson is the sum of processes with one, two, three, etc., common scattering events. We now calculate the building block of such processes: the ladder kernel. It describes one single common scattering event. In this section we assume a thick slab and use the diffusion approximation for the boundary behavior. In Fig. 2.4 we show part of the diffuson diagram. The ladder kernel $U$ corresponds to the expression

$$U = n t \bar{t} \int \frac{\mathrm{d}^3 \mathbf{p}}{(2\pi)^3} G(\mathbf{p} + \mathbf{q}; \omega + \frac{1}{2}\Delta\omega) G^*(\mathbf{p}; \omega - \frac{1}{2}\Delta\omega) \tag{2.29}$$

Here we included the momentum difference $\mathbf{q}$ and the frequency difference $\Delta\omega$. After Fourier transformation they give the spatial and the time dependence of the diffuson. In the diffusion approximation variations of the intensity are small. This corresponds to expanding the Green's in small momenta and frequency difference

$$\begin{aligned} G(\mathbf{p} + \mathbf{q}; \omega + \frac{1}{2}\Delta\omega) &\approx \left[p^2 + 2\mathbf{p} \cdot \mathbf{q} + q^2 - \frac{\omega^2}{c^2} - \Delta\omega\frac{k}{c} - nt\right]^{-1} \\ &\approx G + (-2\mathbf{p} \cdot \mathbf{q} - q^2 + \Delta\omega\frac{k}{c})G^2 + 4(\mathbf{p} \cdot \mathbf{q})^2 G^3 \end{aligned} \tag{2.30}$$

with $G = G(\mathbf{p}; \omega)$. We define the "standard" integrals

$$I_{kl} = \int \frac{\mathrm{d}^3 \mathbf{p}}{(2\pi)^3} G^k(\mathbf{p}) G^{*l}(\mathbf{p}). \tag{2.31}$$

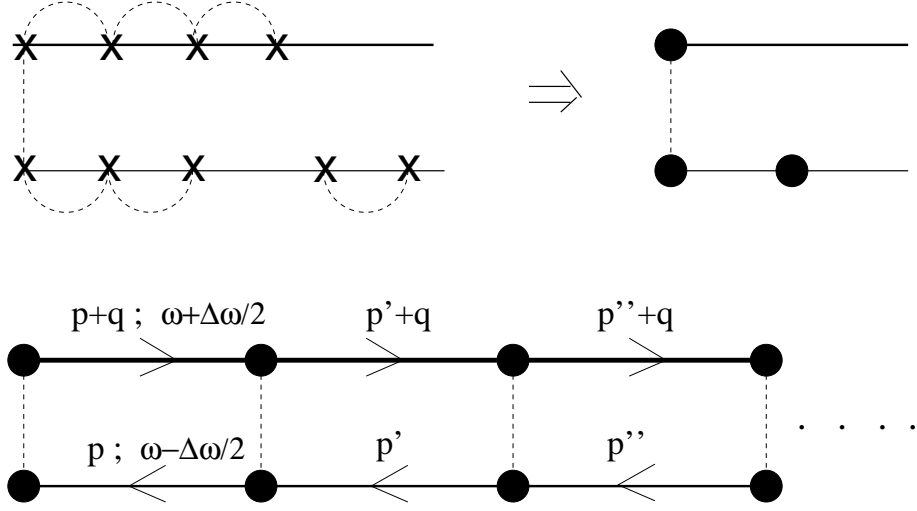

Figure 2.4: Detailed picture of a part of the diffuson. On the top line we have drawn one possible scattering contribution to the ladder kernel. We depicted the scattering potential with 'x'. By its definition the full $t$−matrix, the circle, contains all such repeated scattering events. The Green's functions are bare in the upper line. The bottom line is part of the product of ladder kernels. The sum of diagrams with an arbitrary number of kernels leads to the full ladder or diffuson. Note that the Green's function are dressed now.

We calculate these integrals in section 3.4; their pole at $p \approx k$ dominates them. We expand in second order of $\mathbf{q}$. Due to symmetry of the integral, terms linear in $\mathbf{q}$ cancel, and we are left with

$$U = nt\overline{t}\left(I_{11} + 2\Delta\omega\frac{k}{c}I_{21} - q^2 I_{21} + \frac{4}{3}k^2q^2 I_{31}\right) \tag{2.32}$$

$$= \left(1 - \frac{1}{3}\ell^2(q^2 + \kappa^2 - i\frac{3}{\ell c}\Delta\omega)\right). \tag{2.33}$$

The $q^2 I_{21}$-term was neglected as it is proportional to $1/(k\ell)$. The absorption length $L_{\text{abs}}$ gives the decay length of the intensity due to absorption. However, the inverse absorption length $\kappa = L_{\text{abs}}^{-1}$ is more convenient for us. It relates to the albedo $a$ as

$$\kappa^2\ell^2/3 = 1 - a, \tag{2.34}$$

when absorption is small, i.e. $\kappa\ell \ll 1$, or $a \approx 1$. The geometric sum of the kernel, the ladder-sum, gives the full diffuson

$$\mathcal{L}(\mathbf{q}) = nt\overline{t}\sum_{j=0}^{\infty} U^j = \frac{nt\overline{t}}{1 - U}, \tag{2.35}$$

or, in the approximation of Eq. (2.33),

$$\left[q^2 + \kappa^2 + i\Omega\right]\mathcal{L}(\mathbf{q}) = \frac{12\pi}{\ell^3}. \tag{2.36}$$

Thus $\mathcal{L}$ obeys the well known *diffusion equation*, written here in momentum space. Note that we have multiplied Eq. (2.35) with one extra factor $nt\overline{t}$ to let the ladder end on a scatterer (conform Fig. 2.4). In taking the lowest value $j = 0$, we also include the single scattering event. We have defined the reduced frequency difference $\Omega$ as

$$\Omega = -\frac{\Delta\omega}{D}, \qquad (2.37)$$

with the diffusion constant $D = c\ell/3$.

It was shown that $D = \frac{1}{3}v_E\ell$ instead of $\frac{1}{3}c\ell$, where $v_E$ is transport speed [27, 52, 46]. This energy transport velocity is not derived here. It is the result of the resonant scatterers, which makes the $t-$matrix strongly frequency dependent. If now $G$ and $G^*$ have a somewhat different frequency, they "see" a different $t-$matrix. This results in a much lower diffusion constant in dynamical problems than expected from $D = c\ell/3$. The physical picture is that the light gets trapped inside and close to the resonating scatterer. The discovery of this effect limited the excitement that, concluding from the diffusion constant, the mean free path of light might be close to the wavelength in experiments. According to the Ioffe-Regel criterion, Eq. (2.1), localization should then be within reach. However, the small diffusion constant was the consequence of $v_E$, which can be much smaller than $c$, and the actual mean free path was still large.

Let us return to the diffusion equation 2.36. Let us by $\mathcal{L}_{\text{int}}$ denote the corresponding diffusion propagator from $\mathbf{r}$ to $\mathbf{r}'$. In an infinite medium it obeys

$$[-\nabla^2 + \kappa^2 + i\Omega]\mathcal{L}_{\text{int}}(\mathbf{r}) = \frac{12\pi}{\ell^3}\delta(\mathbf{r} - \mathbf{r}'). \qquad (2.38)$$

In the slab geometry we can perform a Fourier-transformation in the $x, y$-directions, and the diffusons then obey

$$-\frac{d^2}{dz^2}\mathcal{L}_{\text{int}}(z, z'; M) + M^2\mathcal{L}_{\text{int}}(z, z'; M) = \frac{12\pi}{\ell^3}\delta(z - z'). \qquad (2.39)$$

Here we have defined the decay rate

$$M^2 = Q^2 + \kappa^2 + i\Omega, \qquad (2.40)$$

where $Q$ is the 2D transversal momentum[6]. The complex parameter $M$ describes the exponential decay $\mathcal{L} \sim \exp(-\text{Re}(M)z)$ of the diffuse intensity in the $z$-direction. This decay can occur either as a result of absorption, or as a result of de-phasing caused by a frequency difference or a transverse momentum difference. In analogy with the range of the interaction particles of the fundamental forces, we sometimes call this the *mass*. In our previous derivation we assumed that $|M|\ell \ll 1$, also below we will work only in first order of $M\ell$. For most calculations we will first consider the $M = 0$ case, to estimate the magnitude of the effect.

By realizing that the diffusion equation (2.39) is just a wave equation with complex frequency, one sees that the solutions to the diffusion equation are a linear combination of hyperbolic sines and cosines. The solution of the diffusion equation in a slab geometry can also be obtained using the method of "image charges" as known from electrostatic problems. We then fulfill the boundary conditions by summing the original and mirror solutions of the infinite slab. Here, however, that method is more involved. Introducing a yet undetermined "extrapolation length" $z_0$, we can write [53, 54]

$$\mathcal{L}_{\text{int}}(z, z'; M) = \frac{12\pi}{\ell^3} \frac{[\sinh Mz + Mz_0 \cosh Mz][\sinh M(L-z') + Mz_0 \cosh M(L-z')]}{(M + M^3 z_0^2) \sinh ML + 2M^2 z_0 \cosh ML} \tag{2.41}$$

where we assumed that $z < z'$, otherwise $z$ and $z'$ must be interchanged on the r.h.s. The diffuson $\mathcal{L}$ describes the diffuse propagation from one point in the slab to another. With roughly equal indices of refraction inside and outside the sample, the extrapolation length $z_0$ is a couple of mean free paths and thus the terms involving $z_0$ yield contributions of the order $\ell/L$. For optically thick samples ($L \gg \ell$), this is negligible and one has

$$\mathcal{L}_{\text{int}}(z, z'; M) = \frac{12\pi}{\ell^3} \frac{\sinh Mz \ \sinh M(L-z')}{M \sinh ML}. \tag{2.42}$$

---

[6]We denote 2D vector with capitals, 3D vectors with small bold faces

Consistent with this expression the boundary conditions become in that case $\mathcal{L}_{\text{int}}(0, z') = 0$, $\mathcal{L}_{\text{int}}(z, L) = 0$.

The case of amplifying media is also of some practical interest. The diffuse intensity then increases at each scattering (this can be done by pumping the sample with light of another frequency). The albedo is now larger than one, or, equivalently, $\kappa$ becomes imaginary. If $Q = 0$ and $\Omega = 0$, the diffuse intensity has a negative curvature and consists of sines and cosines instead of hyperbolic functions. On the top of the enhanced backscatter cone (see below) there appears an extra sharp peak [55, 56].

### 2.3.1 The incoming and outgoing beams in the diffusion approach

The diffusion equation does not hold if the intensity gradient is steep, i.e. $q\ell \sim 1$. It can therefore not describe properly how the intensity couples into and out of the medium. Nevertheless, in a heuristic manner one makes often the following assumptions in the diffusion approach, see Ishimaru [57]:

- The diffuse intensity from an outside plane wave source is assumed to be given by substituting $z = \ell$ in Eq. (2.41). As if all diffuse intensity originates from this $z = \ell$ plane. Likewise, the coupling outwards is obtained by taking $z' = L - \ell$. In this way two extra propagators, the incoming: $z < 0, 0 < z' < L$; and the outgoing $0 < z < L, z' > L$ are constructed from the internal diffuson ($0 < z, z' < L$).

- The boundary conditions are [54]

$$\left. \frac{\mathcal{L}(z, z'; M)}{\partial_z \mathcal{L}(z, z'; M)} \right|_{z=0+, L-} = z_0. \tag{2.43}$$

Solution (2.41) fulfills these conditions.

The extrapolation length $z_0$ is determined by the reflectance at the surface. In the next section we show how we do this properly. Here we treat the diffusion approach. With non-reflecting boundaries the boundary condition on the diffuse intensity on the surface is the absence of diffuse intensity propagating into the medium. Physically this is immediately clear, as there can be no diffuse source outside the medium. In the index matched case $z_0$ is $2\ell/3$. This well known result is, for instance, derived in the book of Ishimaru [57].

However, in realistic systems there can be internal reflection. One of the first treatments of internal reflection in the diffusion approach was given by Lagendijk, Vreeker and De Vries [58]. Here we briefly summarize the result of Zhu, Pine and Weitz [54]. The reflection of a scalar wave on an index mismatched boundary is [59] (define $\mu$ as the cosine of the angle of the wave to the $z-$axis)

$$R(\mu) = \left( \frac{\mu - \sqrt{\mu^2 - 1 + 1/m^2}}{\mu + \sqrt{\mu^2 - 1 + 1/m^2}} \right)^2, \qquad (2.44)$$

where $m = n_0/n_1$ is again the ratio of the refractive index of the sample, $n_0$, and it's surrounding, $n_1$. Considering the current of diffuse intensity at the boundary of the slab, one finds [54]

$$z_0 = \frac{2}{3} \ell \frac{1 + 3R_2}{1 - 2R_1} \qquad (2.45)$$

where the angular averaged reflectance for the intensity and current, respectively, are

$$R_1 = \int_0^1 \mathrm{d}\mu \; \mu R(\mu) \qquad R_2 = \int_0^1 \mathrm{d}\mu \; \mu^2 R(\mu). \qquad (2.46)$$

If there is no internal reflection, one has $m = 1$ and $R(\mu) = 0$, now $R_1$ and $R_2$ vanish and one retrieves $z_0 = 2/3\ell$. For $m \neq 1$ the extrapolation length increases. If the index mismatch gets larger and larger, the gradient of the intensity near the surface will be less steep. Then the diffusion approach, which is good for the bulk, works also fine near the surface. One can also say: if the boundary strongly reflects, it is invisible for the diffuse light from inside the slab, since the diffuse intensity inside the slab is everywhere the same. Therefore, the bulk behavior applies everywhere. The correctness of the diffusion approach for this case was confirmed on an analytical basis in Ref. [59].

## 2.4   Radiative transfer

Although the diffusion equation describes the transport deep in the bulk accurately, it breaks down near the surface as the intensity is not longer slowly varying as function of $z$. Nor is the angular average representative, as the incoming waves have undergone only few scatterings. The precise behavior near the (reflecting) surface has to be derived from the Schwarzschild-Milne integral-equation. This equation has been known in the literature for a long time, especially in the astrophysics community [2, 1]. Particularly in the book of Van de Hulst the diffuse intensity is studied in this approach for very many situations, including absorption, the $\ell \approx L$ regime, and anisotropic scattering. Yet in those works the average indices of refraction of the scattering medium and its surrounding are assumed to match. This is fully justified for interstellar and meteorological clouds, which are very dilute and hardly refract. (The change in the imaginary part of the dielectric constant in going from one medium to another is important for describing the scattering, but in those systems the change in the real part is negligible.) Nieuwenhuizen and Luck [59] extended the transport equation to the case where there is a mismatch between the index of refraction in the scattering medium and outside, causing internal reflection. We first consider the $M = 0$ case, see Eq. (2.40).

For the moment we take our slab semi-infinite, filling the space $z > 0$. Consider a plane wave with unit flux impinging on the sample under angle $\theta_a$

$$\psi_{in}^a(r) = \frac{1}{\sqrt{Ak_1 \cos \theta_a}} \, e^{iQ_a \rho + ik_1 \cos \theta_a z}, \; z < 0, \tag{2.47}$$

where $\rho = (x, y)$ is the transversal coordinate and $Q_a = k_1 \sin \theta_a(\cos \phi_a, \sin \phi_a)$ is the two-dimensional transverse momentum of the incoming beam. At the $z = 0-$plane part of this beam is transmitted, and part is specularly reflected. Requiring a smooth behavior of $\psi$ gives

$$\begin{aligned}
\psi_{in}^a(r) &= \frac{1}{\sqrt{Ak_1 \cos \theta_a}} \left[ e^{iQ_a \rho + ik_1 \cos \theta_a z} - \frac{P_a - p_a}{P_a + p_a} e^{iQ_a \rho - ik_1 \cos \theta_a z} \right], \; (z < 0) \\
&= \frac{1}{\sqrt{Ak_1 \cos \theta_a}} \frac{2p_a}{P_a + p_a} e^{iQ_a \rho + ik \cos \theta_a' z - z/(2\ell \cos \theta_a')}, \; (z > 0), \tag{2.48}
\end{aligned}$$

where we defined

$$p_a^2 = k_1^2 - Q_a^2, \qquad P_a^2 = k^2 + nt - Q_a^2. \tag{2.49}$$

The angle $\theta_a'$ is the angle of the refracted beam with respect to the $z$-axis.

Inside the slab the unscattered part of the intensity decays over one mean free path, see Eq. (2.24). The source of diffuse intensity is the first scattered intensity. The first scattered intensity is the intensity "just after" it has hit the first scatterer. We choose this definition because the next scattering event can now also easily be chosen to end on a (different) scatterer. This is consistent with our previous derivation of the diffuse intensity, see Fig. 2.4. In leading order of $(1/k\ell)$ the source $S$ equals

$$\begin{aligned}
S_a(z) &= nt\bar{t} \, |\psi_{in}^a(r)|^2 \\
&= \frac{16\pi p_a^2}{Ak_1 \mu_a \ell (P_a + p_a)^2} \, e^{-z/(\ell \mu_a)} = \frac{4\pi}{Ak \mu_a \ell} \, T(\mu_a) \, e^{-z/(\ell \mu_a)}, \tag{2.50}
\end{aligned}$$

where $\mu_a = \cos \theta_a'$. The source is proportional to the intensity transmission coefficient of the boundary between the two dielectrics

$$T(\mu) \equiv \frac{4p_a P_a}{(P_a + p_a)^2} = \frac{4\mu \sqrt{\mu^2 - 1 + 1/m^2}}{[\mu + \sqrt{\mu^2 - 1 + 1/m^2}]^2}. \tag{2.51}$$

which obeys $T(\mu) + R(\mu) = 1$. If $\mu^2 < 1 - 1/m^2$, we see that $R = 1$, corresponding to the case of total internal reflection.

From $S_a$ the twice scatterered intensity is generated, $nt\bar{t} \int d\mathbf{r}' |G(\mathbf{r}, \mathbf{r}')|^2 S_a(\mathbf{r}')$. Likewise, we obtain the intensity scattered three times, etc. The geometric sum of first, second, third,...., times scattered intensity is again the multiple scattered or diffuse intensity $\mathcal{L}^a$. This is exactly the same sum, or the same diagrams, as we used in the

derivation of the diffuson equation, but the assumption of slowly varying functions is not necessary and we treated the coupling into the medium properly. The diffuse intensity now is calculated from the first scattered intensity by the Schwarzschild-Milne equation

$$\mathcal{L}^a(z) = S_a(z) + \int_0^L dz' M_{\mathrm{SM}}(z,z') \mathcal{L}^a(z'), \tag{2.52}$$

which is a self-consistent transport equation. By inserting $\mathcal{L} = 0$ at the r.h.s. and iterating the solution, one recovers the original ladder sum.

The kernel $M_{\mathrm{SM}}$ describes the transport between two scatterings

$$M_{\mathrm{SM}}(z,z') = M_B(z-z') + M_L(z+z') + M_L(2L-z-z'), \tag{2.53}$$

that decomposes into the bulk term $M_B$ and into layer terms describing internal reflections at the interfaces at $z = 0$ and $z = L$. The bulk term is given by integrating an intensity in the $\rho' = (x', y')$ plane at $z'$ propagating unscattered towards a point in the plane at $z$. By putting $\rho'$ and $z'$ temporarily zero one obtains,

$$\begin{aligned} M_B(z,0) &= 4\pi \int \mathrm{d}x \mathrm{d}y |G(z,\rho;0,0)|^2 \\ &= \int \mathrm{d}x \mathrm{d}y \frac{\mathrm{e}^{-\sqrt{x^2+y^2+z^2}/\ell}}{4\pi(x^2+y^2+z^2)} = \int_0^1 \frac{\mathrm{d}\mu}{2\mu} \mathrm{e}^{-z/\ell\mu}, \end{aligned} \tag{2.54}$$

Thus

$$M_B(z-z') = \int_0^1 \frac{\mathrm{d}\mu}{2\mu} \mathrm{e}^{-|z-z'|/\ell\mu} = \frac{1}{2} E_1(|z-z'|), \tag{2.55}$$

with $E_1$ being the exponential integral. Notice again the exponential decay of the intensity between scatterings.

Similarly, we derive the layer kernel. It is proportional to the reflection coefficient $R(\mu)$. Now the argument in the exponent is the total path length from $z'$ to the surface at $z = 0$, and back again to the final point $z$. The kernel therefore reads

$$M_L(z+z') = \int_0^1 \frac{d\mu}{2\mu} [1 - T(\mu)] \mathrm{e}^{-(z+z')/\mu\ell}. \tag{2.56}$$

Here we assumed specular reflection at the surface: $\mu_{\mathrm{in}} = \mu_{\mathrm{out}}$. [7]

The precise solution of Eq. (2.52) can be obtained numerically (see Refs. [2, 60] for details). We have plotted it in Fig. 2.5 for a relatively thin slab ($L = 4\ell$). We took the data from table 17 in Ref. [2]. The albedo was unity, and there was no index mismatch. The form of the solution near the incoming surface is quite different for the three cases drawn: $\mu = 1$, or perpendicular incidence; $\mu = 0.1$, the angle with the $z-$ axis is then 84 degrees; and diffuse incidence uniformly distributed over all angles. At the outgoing surface all solution behave the same: there is only a small deviation from the straight line crossing zero at $L + z_0$.

Generally we can split the solution of Eq. (2.52) in a special and a homogenous solution, $\mathcal{L}_S$ and $\mathcal{L}_H$ respectively, which correspond to the solution with and without the source $S$, respectively. For thick slabs the intensity propagator in the region, say $0 < z < L - 5\ell$, can thus be written as

$$\mathcal{L}^a(z) = \alpha \mathcal{L}_H(z) + \mathcal{L}_S^a(z). \tag{2.57}$$

---

[7]For non-specular reflection one has in general: $\int \frac{\mathrm{d}\mu}{2\mu} \int \mathrm{d}\mu' R(\mu,\mu') \mathrm{e}^{-z/\mu\ell - z'/\mu'\ell}$. One recovers specular reflection by putting $R(\mu,\mu') = \delta(\mu-\mu')R(\mu)$. Lambert reflection corresponds to $R(\mu,\mu') \propto \mu$.

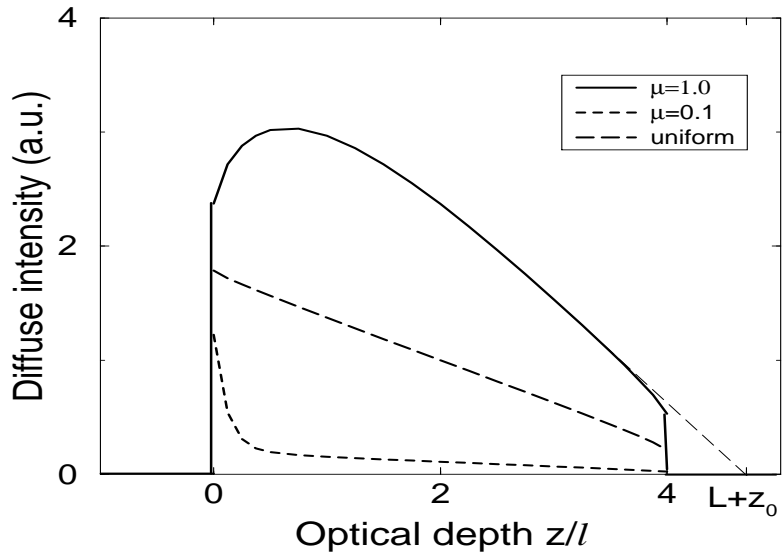

Figure 2.5: The solution of the Schwarzschild Milne equation as taken from the table 17 in Van de Hulst [2] for slab of thickness $L = 4\ell$ with no internal reflection, $a = 1$. The precise solution near the incoming plane strongly depends on the cosine of the incoming angle, $\mu$. The bulk behavior is apart from a multiplication factor the same, they all extrapolate to $L + z_0$ (thin dashed line).

Apart from a linear component there are exponential corrections in these solutions, as is clear from Fig. 2.5. Far from the surface, these corrections of order $e^{-z/\ell}$ vanish. We can impose the asymptotic behavior $\mathcal{L}_S(z) \to const.$, $\mathcal{L}_H(z) \sim (z + z_0)$ and the solution behaves for, say, $5\ell < z < L - 5\ell$, as

$$\mathcal{L}^a(z) = \frac{4\pi T(\mu_a)\tau_1(\mu_a)}{k\ell \, A\mu_a} - \alpha(z + z_0) \, . \tag{2.58}$$

On the other hand, the propagator on the outgoing region, say $5\ell < z < L - 5\ell$, obeys

$$\mathcal{L}(z, M) = \beta \mathcal{L}_H(L - z, M) = \beta(L - z + z_0) \, . \tag{2.59}$$

Matching of the two solutions in the middle region of the slab determines $\alpha$ and $\beta$. In the bulk of the slab (at least few mean free paths away from the boundaries) the diffuse intensity thus has a simple behavior

$$\mathcal{L}_{\text{in}}^a(z) = \frac{4\pi T(\mu_a)\tau_1(\mu_a)}{k\ell \, A\mu_a} \frac{L + z_0 - z}{L + 2z_0}. \tag{2.60}$$

Note that in this region the intensity again satisfies the diffusion equation $\nabla^2 \mathcal{L}_{\text{in}}^a = 0$. Indeed, using Taylor expansion of $\mathcal{L}$, it can be shown that the Schwarzschild-Milne equation reduces to the diffusion equation if the diffuse intensity is slowly varying with respect to the mean free path, i.e. deep inside a thick slab. We only use the more complicated Schwarzschild-Milne equation to fix the parameters of its solution, i.e. $\tau_1$ en $z_0$. In principle one could use the precise form of solution resulting from the Milne equation, but this would be much too complicated for further calculations. For thick slabs the contributions from the boundary corrections are exponentially small anyhow.

The $\tau_1$ in Eq. (2.60) is the limit intensity of a semi-infinite slab [59]; of course, it depends on the incident angle. For matched refractive indices (m=1) and perpendicular incidence the numerical value is $\tau_1(1) = 5.04$ [59]. In the calculations of the correlation functions in the next chapters, this pre-factor will drop out as we always normalize to the average. The extrapolation length $z_0$ in Eq. (2.60) is more important as it influences the functional form of the diffuse intensity. For isotropic scattering and matched refractive indices one has [2] $z_0 = 0.7104\ell$. However, it becomes larger when there is a mismatch of the indices of refraction between the scattering medium and the surroundings. The parameters $z_0$ and $z_1$ ($z_1$ is defined as $z_1 = \tau_1\ell$) are calculated numerically in Ref. [59] for various values of the index mismatch $m$. For given $m$ both $z_0$ and $z_1$ take values comparable to the mean free path. Note that the diffusion approximation does quite well in the estimate of $z_0 = (2/3)\ell$ as compared to the more detailed Schwarzschild-Milne approach that yielded $z_0 = 0.71\ell$. In the limit of large index mismatch, $m \to 0$, or $m \to \infty$, the extrapolation length tends to infinity. The estimate of $z_0$ using the diffusion approximation of Zhu et al. [54] becomes then asymptotically exact [59]. The asymptotic value is already quite good at moderate index mismatch, say $m \approx 3/2$.

However, the index of refraction of a diffuse medium is experimentally and theoretically not easily determined. The sample typically is a dense packing, say 30%, of irregular shaped particles suspended in a liquid or solid. The average index of refraction of a binary mixture is a classical problem [61, 62, 63]. Den Outer and Lagendijk [64] have this studied problem both experimentally and theoretically by considering the maximum width of the enhanced backscatter cone, see section 2.7. It turns out that for a typical system the average index of refraction is best described by independent Mie scatterers. It also possible to measure the refractive index by Brewster angle measurements, although this method is only sensitive to a very thin surface (few wavelengths) of the sample, which may be not representative for the rest of the sample [65]. Both the treatment in the diffusion approximation and the Schwarzschild-Milne equation assumed specular reflecting boundaries. Due to the roughness of the surface, however, this is experimentally not fully justified. The surface of typical samples scatters over some 40 degrees [66]. The opposite limit, diffuse surface scattering, is called Lambert reflection. It is studied for a full reflecting boundary in the book of Kagiwada et al. [60]. In Ref. [2] also the thickness dependence of $z_0$ was studied, yet for already moderate values of $L/\ell \sim 4$, $z_0$ is close to its value for infinitely thick slabs. Concluding from the above remarks, we consider the extrapolation length $z_0$ as an extra experimental parameter. For small $z_0/L$ we have

$$\mathcal{L}_{\text{in}}^a(z) = \frac{4\pi\tau_1(\mu_a)T(\mu_a)}{k\ell A\mu_a}\frac{L-z}{L}. \qquad (2.61)$$

### 2.4.1   The $M \neq 0$ case

Similar to the $M = 0$ case, one can consider the $M$-dependent solution of the Schwarz-schild-Milne equation. For our optically thick samples ($L \gg \ell$) somewhat away from the surface, we have homogeneous and special solutions of equation (2.52)

$$
\begin{aligned}
\mathcal{L}_H(z, M) &= \frac{\sinh M(L - z)}{M\ell} + \frac{z_0(M; m)e^{-Mz}}{\ell} , \\
\mathcal{L}_S(z, M) &= \frac{4\pi T(\mu_a)\tau_1(M; \mu_a)}{k\ell\, A\mu_a} e^{-Mz} ,
\end{aligned}
\tag{2.62}
$$

where we used the analogy with the diffusion equation for the bulk behavior. The extrapolation length $z_0$ depends on $M$. Van de Hulst and Stark treated the dependence of the extrapolation length on the albedo [67] (where a different definition of $z_0$ is used). Again the $M\ell \ll 1$ simplifies the situation here. In a way similar to the treatment of large index mismatch in reference [59], it can be shown that in the limit of large index mismatch $z_{0,1} \to 1/M$, as $m \to \infty$, where $M\ell \ll 1$. A useful interpolation formula is [68]

$$
z_{0,1}(M; m) = \frac{z_{0,1}(0; m)}{1 + Mz_{0,1}(0; m)} .
\tag{2.63}
$$

We denote $z_{0,1}(0; m)$ as $z_{0,1}$. We have checked this formula for $M = \kappa$ numerically to be correct within a few percent up to $M\ell = 1/2$.

Using the same matching procedure as before, and assuming $M-$independence of $T(\mu)$, we find the incoming diffuse intensity needed for our further calculations

$$
\mathcal{L}_{\text{in}}^a(z, M) = \frac{4\pi T(\mu_a)\tau_1(\mu_a)}{k\ell\, A\mu_a} \frac{\sinh(ML - Mz) + Mz_0\cosh(ML - Mz)}{(1 + M^2 z_0^2)\sinh(ML) + 2Mz_0\cosh(ML)} .
\tag{2.64}
$$

This formula is similar to the diffuse intensity used in the diffusion approximation above, but the parameters $z_0$ and $z_1$ are now precisely prescribed.

## 2.5   Transmission quantities

Untill now we considered an incoming wave from a fixed direction. In optical systems there is also the possibility to sum over incoming or outgoing directions. We distinguish three quantities

- The angular resolved transmission $T_{ab}$
- The total transmission $T_a = \sum_b T_{ab}$
- The conductance $T = \sum_{ab} T_{ab}$

The calculation of the auto-correlations in these transmission quantities will be a large part of this study. In this section we just calculate the average values. The used formalism is also known as the Landauer approach, after the Landauer formula for the conductance, [69, 70]

$$
G = \frac{2e^2}{h} \sum_{ab} T_{ab},
\tag{2.65}
$$

where one sums the transmission coefficients $T_{ab}$ of waves with unit flux coming in channel $a$ and going out to channel $b$. (In the optical case "one channel" means that one incoming or outgoing direction is probed.) In optics the Landauer approach is very natural, as the basic quantity is the angular resolved transmission $T_{ab}$, yet a summation over the channels is possible. In optical experiments one carries out this summation by placing an integrating sphere, which is a sphere whose inside is coated with white paint. It has two small holes, one for placing the sample, and one for the detector. Using this technique total transmission measurements were performed by integrating over all outgoing directions [71, 72, 73]. Conductance measurements are principle possible with integrating spheres on both sides of the sample. (We don't know of published conductance measurement in optical systems, neither by using integrating spheres, nor by other methods, see Ref. [74] for experimental considerations.)

In the previous sections we already derived the diffuse intensity inside the slab relevant for the angular resolved and total transmission measurements, when one incoming angle is used. Conductance measurements involve waves coming in from all directions. Summing the source term of the previous section, Eq. (2.50) over the channels $a$ yields a source for the diffuse intensity

$$\begin{aligned} S(z) &= \sum_a S_a = A \int_0^k \frac{\mathrm{d}Q_a}{2\pi} Q_a S_a \\ &= \frac{2k}{\ell} \int_0^1 \mathrm{d}\mu\, T(\mu) \mathrm{e}^{-z/\ell\mu}. \end{aligned} \tag{2.66}$$

This is again the input in the Schwarzschild-Milne equation (2.52). Defining

$$\epsilon_a = \frac{\pi T(\mu_a)\tau_1(\mu_a)}{kA\mu_a}, \tag{2.67}$$

and by comparing the pre-factors, the intensity in the bulk now has the diffusive behavior $\mathcal{L}_{\mathrm{in}}(z) = [\sum_a \frac{4k\epsilon_a}{\ell}] \frac{L+z_0-z}{L+2z_0}$. One thus has

$$\mathcal{L}_{\mathrm{in}}^a(z) = \epsilon_a \mathcal{L}_{\mathrm{in}}(z). \tag{2.68}$$

We can use the sum rule

$$\sum_a \epsilon_a = \int \frac{\mathrm{d}^2 Q}{(2\pi)^2} \frac{\pi T(\mu)\tau_1(\mu)}{k\mu} = \frac{1}{2} \int \mathrm{d}\mu\, T(\mu)\tau_1(\mu) = 1. \tag{2.69}$$

Here we used that $\mu = \sqrt{1 - Q^2/k^2}$. The last equation is a sum rule on the diffuse intensity, see Eq.(2.30) of Ref. [59]. We find

$$\mathcal{L}_{\mathrm{in}}(z) = \frac{4k}{\ell} \frac{L+z_0-z}{L+2z_0}. \tag{2.70}$$

We term this object the incoming *total-flux diffuson*. Although it has a different prefactor, it has the same depth dependence as $\mathcal{L}_{\mathrm{in}}^a$. This can be understood as the dependence of the solution of the Schwarzschild-Milne equation on the incoming angle only enters in terms that describe low order scattering. These are only relevant near the surface, see Fig. 2.5. Deep inside the bulk, these exponential corrections vanish and the only angular dependence is in the pre-factor.

Having calculated the coupling into the sample, we also want the radiation coming of the slab on the outgoing side. On the outgoing side, radiation emitted at a point $r = (\rho, z)$ inside the slab propagates to a point $(Z, \rho')$ outside the sample $(Z > L)$ as described by the Green's function of semi-infinite medium with dielectric function $\epsilon(\mathbf{r}) = \epsilon_0$ for $z < L$ and $\epsilon(\mathbf{r}) = \epsilon_1$ for $z > L$

$$G(\rho, z; \rho', Z) = \frac{1}{A} \sum_Q G(z; Z; Q) e^{iQ(\rho - \rho')}, \qquad (2.71)$$

in which $G(z; Z; Q)$ is the 1-d Fourier transform of Eq. (2.22),

$$\begin{aligned} G(z; Z; Q) &= \frac{i}{P + p} e^{iP(L-z) + ip(Z-L)}, \\ P &= \sqrt{k^2 - Q^2 + nt}, \qquad p = \sqrt{k_1^2 - Q^2}. \end{aligned} \qquad (2.72)$$

In the far field $(Z \gg L)$ the total transmitted intensity reads

$$\int d^2\rho' |G(\rho, z; \rho', Z)|^2 = \frac{1}{A} \sum_Q |G(z; Z; Q)|^2. \qquad (2.73)$$

Since $p_z = k_1 \cos\theta_a = k\mu_a$, the according flux is

$$\Phi(z) = \frac{1}{A} \sum_Q k\mu |G(z; Z; Q)|^2 = \frac{k}{8\pi} \int_0^1 d\mu\, T(\mu) e^{-(L-z)/\ell\mu}. \qquad (2.74)$$

So the outgoing intensity is generated near the outgoing surface $z = L$, as is obvious. It leads to a source $S(z) = 4\pi\Phi(z)/\ell$ in the Schwarzschild-Milne equation. Apart from a reflection in $z$, this expression differs by a factor 4 from (2.66). The pre-factors would be the same if our Green's functions were multiplied by a factor 2; this amounts to the same as taking a kinetic term $\nabla^2/2$ rather than $\nabla^2$, such as occurs in electronics in units where $\hbar = m = 1$. The corresponding diffuson is the outgoing total-flux diffuson in transmission. It reads

$$\mathcal{L}_{\text{out}}(z) = \frac{k}{\ell} \frac{z + z_0}{L + 2z_0}. \qquad (2.75)$$

If $z_0/L$ is small we may use

$$\mathcal{L}_{out}(z) = \frac{k}{\ell} \frac{z}{L}. \qquad (2.76)$$

The outgoing diffuson gives the diffuse intensity inside the slab that will finally come out at the outgoing side.

For angular resolved measurements the outgoing intensity in a certain direction $b$ is, in analogy with (2.68),

$$\mathcal{L}_{\text{out}}^b(z) = \epsilon_b \mathcal{L}_{\text{out}}(z) \qquad (2.77)$$

We call the incoming and outgoing diffusons, *external* diffusons, because they connect to the outside of sample. This in contrast to the *internal* diffusons that begin and end inside the medium (for instance at an interference vertex).

We find the conductance $\langle T \rangle$ by integrating the outgoing diffuson with source of the incoming intensity, Eq. (2.66). Equivalently, one can integrate the incoming diffuson with the outgoing source, Eq. (2.74). For this calculation the precise behavior of the diffuson near the boundary is important and Eq. (2.75) cannot be used. Instead, we use that

$$\int_0^\infty dz \mathcal{L}_{\text{out}}(z) \int d\mu T(\mu) e^{-z/\ell_\mu} = 2k/3\ell.$$

The derivation of this sum-rule falls outside the scope of this chapter, see however Eq. (2.19) of Ref. [59]. We obtain

$$g \equiv \langle T \rangle = \int d^3\mathbf{r} \mathcal{L}_{\text{out}}(z) S(z)$$
$$= \frac{k^2 A \ell}{3\pi(L + 2z_0)} = \frac{4N\ell}{3(L + 2z_0)}. \qquad (2.78)$$

For the electronic case one has for the average conductance $\langle G \rangle = \frac{2e^2}{h} \frac{k^2 A \ell}{3\pi(L+2z_0)}$. In Eq. (2.78) the number of modes $N$ is $N = k^2 A / 4\pi$, which is also known as the Weyl formula. It corresponds to dividing the total beam area into small coherence regions, or channels, of area $\lambda^2$. The conductance can then be interpreted as the sum of $N$ channels, on the average a channel contributes $4\ell/3(L + 2z_0)$ to the conductance. When the diameter of the incoming beam profile becomes comparable to the thickness, the diffuse intensity broadens inside the slab. The number of modes becomes then ill-defined. (We study such effects in chapter 4.) The number of modes is also one of the few parameters where the vector character comes in: The two polarization directions travel independently through the slab, so the number of available independent channels effectively doubles. In our approach only in $1/N$-correction this effect is important, else it just changes the average transmission and does not change our (normalized) results.

The angular and total transmission coefficients read

$$\langle T_{ab} \rangle = \epsilon_a \epsilon_b \langle T \rangle = \frac{\pi \tau_1(\mu_a) \tau_1(\mu_b) T(\mu_a) T(\mu_b) \ell}{3\mu_a \mu_b A (L + 2z_0)}, \qquad (2.79)$$

$$\langle T_a \rangle = \frac{\tau_1(\mu_a) T(\mu_a) \ell}{3\mu_a (L + 2z_0)}. \qquad (2.80)$$

Transmission in the diffusion approximation is proportional to the value of the diffuson at $z = \ell$ and $z' = L - \ell$. The internal diffuson always ends on scatterers. Yet for transmission quantities, coupled to a source/detector outside, the diagram should end on propagators. As it is not clear how to obtain this in the diffusion approach, on often estimates the transmission diagram ending with (four) propagators by dividing by $(nt\bar{t})^2$. The total transmission is thus in the $M = 0$ case

$$\langle T_a \rangle \stackrel{\text{diff}}{=} \frac{12\pi}{\ell^3} \frac{\ell^2}{L + 2z_0} \left( \frac{4\pi}{\ell} \right)^{-2} = \frac{3\ell}{4\pi(L + 2z_0)} \qquad (2.81)$$

One observes that the dependence on $\ell$ and $L$ is correct, but the pre-factor is not correct and all angular dependence is lost. Note that the region where the diffuse intensity is non-zero, seems to exceed the slab on both sides by a length $z_0$, see Eq. (2.41), Eq. (2.64).

This is seen back in the transmissions Eq. (2.81), but also in Eq. (2.80). The resistance of the sample is inversely proportional to the conductance, thus $R \propto L + 2z_0$. The first term is the thickness dependence known from Ohm's law, the second term can be interpreted as contact resistances at both sides.

For the total transmission and the conductance measurements one integrates over the outgoing surface. If on the incoming side, two or more different plane waves are present, they can combine into diffusons with non-zero transverse momentum, if the plane waves have different incoming angles. In that case the outgoing diffusons have a phase factor $\exp(-iQ \cdot \rho)$, which depends on the position on the outgoing surface $\rho$. The integration over the outgoing surface yields a delta function $\delta(Q)$. In other words, only the momentum independent diffusons, where the phases precisely cancel, survive this averaging. The same holds for the frequency difference: The integrating sphere acts also as a time integrator due the random scattering inside it, yielding a factor $\delta(\Omega)$. In conclusion, the only relevant outgoing diffusons have mass $M = \kappa$. This observation is especially of importance in considering correlation functions. There the diffusons interfere, such that the mass of outgoing diffusons can differ from the incoming ones. In section 4.3.2 we will also calculate small $1/N$ deviations to this "selection rule".

## 2.6 Scattering from one extra, static scatterer

Following Berkovits and Feng [75], we consider the influence of one additional point scatterer on the diffuse intensity. This is a problem of practical importance as researchers try to locate an object in a strongly scattering medium by looking at the transmitted diffuse beam [76]. [8] We treat this problem within our diagrammatic approach. The problem can be rephrased as: 'how can two diffusons be tied together ?' In Fig. 2.6 we present all relevant contributions to the scattering. The two rightmost lower diagrams are self-energy diagrams that are not taken into account in the diffuson $\mathcal{L}$. These diagrams were overlooked in Ref. [75]. However, their contributions are important as they cancel the leading term of the first r.h.s. (lower) diagram.

The extra scatterer is described by a $t$-matrix $t_e$, with an albedo $a_e$. It has position $r_e = (0, 0, z_e)$. The diagrams again calculated using the tables in section 3.4. Let the diffuson propagators connected to the extra scatterer have momenta $q_1$ and $q_2$. Both momenta are pointing towards the extra scatterer. In lowest order of these momenta, we find for the lower diagrams of Fig. 2.6:

$$
\begin{aligned}
K_1 &= t_e \overline{t}_e \int \mathrm{d}^3 \mathbf{p} \, G(\mathbf{p} + \tfrac{1}{2}\mathbf{q}_1) G^*(\mathbf{p} - \tfrac{1}{2}\mathbf{q}_1) \int \mathrm{d}^3 \mathbf{p}' \, G(\mathbf{p}' - \tfrac{1}{2}\mathbf{q}_2) G^*(\mathbf{p}' + \tfrac{1}{2}\mathbf{q}_2) \\
&\approx t_e \overline{t}_e [I_{1,1} - \tfrac{1}{3} k^2 q_1^2 I_{2,2} + \tfrac{2}{3} k^2 \mathbf{q}_1^2 I_{1,3} + \tfrac{1}{3} k\ell\Omega I_{2,1} - \tfrac{1}{3} k\ell\Omega I_{1,2}] \times [\mathbf{q}_1 \to \mathbf{q}_2] \\
&= t_e \overline{t}_e \frac{\ell^2}{16\pi^2} \left[ 1 - \tfrac{1}{3}(q_1^2 + q_2^2)\ell^2 + \tfrac{2}{3} i\Omega\ell^2 - 2(1 - a_e) \right]
\end{aligned}
$$

---

[8] A different method for locating objects in diffusely scattering media is to measure the time-resolved transmission. The light that comes out first has scattered few times and mostly in the forward direction and is therefore not diffusely broadened, see references in Ref. [76, 77].

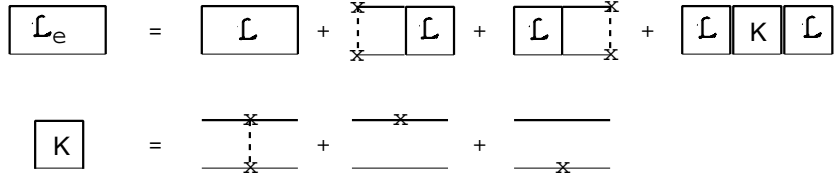

Figure 2.6: Contributions from an additional scatterer to the diffuson intensity. $\mathcal{L}$ denotes the diffuson without the extra scatterer, $\mathcal{L}_e$ the resulting diffuson with contributions from the extra scatterer taken into account. Thick lines denote particle propagators, thin lines denote hole propagators. The dashed line indicates that both the particle and the hole line interact with the extra scatterer. The $t_e$−matrix of the extra scatterer is denoted by a cross. Its total effect is denoted by $K$.

And for the second and third term one finds

$$
\begin{aligned}
K_2 = K_3^* &= t_e \int \mathrm{d}^3\mathbf{p}\, G^*(\mathbf{p}) G(\mathbf{p}+\mathbf{q}_1) G(\mathbf{p}-\mathbf{q}_2) \\
&\approx t_e \left[ I_{2,1} + \frac{4}{3}(q_1^2 + q_2^2 - \mathbf{q}_1 \cdot \mathbf{q}_2)k^2 I_{4,1} - \frac{2}{3}k\ell\Omega(-I_{2,2}+2I_{1,3}) \right] \\
&= -\frac{it_e\ell^2}{8\pi} \left[ -1 + \frac{1}{3}(q_1^2 + q_2^2 - \mathbf{q}_1 \cdot \mathbf{q}_2)\ell^2 - \frac{2}{3}i\Omega\ell^2 + 2(1-a_e) \right]
\end{aligned}
$$

The sum of the diagrams brings $K = K_1 + K_2 + K_3$:

$$
K = -\frac{t_e \bar{t}_e \ell^2}{48\pi^2}[\, \mathbf{q}_1 \cdot \mathbf{q}_2 \, \ell^2 + 3(1-a_e)] \,. \tag{2.82}
$$

Corresponding to the first line of Fig. 2.6, the expression for the diffuson with the extra scatterer included, reads in spatial coordinates:

$$
\begin{aligned}
L_e(r,r') &= L(r,r') + \frac{t_e \bar{t}_e}{nt\bar{t}}[\delta(r'-r_e) + \delta(r-r_e)]L(r,r') \\
&\quad + \frac{t_e \bar{t}_e \ell^2}{48\pi^2} \int dr_1 dr_2 [\ell^2 \nabla_1 \cdot \nabla_2 - 3(1-a_e)]L(r,r_1) \\
&\quad \times L(r_2,r')\delta(r_1 - r_e)\delta(r_2 - r_e) \,. \tag{2.83}
\end{aligned}
$$

We assume that the extra scattering is weak as compared to the total scattering $n_e t_e \bar{t}_e \ll nt\bar{t}$, with $n_e = 1/V$, $V$ is the slab volume. First we average the position of the extra scatterer over the whole slab. We expect to recover the diffusion equation (2.36) for a density of $n + n_e$ scatterers. Indeed, we find a reduced mean free path $\ell_e = 4\pi/(nt\bar{t} + n_e t_e \bar{t}_e)$. Starting from equation (2.83) and substituting formula (2.36) we find to leading order in $n_e$

$$
\begin{aligned}
L_e(q) &= L(q) + 2\frac{n_e t_e \bar{t}_e}{nt\bar{t}}L(q) + [q^2\ell^2 - 3(1-a_e)]\frac{n_e t_e \bar{t}_e \ell^2}{48\pi^2}L^2(q) \\
&\approx \frac{12\pi}{\ell_e^3}(q^2 + i\Omega_e + \kappa_e^2)^{-1} \,, \tag{2.84}
\end{aligned}
$$

with $\Omega_e \equiv \Omega\ell/\ell_e$, consistent with our definition of $\Omega = -3\Delta\omega/\ell v_E$. Further, $\kappa_e^2 \ell_e \equiv 3[nt\bar{t}(1-a) + n_e t_e \bar{t}_e(1-a_e)]/4\pi$, involves a weighted sum of albedos.

The situation with the scatterer fixed is more interesting. We consider the transmission of a plane wave through a non-absorbing medium. The transmission is (roughly) $T = \ell \frac{dL(L,z)}{dz}|_{z=L}$ (we are only interested in functional form here). We calculate $L_e$ with the method of images [76]. Neglecting the second and third upper r.h.s. diagrams, the transmission in near field reads

$$
\begin{aligned}
T(\rho) &= \frac{\ell}{L} + 2q\ell \sum_{n=-\infty}^{\infty} \frac{(2n+1)L - z_e}{(\rho^2 + [(2n+1)L - z_e]^2)^{3/2}} \\
&\quad + 2p\ell \sum_{n=-\infty}^{\infty} \frac{\rho^2 - 2[(2n+1)L - z_e]^2}{(\rho^2 + [(2n+1)L - z_e]^2)^{5/2}} ,
\end{aligned}
\tag{2.85}
$$

in which $\rho = (x, y)$, and

$$
\begin{aligned}
q &= -\frac{L - z_e}{L} \frac{3 t_e \overline{t}_e}{16 \pi^2 \ell} (1 - a_e) , \\
p &= \frac{\ell}{L} \frac{t_e \overline{t}_e}{16 \pi^2} .
\end{aligned}
\tag{2.86}
$$

Two cases are to be considered. If the extra scatterer does not absorb, only the $p-$term is present. This is analogous to a dipole in a static electric field between two capacitor plates. In the transmitted beam as function of $\rho$ we see a wiggle if $z_e > L/2$, else we see a dip in the transmission. If the extra scatterer absorbs, the $q-$term is dominant and the scatterer acts as a drain for the intensity. The result is a dip in the transmitted beam. This would be equivalent to a (negative) charge in electrostatics.

It is instructive to compare our approach with calculations within diffusion approximation. Den Outer et al. performed experiments where they located a pencil or glass fiber in a diffusive medium [76]. They found that the experiments are well described by a diffusion approximation. Diffusion was assumed both in the medium and inside the scatterer. The scatterer was for instance a sphere with radius $R$, inverse absorption length $\kappa_2$, and different diffusion constant $D_2$. Consider the situation of weak uniform absorption and almost equal diffusion constants inside ($D_2$) and outside the sphere ($D$). Den Outer et al. find then: $q = -\kappa_2^2 R^3 (L - z_e)/3L$ and $p = R^3(D - D_2)/3LD$. We think of our physical scatterer as a sphere of radius $R$ with a density $n_e = 3/4\pi R^3$ of extra scatterers. The extra scattering is weak as compared to the total scattering, if $R$ is not too small, viz. $R \gg (\ell t_e \overline{t}_e)^{1/3}$. The diffusion constant inside the sphere is $D_2 = \frac{1}{3} v_E \ell_e$. With this identification, our $p$ and $q$ values agree to leading order with the results of den Outer et al.

In reference [75] the self-energy contributions presented in Fig. 2.6 are not taken into account. As a result the unphysical result is found that an extra scatterer (without absorption) act as a source of intensity. Obviously a Ward identity has not been satisfied in their approach.[9]

The vertex describing the influence of one extra scatterer can also be used to tie two diffusons together. This is sometimes useful, for instance when calculating the transmission the incoming diffuson can be attached to the outgoing one. For this purpose we take $t_e = t$. The vertex now obeys

$$\int dz_2 \mathcal{L}(z_1, z_2) \hat{K} \mathcal{L}(z_2, z_3) = -\frac{\ell^3}{12\pi} \int dz_2 [\partial_{z_2}^2 \mathcal{L}(z_1, z_2)] \mathcal{L}(z_2, z_3) = \mathcal{L}(z_1, z_3), \qquad (2.87)$$

which is indeed the wanted behavior.

---

[9]Again one easily checks that cutting propagators in the dressed amplitude propagator yields the intensity vertices that we used and that respect the conservation laws.

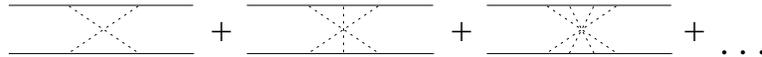

Figure 2.7: Most crossed diagrams

## 2.7  Most crossed diagrams

Another class of diagrams is also of some importance to us. They are the so called most crossed diagrams. We have drawn some elements in Fig. 2.7. These diagrams describe the enhanced return probability of diffuse intensity. They are unique to diffusion of waves (a particle cannot split itself in two, each part following a different path). The interference doubles the return probability. This process is the most important for the eventual Anderson localization, as was shown in Refs. [16, 19, 20]. In an electronic system, the presence of a magnetic field changes the phase of the loop. The return probability is lower, increasing the conductance. This effect is known as the negative magneto-resistance, see [31, 30].

The angular resolved measurements in optical systems allow for a direct measurement of the most crossed diagrams. If a beam impinges on a sample, the backscattered intensity in the incoming direction is almost twice as high as in other directions. The intensity as function of angle has a conical form with a width $1/(k\ell)$. The extra intensity comes from the most crossed diagrams. The effect is also known as the *enhanced backscatter cone*. It provides a direct measurement of the most crossed diagrams. Due to the presence of single scattering contributions the peak value is somewhat lower than twice the diffuse background.

Measurements of this effect were done by Kuga and Ishimaru [78], Van Albada and Lagendijk [79] and Wolf and Maret [80]. For theoretical work, see [81], and Refs. [76, 59, 82, 83]. In this work, however, we do not treat the backscatter cone and only occasionally encounter most crossed diagrams.

# 3

# Interference of diffusons: Hikami vertices

After we have studied the diffuse transport in the previous chapter, we introduce in this chapter the other main ingredient for our calculations: the Hikami vertices. These vertices describe the interaction between diffusons. The Hikami vertices are named after Hikami, who used them in 1981 [24] for the calculation of weak localization corrections on the conductance. Interestingly enough their original introduction dates back two years earlier from the work of Gor'kov, Larkin, and Khmel'nitskii [84]. In the vertices diffusons exchange amplitudes, leading to the correlations of diffusons.

## 3.1 Calculation of the Hikami vertices

In this part of the calculation we again assume that the diffusons are slowly varying. The technique for calculating diagrams is well known [42]: Firstly, draw the diagrams; secondly, write down a momentum for each line; thirdly, use momentum conservation at each vertex; and fourthly, integrate over the free momenta left.

The first step, writing down all leading diagrams requires some care. In Fig. 3.1 we have drawn the diagrams in the Hikami box in second order Born as already presented by Gor'kov *et al.* [84]. In second order Born one neglects scattering more than twice from the same scatterer. Therefore, the (dashed) interaction line connects two potentials. Yet we have not drawn all diagrams, one can imagine. Firstly, the box is attached to a diffuson ending on scatterer (which is consistent with the way we have defined the diffusons in chapter 2). As a result, common scatterings between propagators on the same leg are not allowed within the box. Secondly, diagrams with two or more scatterers and parallel dashed lines are sub-leading. They contain loops with two propagators of the *same* type, i.e. integrals like $\int d^3 \mathbf{p} G(\mathbf{p} + \mathbf{q}) G(\mathbf{p})$, which are sub-leading. Thirdly, diagrams with two crossed dashed lines are also sub-leading. Finally, we again work in the independent scatterer approximation. It is important to consider all corrections to the first term of the figure, as there is again a cancellation of leading terms. These cancellations are imposed by energy conservation.

### 3.1.1 The $H_4$ four point vertex

Let us calculate the vertex. As an example we calculate second r.h.s. diagram of Fig. 3.1 in second order Born. As usual in the diffusion approximation where small derivatives are assumed, we expand the Green's for small momentum and frequency difference. We use the same expansion as in the derivation of the diffusion equation.

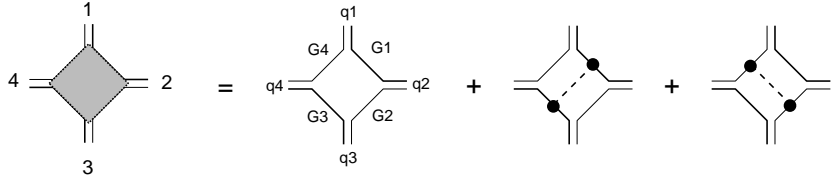

Figure 3.1: The Hikami four point vertex. It describes the exchange of amplitudes of two incoming diffusons 1 and 3 into two outgoing diffusons 2 and 4. The dots linked with the dashed line denote the dressing with an extra scatterer. The solid lines are dressed amplitude propagators.

$$
\begin{aligned}
G(\mathbf{p}+\mathbf{q};\omega+\tfrac{1}{2}\Delta\omega) \quad &\approx \quad \left[p^2+2\mathbf{p}\cdot\mathbf{q}+q^2-\frac{w^2}{c^2}-\Delta\omega\frac{\omega}{c^2}-nt\right]^{-1} \\
&\approx \quad G+(-2\mathbf{p}\cdot\mathbf{q}-q^2+\Delta\omega\frac{\omega}{c^2})G^2+4(\mathbf{p}\cdot\mathbf{q})^2G^3, \quad (3.1)
\end{aligned}
$$

with $G=G(\mathbf{p};\omega)$. The momenta, numbered according to Fig. 3.1, point towards to vertex. We number the Green's functions also according to the figure. To lowest order in $q^2$ we have

$$
\begin{aligned}
H_4^{(b)} \quad &= \quad nV^2\int\frac{d^3\mathbf{p}}{(2\pi)^3}G_1(\mathbf{p}+\mathbf{q}_1,\omega+\omega_1)G_3(\mathbf{p}-\mathbf{q}_4,\omega+\omega_3)G_4^*(\mathbf{p},\omega+\omega_4) \\
&\quad\times\int\frac{d^3\mathbf{p}'}{(2\pi)^3}G_1(\mathbf{p}'-\mathbf{q}_2,\omega+\omega_1)G_3(\mathbf{p}+\mathbf{q}_3,\omega+\omega_3)G_2^*(\mathbf{p},\omega+\omega_2) \\
&= \quad nV^2\left[I_{2,1}+\frac{4k^2}{3}(q_2^2+q_3^2-\mathbf{q}_2\cdot\mathbf{q}_3)I_{4,1}+\frac{k}{c}(\omega_1+\omega_3-2\omega_4)I_{3,1}\right] \\
&\quad\times[I_{2,1}+\ldots], \quad (3.2)
\end{aligned}
$$

where the $I-$ integrals are given in the appendix. We eliminate $nV^2$, in second order Born it holds that $nV^2=4\pi/\ell$. We take absorption terms into account in lowest order, i.e. only in the leading contributions, the $I_{2,1}$ terms. This yields

$$
H_4^{(b)}=\frac{-\ell^3}{16\pi k^2}+\frac{\ell^5}{48\pi k^2}(-\mathbf{q}_2\cdot\mathbf{q}_3-\mathbf{q}_1\cdot\mathbf{q}_4+\sum q_i^2+\kappa^2)+\frac{-i\ell^4}{16\pi k^2c}(\omega_1+\omega_3-\omega_2-\omega_4). \quad (3.3)
$$

A similar calculation gives

$$
\begin{aligned}
H_4^{(a)} \quad &= \quad \frac{\ell^3}{8\pi k^2}+\frac{\ell^5}{24\pi k^2}(-\mathbf{q}_1\cdot\mathbf{q}_3-\mathbf{q}_2\cdot\mathbf{q}_4-\sum q_i^2+\kappa^2) \\
&\quad+\frac{3i\ell^4}{32\pi k^2c}(\omega_1+\omega_3-\omega_2-\omega_4), \\
H_4^{(c)} \quad &= \quad \frac{-\ell^3}{16\pi k^2}+\frac{\ell^5}{48\pi k^2}(-\mathbf{q}_1\cdot\mathbf{q}_2-\mathbf{q}_3\cdot\mathbf{q}_4+\sum q_i^2+\kappa^2) \\
&\quad+\frac{-i\ell^4}{16\pi k^2c}(\omega_1+\omega_3-\omega_2-\omega_4). \quad (3.4)
\end{aligned}
$$

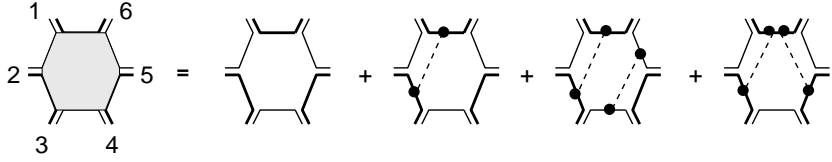

Figure 3.2: Diagrams contributing to the interaction of six diffusons: $H_6$. We did not draw possible rotations of the three rightmost diagrams, in total there are sixteen diagrams.

Furthermore we define the reduced frequency of the legs as, $\Omega_1 = -(\omega_1 - \omega_4)/D$, $\Omega_2 = -(\omega_1 - \omega_2)/D$, $\Omega_3 = -(\omega_3 - \omega_2)/D$ and $\Omega_4 = -(\omega_3 - \omega_4)/D$. The sum of the three diagrams yields

$$
\begin{aligned}
H_4 &= H_4^{(a)} + H_4^{(b)} + H_4^{(c)} \\
&= \frac{\ell^5}{96\pi k^2}\left[-2\mathbf{q}_1\cdot\mathbf{q}_3 - 2\mathbf{q}_2\cdot\mathbf{q}_4 + \sum_{i=1}^{4}(\mathbf{q}_i^2 + \kappa_i^2 + i\Omega_i)\right]\delta(\textstyle\sum \mathbf{q}_i). \tag{3.5}
\end{aligned}
$$

This is the main result of this section. Note that the leading, constant terms proportional to $\ell^3/k^2$ have cancelled. In generalizing the earlier results, of for instance Hikami [24] or Stephen [85], we have included frequency and absorption terms.

It is important to keep track of the $q^2$ terms. When the vertex is attached to a diffuson, the $q_j^2$ terms yield according to the diffusion equation, $(q_j^2 + \kappa^2 + i\Omega_j)\mathcal{L}_j = 1$, a delta-function in real space. If we attach *external* diffusons to the Hikami box, this delta function is roughly one mean free path from the surface, see section 2.3.1 and can be neglected. This simplifies the expression for the Hikami box. As a result the absorption part and the frequency dependent part vanish. After the cancellations the expression is:

$$
H_4(\mathbf{q}_1, \mathbf{q}_2, \mathbf{q}_3, \mathbf{q}_4) = \frac{-\ell^5}{48\pi k^2}(\mathbf{q}_1\cdot\mathbf{q}_3 + \mathbf{q}_2\cdot\mathbf{q}_4)\,. \tag{3.6}
$$

### 3.1.2 Six point vertex: $H_6$

Below, we also need one higher order diagram: the six-point vertex $H_6$, six diffusons are connected to this diagram. We depicted it in Fig. 3.2. This diagram was also already calculated by Hikami [24]. Also here, the dressings of the diagrams have to be added to the bare diagrams. Taking rotations of the depicted diagrams into account, there are 16 diagrams in second order Born approximation. It is not allowed to dress the bare six-point vertex (leftmost r.h.s. diagram in Fig. 3.2) with a scatterer that connects two opposite propagators. This dressing gives also a leading contribution, even if the dressing done with an arbitrary number of scatterers, but the resulting diagram is the same as the composed diagram with two four point vertices. Therefore, it should not be included in $H_6$, but it will enter in diagrams with $H_4$ vertices. A calculation similar to the one above gives [24],

$$
\begin{aligned}
H_6 &= \frac{-\ell^7}{96\pi k^4}\big[\mathbf{q}_1\cdot\mathbf{q}_2 + \mathbf{q}_2\cdot\mathbf{q}_3 + \mathbf{q}_3\cdot\mathbf{q}_4 + \mathbf{q}_4\cdot\mathbf{q}_5 + \mathbf{q}_5\cdot\mathbf{q}_6 + \mathbf{q}_6\cdot\mathbf{q}_1 \\
&\quad + \sum_{i=1}^{6}(\mathbf{q}_i^2 + \tfrac{1}{2}\kappa_i^2 + \tfrac{i}{2}\Omega_i)\big]\delta(\textstyle\sum \mathbf{q}_i). \tag{3.7}
\end{aligned}
$$

Apart from the generalization which includes frequency and absorption, Hikami's original expressions can be recovered from this using momentum conservation.

Other forms of the Hikami-boxes are sometimes more convenient, especially the cancellation of short distance terms depends on the explicit form. That is to say, the final result is of course always the same, but the intermediate results can be much simpler. We will encounter such a situation in chapter 6. Using momentum conservation, $\sum_i \mathbf{q}_i = 0$, the different forms can be transformed into each other. A useful formula is

$$\left(\sum_i \mathbf{q}_i\right)^2 = 2\sum_{i<j} \mathbf{q}_i \cdot \mathbf{q}_j + \sum_i q_i^2 = 0. \tag{3.8}$$

### 3.1.3   Manual for Hikami boxes

In our calculations we do not use the form of Eq.(3.5) for the Hikami boxes. Instead we use a Fourier transform in the $z-$ direction, because in our slab geometry the $(q_x, q_y, z) = (Q, z)$ representation is the most convenient. Now all $q_z$ become differentiations. The Hikami-four point box is

$$H_4 = \frac{\ell^5}{96\pi k^2}\left[2\partial_{z_1}\partial_{z_3} + 2\partial_{z_2}\partial_{z_4} - 2Q_1Q_3 - 2Q_2Q_4 + \sum_{i=1}^{4}(-\partial_{z_i}^2 + Q_i^2 + \kappa_i^2 + \Omega_i)\right]. \tag{3.9}$$

The differentiations work on the corresponding diffusons and afterwards $z_i$ should be put $z$. Formally one has

$$\begin{aligned}
&H_4(z; z_1, z_2, z_3, z_4)\mathcal{L}_1(z_1)\mathcal{L}_2(z_2)\mathcal{L}_3(z_3)\mathcal{L}_4(z_4)\\
&= \frac{\ell^5}{48\pi k^2}\left[\partial_{z_1}\partial_{z_3} + \ldots\right]\mathcal{L}_1(z_1)\mathcal{L}_2(z_2)\mathcal{L}_3(z_3)\mathcal{L}_4(z_4)\big|_{z_i = z},
\end{aligned} \tag{3.10}$$

but most of the times we just write this as

$$H_4(z)\mathcal{L}_1(z)\mathcal{L}_2(z)\mathcal{L}_3(z)\mathcal{L}_4(z). \tag{3.11}$$

The vertices thus give the spatial derivatives of the diffuson, that is to say the fluxes of the diffusons. The observation is helpful to estimate the influence of internal reflection, which lowers the spatial derivatives of the diffusons. Note that all leading terms in $H_4$ and $H_6$ dropped out; this is again closely related to the conservation laws[86].

## 3.2   Beyond second order Born

Going beyond second order Born approximation in the calculation of the diffuson diagrams resulted in a simple replacement of the mean free path. We recall that it became $\ell \equiv nt\bar{t}/4\pi$, instead of $\ell \equiv n|V|^2/4\pi$. The diagrams of the diffuson were slightly different but this was only a superficial problem, as the same calculation remained applicable. However, for the vertices the situation is more subtle. Previous section we only worked in second order of the scattering potential. Therefore, we calculated the box up to the same order, leading to the first three contributions of Fig. 3.3. As we work in all orders of the potential all contributions have to be calculated. To leading order in $1/k\ell$, these are the eight diagrams depicted in Fig. 3.3.

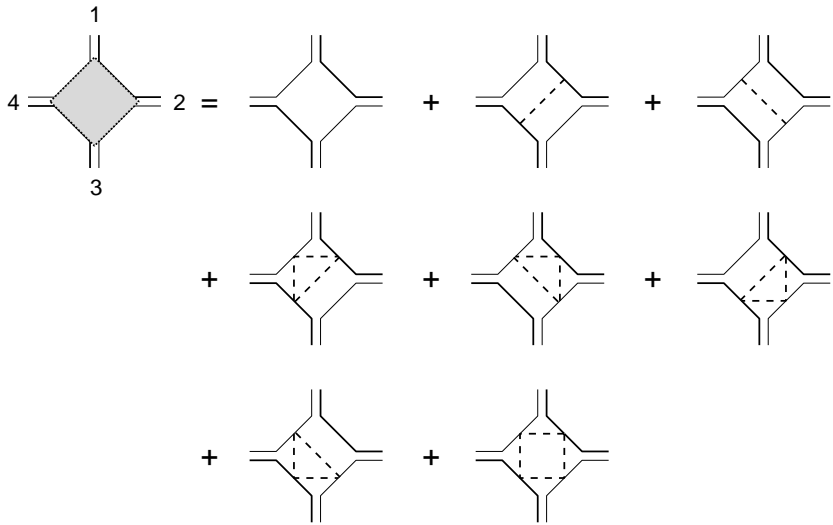

Figure 3.3: The Hikami four point vertex, the Hikami box, beyond the second order Born approximation. Instead of three, there are now eight diagrams to be calculated. The resulting, however, is apart from a renormalization of the mean free path the same as in second order Born approximation.

The calculation goes very similar to the second order Born above. The only difference is that instead of $V$, the $t-$matrices $t$ and $\bar{t}$ occur ($t$ on $G$'s, $\bar{t}$ on $G^*$'s). Already for the cancellation of the leading term, the three diagrams are not sufficient, but one needs all eight diagrams. One finds

$$
\begin{aligned}
H_4 &= H_4^{(a)} + H_4^{(b)} + \ldots + H_4^{(h)} \\
&= \frac{\ell^5}{96\pi k^2}\left[-2\mathbf{q}_1\cdot\mathbf{q}_3 - 2\mathbf{q}_2\cdot\mathbf{q}_4 + \sum_{i=1}^{4}(\mathbf{q}_i^2 + \kappa_i^2 + \Omega_i)\right]\delta(\sum\mathbf{q}_i). \quad (3.12)
\end{aligned}
$$

This is the same expression as found in the second order Born approximation ! Note, however, that now the definition of the mean free path is different: $\ell = nt\bar{t}/4\pi$ instead of $\ell = nV^2/4\pi$. One can conclude that we only renormalize the mean free path if we go from second order Born to the full Born series. This can also be shown on a much more general level in the non-linear sigma model [87]. The number of diagrams increases substantially as compared to second order Born. Most of the times this is not a problem as once we did the calculation, we only need the sum of diagrams, Eq. (3.12). (The calculation in chapter 5 is a different story as we will undress the boxes again. There we will stick to second order Born.) For the six-point vertex we have not checked whether the second and full Born approximation give the same result; there are (at least) 64 diagrams (excluding rotations)!

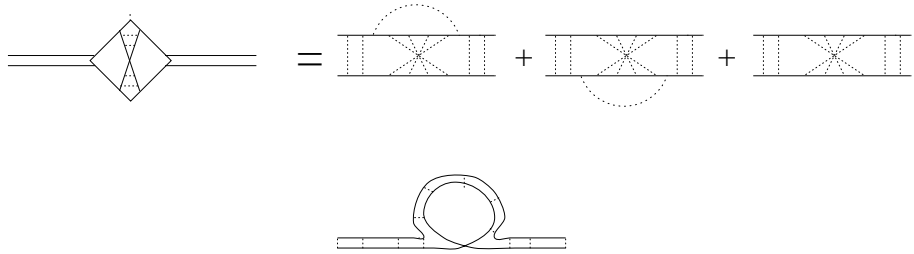

Figure 3.4: The first order correction to the conductivity in the Hikami formalism. On the right we have written out the vertices in detail in second order Born approximation. On the second line the first r.h.s. diagram is redrawn, one sees that it is a loop effect.

## 3.3    Corrections to the conductivity

In the original works of Gor'kov *et al.* and Hikami the vertices are introduced to calculate to weak-localization corrections to the conductivity. We already said that the most crossed diagrams give the largest correction to the conductivity. They increase the return probability of the intensity above the value obtained from diffusion theory. In the Hikami vertex formalism we have drawn these diagrams in Fig. 3.4. Hikami calculated these corrections up to second order, i.e. with two Hikami-boxes. In this work we do not consider this type of higher order loop effects. We suppose that we are far from localization, so that we can restrict ourselves to the leading processes. Then the vertices only show up in the interaction between separate diffusons.

## 3.4    Appendix

We calculate the integrals

$$I_{k,l} \equiv \int \frac{\mathrm{d}^3 \mathbf{p}}{(2\pi)^3} G^k(\mathbf{p}) G^{*\,l}(\mathbf{p}), \tag{3.13}$$

where

$$G(\mathbf{p}) = [\mathbf{p}^2 - k^2 - nt]^{-1} \qquad G^*(\mathbf{p}) = [\mathbf{p}^2 - k^2 - n\overline{t}]^{-1}. \tag{3.14}$$

For the simplest integral, $I_{1,1}$ we find

$$I_{1,1} = \int \frac{\mathrm{d}^3 \mathbf{p}}{(2\pi)^3} G(\mathbf{p}) G^*(\mathbf{p}) = \int_{-\infty}^{\infty} \frac{\mathrm{d}p}{(2\pi)^2} \frac{p^2}{(p^2 + \mu^2)(p^2 + \overline{\mu}^2)}. \tag{3.15}$$

The sum of the residues yields $1/4\pi(\mu + \overline{\mu})$, and the optical theorem brings

$$\mu + \overline{\mu} = \frac{-in(t - \overline{t})}{2k} = \frac{n\,\mathrm{Im}t}{k} = \frac{n\,t\,\overline{t}}{4\pi} = \frac{1}{\ell}. \tag{3.16}$$

Therefore,

$$I_{1,1} = \frac{\ell}{4\pi}.$$

Or, with absorption, $\mu + \overline{\mu} = [\ell(1 - \kappa^2\ell^2/3)]^{-1}$, yielding

$$I_{1,1} = \frac{1}{4\pi(\mu + \overline{\mu})} = \frac{\ell}{4\pi}\left(1 - \frac{1}{3}\kappa^2\ell^2\right) \tag{3.17}$$

The $I_{k,l}$ can be found from the $I_{1,1}$ integral as

$$I_{k+1,l} = \frac{-1}{2k\mu}\frac{\mathrm{d}}{\mathrm{d}\mu}I_{k,l} \; ; \qquad I_{k,l+1} = \frac{-1}{2l\overline{\mu}}\frac{\mathrm{d}}{\mathrm{d}\overline{\mu}}I_{k,l}. \tag{3.18}$$

With differentiation we now find without absorption

$$
\begin{array}{llll}
I_{1,1} = & \frac{1}{4\pi(\mu+\overline{\mu})} = \frac{\ell}{4\pi}, & I_{2,2} = & \frac{1}{8\pi\mu\overline{\mu}(\mu+\overline{\mu})^3} \approx \frac{\ell^3}{8\pi k^2} \\
I_{1,2} = & \frac{1}{8\pi\overline{\mu}(\mu+\overline{\mu})^2} \approx \frac{-i\ell^2}{8\pi k}, & I_{2,1} = & \frac{1}{8\pi\mu(\mu+\overline{\mu})^2} \approx \frac{i\ell^2}{8\pi k} \\
I_{1,3} = & \frac{1}{16\pi\overline{\mu}^2(\mu+\overline{\mu})^3} \approx \frac{-\ell^3}{16\pi k^2}, & I_{3,1} = & \frac{1}{16\pi\mu^2(\mu+\overline{\mu})^3} \approx \frac{-\ell^3}{16\pi k^2} \\
I_{2,3} = & \frac{3}{32\pi\mu^1\overline{\mu}^2(\mu+\overline{\mu})^4} \approx \frac{-3i\ell^4}{32\pi k^3}, & I_{3,2} = & \frac{3}{32\pi\mu^2\overline{\mu}^1(\mu+\overline{\mu})^4} \approx \frac{3i\ell^4}{32\pi k^3} \\
I_{1,4} = & \frac{1}{32\pi\overline{\mu}^3(\mu+\overline{\mu})^4} \approx \frac{i\ell^4}{32\pi k^3}, & I_{4,1} = & \frac{1}{32\pi\mu^3(\mu+\overline{\mu})^4} \approx \frac{-i\ell^4}{32\pi k^3} \\
I_{2,4} = & \frac{1}{16\pi\mu\overline{\mu}^3(\mu+\overline{\mu})^5} \approx \frac{-\ell^5}{16\pi k^4}, & I_{4,2} = & \frac{1}{16\pi\mu^3\overline{\mu}(\mu+\overline{\mu})^5} \approx \frac{-\ell^5}{16\pi k^4} \\
I_{3,3} = & \frac{3}{32\pi\mu^2\overline{\mu}^2(\mu+\overline{\mu})^5} \approx \frac{3\ell^5}{32\pi k^4}
\end{array}
$$

We will also use

$$\int \frac{\mathrm{d}^3\mathbf{p}}{(2\pi)^3}(\mathbf{p}\cdot\mathbf{q})^2 G^k(\mathbf{p})G^{*l}(\mathbf{p}) = \frac{1}{3}k^2 q^2 I_{k,l}, \tag{3.19}$$

that has a factor $1/3$ from the angular average.

With absorption there is generally a prefactor

$$I_{n,m} \propto (1 - \kappa^2\ell^2/3)^{n+m-1}, \tag{3.20}$$

where we have assumed that $\kappa\ell \ll 1$. However in our approximation this prefactor only has to be included for terms not proportional to $q$ or $\Omega$, as we work in first order of $M\ell$.

### 3.4.1 Beyond the diffusion limit

In the previous part we assumed that $q\ell \ll 1$, in real space this assumption means that variations of the diffusons are small on the length scale of one mean free path. (This is the same approximation that we made for the diffusons ). Yet this inequality is not always fulfilled, sometimes variations are large on mean free path. This is can happen if we integrate over the distance between two Hikami boxes, this we will see in chapter 5. The diffusion approximation results there in a divergent term. (In principle the diffusion approximation does also not hold near the boundary, but we won't consider these small effects.) We will need the more general integrals $I_{kl}^{mn}$, which we define as

$$I_{k,l}^{m,n}(q) \equiv \int \frac{\mathrm{d}^3\mathbf{p}}{(2\pi)^3}G^k(\mathbf{p})G^{*l}(\mathbf{p})G^m(\mathbf{p}+\mathbf{q})G^{*n}(\mathbf{p}+\mathbf{q}). \tag{3.21}$$

With the properties

$$I_{m,n}^{k,l}(q) = I_{k,l}^{m,n}(q), \quad I_{n,m}^{l,k}(q) = (-1)^{k+l+m+n}I_{k,l}^{m,n}(q). \tag{3.22}$$

| | |
|---|---|
| $I_{0,1}^{1,0} = \frac{\ell}{4\pi^3} A_1$ | $I_{0,2}^{1,0} = \frac{-i\ell^2}{8\pi k} A_2$ |
| $I_{0,2}^{2,0} = \frac{\ell^3}{8\pi k^2} A_3$ | $I_{1,1}^{0,1} = \frac{-i\ell^2}{8\pi k} A_1$ |
| $I_{1,1}^{0,2} = \frac{-\ell^3}{16\pi k^2} A_2$ | $I_{1,1}^{1,1} = \frac{-\ell^3}{8\pi k^2} A_1$ |
| $I_{1,2}^{0,0} = \frac{-i\ell^2}{8\pi k}$ | $I_{1,2}^{0,1} = \frac{-\ell^3}{16\pi k^2} A_1$ |
| $I_{1,2}^{1,0} = \frac{\ell^3}{16\pi k^2}[A_1 + A_2]$ | $I_{1,2}^{2,0} = \frac{i\ell^4}{32\pi k^3}[A_2 + 2A_3]$ |
| $I_{1,2}^{1,1} = \frac{-i\ell^3}{32\pi k^3}[2A_1 + A_2]$ | $I_{1,2}^{2,1} = \frac{\ell^5}{32\pi k^4}[A_1 + A_2 + A_3]$ |

Table 3.1: The short distance diagrams can be factorized in these integrals, defined in Eq.(3.21); $A_i$ is defined in Eq. (3.23).

In the calculation of the diffuson and the Hikami vertices in the previous sections we expanded the integrals in $q\ell$, but if we are after contributions for $q \sim 1/\ell$, this expansion is not allowed. We can still assume that $q \ll k$, as we never will need the physics on length scales comparable to the wavelength, but comparable to one mean free path only.

The integrals needed in the calculation of chapter 5 are given in Table 3.1. We define the angular average $A_i$ as

$$A_i(q) = \frac{1}{4\pi} \int d\hat{\mathbf{p}} \frac{1}{(1 + \ell\mathbf{q} \cdot \hat{\mathbf{p}})^i} \qquad (i = 1, 2, 3).$$ (3.23)

The integral for $i = 1$ yields the diffuson kernel $A_1 = \arctan(q\ell)/q\ell$. The internal diffuson now reads

$$\mathcal{L}_{\text{int}}(q) = \frac{4\pi}{\ell} \frac{1}{1 - A_1}.$$ (3.24)

We recover previous results by expansion in $(q\ell)$. This yields $\mathcal{L}_{\text{int}} = 12\pi/(\ell^3 q^2)$, in agreement with the bulk solution of (2.39).

The integrals relate to the integrals of the previous section when we expand again in $q\ell$

$$I_{kl}^{mn} \overset{q\ell \ll 1}{=} I_{(k+m),(l+n)} + \frac{4}{3}q^2 \left[ mn I_{(k+m+1),(l+n+1)} \right.$$
$$\left. + \frac{1}{2}m(m+1)I_{(k+m+2),(l+n)} + \frac{1}{2}n(n+1)I_{(k+m),(l+n+2)} \right].$$ (3.25)

In the non-linear sigma model the vertices in the expansion beyond $q^2$ terms are often called higher gradient vertices, as they correspond in real space to higher order derivatives. It was noted by Altshuler et al. [88, 23] that they become important in the higher moments of the conductivity, see also chapter 7.

# 4

# Short and long range correlations: $C_1$, $C_2$

## 4.1  Physics of the correlation functions

A nice feature of optical mesoscopic systems above electronic systems is the possibility to measure several transmission quantities. This depends on whether integration over incoming and/or outgoing directions is performed. In electronic systems one usually measures the conductance of the sample. This is done by connecting incoming and outgoing sides to a clean electron "bath". All electrons are collected and all angular dependence is lost.[1] In optical systems one usually measures angular resolved, but angular integration is also possible. The correlation functions of the different transmission quantities, their magnitudes, their decay rates, and even the underlying diagrams are all very different. In this chapter we discuss the angular transmission and total transmission correlation functions.

### 4.1.1  Angular resolved transmission: speckle

If a monochromatic plane wave shines on a disordered sample we see large intensity fluctuations, so called "speckles", in the transmitted beam. The speckle pattern is wildly fluctuating (as function of the frequency of the light or the outgoing angle). Simply shining a laser on the (rough) wall already produces such a speckle pattern. The typical diameter of a spot at the outgoing surface is $\lambda$. In this chapter we are interested in the correlation between speckle patterns of two different beams. The beams may have different incoming angles, different frequencies or different positions.

There are short, long and "infinite" range contributions to the correlation function of the speckles. In a recent review by Berkovits and Feng [90] the different correlations and their physical interpretation are discussed. We denote the transmission from incoming channel $a$ (wave coming in under angles $\theta_a$, $\phi_a$) to outgoing channel $b$ (waves transmitted

---

[1] An exception is the recent measurement of Gao *et al.* [89]. They injected electrons into a two-dimensional sample using a so called quantum point contact. A quantum point contact is a very small connection between two metals, such that only one electron mode can be selected. The electrons were also collected from the sample with a quantum point contact. A fluctuation very similar to the speckle pattern was seen.

into angles $\theta_b$, $\phi_b$) as $T_{ab}$. the correlation functions can then be classified as [90, 91]

$$\frac{\langle T_{ab}(\omega) T_{cd}(\omega + \Delta\omega)\rangle}{\langle T_{ab}(\omega)\rangle\langle T_{cd}(\omega + \Delta\omega)\rangle} = 1 + C_1^{abcd}(\Delta\omega) + C_2^{abcd}(\Delta\omega) + C_3^{abcd}(\Delta\omega). \qquad (4.1)$$

Feng *et al.* [91] originally put this expression forward as an expansion in $1/g$, the $C_1$ being of order one, the $C_2$ of order $1/g$, and the $C_3$ of order $1/g^2$. We present a sketchy picture of the diagrams of the different correlations in Fig. 4.1. The largest part of transport are the independent diffusons, yet correlations are present, mixing diffusons with different frequencies or angles.

The unit contribution in Eq. (4.1) just comes from the uncorrelated product of average intensities. The $C_1$-term in the correlation function is the most important one if the angular resolved transmission $T_{ab}$ is measured. The $C_1$ is namely unity when both the incoming directions $a$ and $c$, and the outgoing directions $b$ and $d$ are pairwise the same, and $\Delta\omega$ is zero. If one changes the frequency of the incoming beam, the speckles of the outgoing beam deform. Eventually the speckle disappears and the correlation vanishes exponentially. This effect is the short ranged $C_1$ contribution. Also of interest is the case of the angular $C_1$ correlation. This time we keep the frequency fixed. If we change the angle of the incoming beam, the speckles change due to two effects. First, the speckle follows the incoming beam. This is also known as the memory effect, which one can even see by the naked eye. Secondly, the speckle pattern also deforms, i.e. it de-correlates, this is again the $C_1$. The $C_1$ is a sharply peaked function non-zero only if the angle of the incoming and outgoing channel are changed by the same amount. Diagrammatically, the $C_1$ factorizes in two disconnected diagrams, see Fig. 4.1. Just as the unit contribution in Eq. (4.1), it is equal to the product of two averages.

Theoretical studies of the $C_1$ were first done by Shapiro [92]. Experimental work on the $C_1$ as function of angle, explicitly showing the memory effect, was done by Freund *et al.* [93]. Albada *et al.* studied the frequency dependence of the $C_1$ [71]. Later, effects of absorption and, especially, internal reflection were studied and turned out to be of practical importance [94, 68].

### 4.1.2   Total transmission correlation: $C_2$

The $C_2$ gives the long range correlations in the speckle pattern. Consider again a single direction in, single direction out experiment. Yet this time we look at the cross correlation of two speckles far apart. As there is an angle difference in the outgoing channel, though not in the incoming beam, $C_1$ is now absent. Instead one sees a much weaker correlation. As the frequency shifts, this correlation decays algebraically. This is the $C_2$ contribution, it describes correlations between speckles far apart. In a single channel in, single channel out experiment it is only possible to see the weak effects of these higher order correlations in very strongly scattering media [95], i.e. if the sample is rather close to Anderson localization. The $C_2$ can be measured more easily in a set-up using one incoming channel and integrating over all outgoing channels. By collecting the outgoing light, one measures the total transmission $T_a = \sum_b T_{ab}$. In this set-up the $C_2$ correlation function, which has a much smaller peak value but is long ranged, contributes for all outgoing angles. The $C_2$ is now the leading correlation as the sharply peaked $C_1$ is overwhelmed. As we already said in section 2.5, only outgoing diffusons with no transverse momentum and no frequency difference are leading in the total transmission. This way the phases of the outgoing amplitudes are exactly opposite and thus cancel. From Fig. 4.1 one sees indeed that it holds for the $C_2$ diagram (outgoing lines of similar style pair), but not for the $C_1$ diagram. The long range character arises due to interference of the diffuse light paths. The $C_2$ correlation, which still depends on the angles of the incoming beams $a$ and $c$, is of order $g^{-1}$. The $C_2$ corresponds to a diagram where the two incoming diffusons interact through a Hikami-vertex. In this vertex the diffusons exchange amplitudes. Rather than thinking of this exchange as a real physical process, we consider this more like something left after subtraction of the average process, see Fig. 4.1.

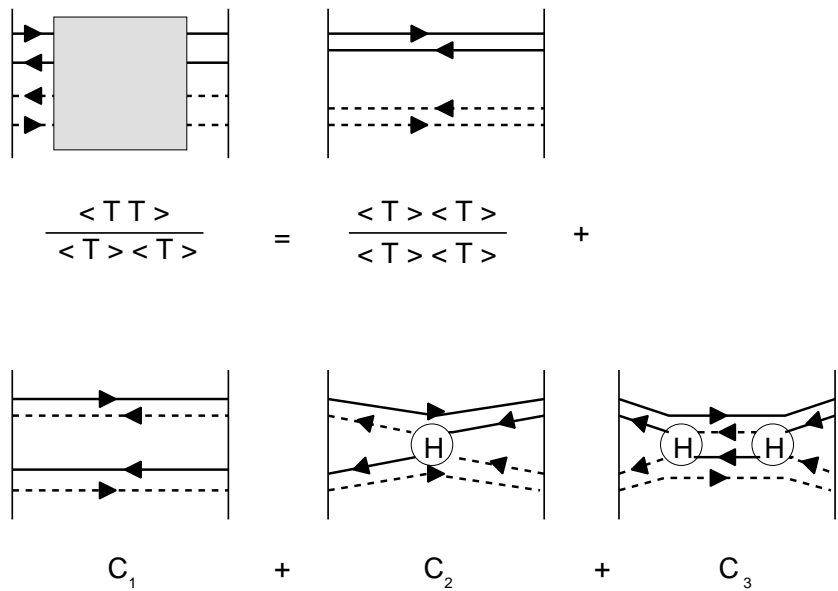

Figure 4.1: A schematic picture of the different correlation functions present in the transmission of two intensities. The arrows distinguish between advanced and retarded propagators. Propagators with equal transverse momentum or frequency have the same style of line (dashed or solid). The black box contains in principle all contributions to $\langle T_{ab}T_{cd}\rangle$. The main part just factorizes, but correlations are present. Below, we will present the diagrams for the $C_2$ and $C_3$ in more detail.

The long range $C_2$ correlation function was first studied by Stephen and Cwilich [96]. Zyuzin and Spivak introduced a Langevin approach to simplify the calculation of correlation functions [97]. Pnini and Shapiro applied this method to calculate the correlation functions of light transmitted through and reflected from disordered samples [98]. The $C_2$ correlation functions were measured in several experiments. For optical systems Albada *et al.* performed measurements [71, 72]). Microwave experiments were performed by Genack *et al.* [99], [100]. In several papers effects of absorption and internal reflection was studied, see Pnini and Shapiro [101], Lisyansky and Livdan [53], Zhu *et al.* [54], and Ref. [68]. They showed that absorption and internal reflection, neglected in the earliest calculations, significantly reduce the correlations.

Finally, the $C_3$ term in Eq. (4.1) is dominant when the incoming beam is diffuse,

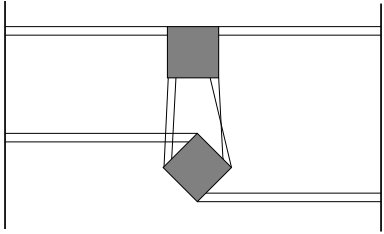

Figure 4.2: A contribution of order $1/g^2$ to the $C_2$ correlation. By following the amplitudes, one can check that the incoming pairings $ij^*$ and $ji^*$, change on the outgoing side into pairings $ii^*$ and $jj^*$, as is required for $C_2$ diagrams. It is thus not contributing to the $C_3$, which is also of order $1/g^2$, but requires diagrams without amplitude exchange.

and one collects all outgoing light. Then one measures, just as in electronic systems, the conductance $g = \langle T \rangle = \sum_{a,b} T_{ab}$. In that measurement contributions to the correlation where $a$ and $c$ are arbitrary far apart are dominant. Though $C_3$ is of order $g^{-2}$, it dominates over the $C_1$ and $C_2$ terms as it has contributions for all incoming and outgoing angles. Therefore it is sometimes called the 'infinite' range correlation. In contrast to the $C_2$, now also the incoming amplitudes must have opposite phase. This occurs in a diagram where the two incoming diffusons interact twice. Note that a loop occurs, see Fig. 4.1. Apart from the normalization to the average in the $C_3$, the $C_3$ describes just the well known Universal Conductance Fluctuations, or UCF [102, 103, 104]. We discuss the $C_3$ in detail in chapter 5.

In principle the $C_2$ and $C_3$ correlations are present as a sub-leading term in the angular correlation function, but there are also other contributions of order $1/g$ and $1/g^2$, respectively, to the correlation Eq.(4.1). The weak localization correction, see section 2.7, is an example of a $1/g$ contribution to the $C_1$. We have drawn a $1/g^2$ contribution to the $C_2$ in Fig. 4.2. In a paper by Garcia *et al.* [100] these corrections are studied. Therefore, the interpretation of the $C$'s as an expansion in $1/g$ is somewhat impractical. Rather we define the $C_1$, $C_2$, and $C_3$ as the leading term in the correlate of $T_{ab}$, $T_a$, and $T$, respectively. Equivalently, the $C_1$ contains disconnected diagrams; incoming and outgoing amplitudes *must* have the same pairing. the $C_2$ contains connected diagrams swapping the initial pairing, and the $C_3$ contains the connected diagrams in which incoming and outgoing pairings are the same.

## 4.2    Calculation of the $C_1$ correlation

The calculation of the $C_1$ is rather simple as it is just the product of two independent diffusons with exchanged partners. Both diffusons consist of one amplitude from one beam and one complex conjugated amplitude from the other beam. Due to momentum conservation in each diffuson, the perpendicular momentum difference of the two incoming beams, $Q$, is equal to the difference of the outgoing momenta. Therefore, the $C_1$ is only non-zero if incoming and outgoing angles are shifted equally during the experiment.

We calculate the normalized product of a diffuson with 'squared mass' $M^2 = Q^2 + \kappa^2 + i\Omega$ and the complex conjugated diffuson. Due to outgoing Green functions, $C_1$ is a

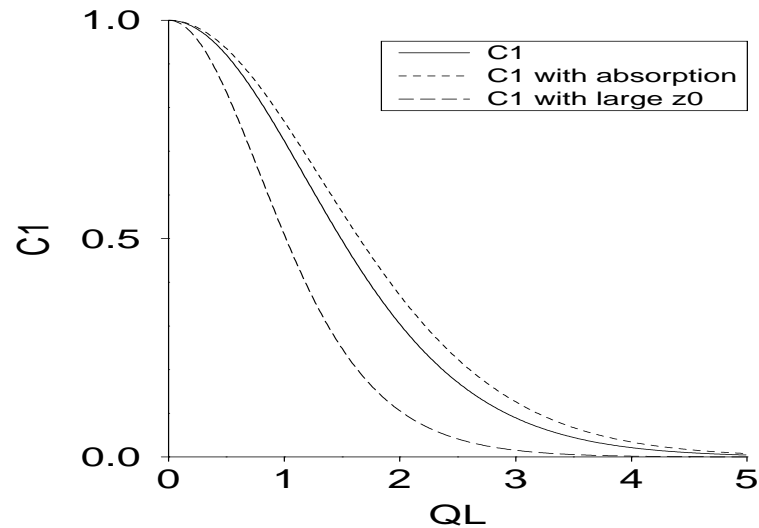

Figure 4.3: $C_1$ angular correlation function plotted against the scaled perpendicular momentum difference. Solid line: small $z_0$, no absorption. Short dashed line: absorption ($\kappa = 2/L$), small $z_0$. Long dashed line: large skin-layers ($z_0 = L/3$), no absorption.

Laplace transform of the diffusons. For small mass ($M\ell \ll 1$) the same integrals occur in the numerator and denominator. The result can be expressed by the values on the boundary $z = L$. This yields:

$$
\begin{aligned}
C_1(M) &= \left| \frac{T_{ab}(M)}{T_{ab}(\kappa)} \right|^2 = \left| \frac{\mathcal{L}(z=L, M)}{\mathcal{L}(z=L, \kappa)} \right|^2 \\
&= \frac{|M|^2}{\kappa^2} \left| \frac{(1 + \kappa^2 z_0^2) \sinh \kappa L + 2\kappa z_0 \cosh \kappa L}{(1 + M^2 z_0^2) \sinh ML + 2Mz_0 \cosh ML} \right|^2 .
\end{aligned}
\tag{4.2}
$$

In Fig. 4.3 we plotted this function for various situations. In thick samples in the limit of $Mz_0 \ll 1$ and no absorption this reduces to the result already known from the diffusion approximation [91]

$$
C_1(M) = \left| \frac{ML}{\sinh ML} \right|^2 .
\tag{4.3}
$$

For large angle or frequency difference ($|M|L \gg 1$), $C_1$ decays exponentially, as $C_1(M) \sim e^{-2\mathrm{Re}(M)L}$. The spatial correlations of beams can be extracted from the Fourier transform of the angular correlation function.

## 4.3 Calculation $C_2$ correlation

The diagram of the long range correlation $C_2$ is depicted in Fig. 4.4, Remember that the $C_1$ is zero if incoming and outgoing angle are changed unequally, but the $C_2$ is connected, allowing momentum flow from one diffuson to the other. It is the simplest diagram that is non-zero if one changes incoming and outgoing angle by a different amount. $C_2$ stems from the interaction between two diffusons, which exchange partners somewhere inside

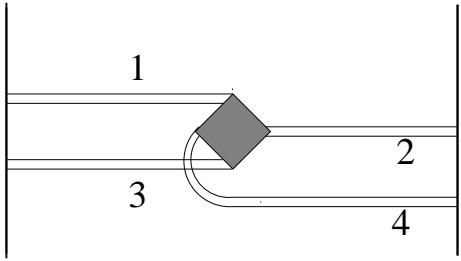

Figure 4.4: The diagram of the long range $C_2$ correlation function. The shaded box denotes the Hikami four-vertex, worked out in Figs. 3.1 and 3.3.

the slab. The Hikami box describes this exchange. The full expression reads

$$
\begin{aligned}
C_2 \; = \; & \frac{1}{\langle T \rangle \langle T \rangle} \int_0^L \mathrm{d}z \mathrm{d}z_1 dz_2 dz_3 dz_4 \, \mathcal{L}_1(z_1, M_1) \mathcal{L}_3(z_3, M_3) H_4(z_1, z_2, z_3, z_4) \times \\
& \delta(z - z_1)\delta(z - z_2)\delta(z - z_3)\delta(z - z_4) \mathcal{L}_2(L - z_2, M_2) \mathcal{L}_4(L - z_4, M_4) \;, \quad (4.4)
\end{aligned}
$$

where we used the fact that also $C_2$ can be expressed as the ratio of the values at the slab boundary. We have labelled the incoming diffusons with odd numbers, and the outgoing diffusons with the even numbers; a convention used throughout this work. Attaching diffusons to the Hikami box allows a simplification of the expression for the box. We calculate the $C_2$ by inserting $M_{2,4}^2 = Q^2 + \kappa^2 \pm i\Omega$, $M_1 = M_3 = \kappa$. In the $z$−coordinate representation the box then reads, see chapter 3,

$$
H_4(z_1, z_2, z_3, z_4) = \frac{\ell^5}{48\pi k^2}(\partial_{z_1}\partial_{z_3} + \partial_{z_2}\partial_{z_4} + Q^2) \;. \tag{4.5}
$$

This gives one contribution to the $C_2$, another is obtained by interchanging incoming and outgoing beams. But as in a typical $C_2$ experiment one integrates the outgoing beam over all directions, this second term averages out.

The above expression for $C_2$ is similar to the one obtained in the Langevin approach, as one can see by comparing formula (39) of reference [98] to (4.5) and (4.4) in this chapter. Indeed, in the Langevin approach one assumes that there is a macroscopic intensity, which describes the average diffusion, and an uncorrelated random noise current superposed. In our approach we have a slowly varying diffuse intensity, while the point-like, uncorrelated random interactions originate from the Hikami-box.

We obtain the resulting general expression for $C_2$ by inserting the appropriate diffusons and the Hikami-box Eq. (4.5) into Eq. (4.4). The result is rather lengthy and given in the appendix.

Let us first study the top of the correlation function. By its definition the top of the correlation functions corresponds to the second cumulant of the total transmission distribution function. In the plane wave limit of the incoming beam, all transverse momenta are absent, we also neglect absorption for the moment. Mathematically these approximations correspond to all $M$ being equal zero. The diffusons are now simple linear functions, given by Eq. (2.61) and (2.76). Neglecting internal reflection, one obtains the known result for the second cumulant [96]

$$
\langle\langle T_a^2 \rangle\rangle \; = \; \frac{\langle T_a^2 \rangle - \langle T_a \rangle^2}{\langle T_a \rangle^2}
$$

$$
\begin{aligned}
&= \langle T_a \rangle^{-2} \int \int \mathrm{d}x \mathrm{d}y \int_0^L \mathrm{d}z H_4 \mathcal{L}_1(z) \mathcal{L}_2(z) \mathcal{L}_3(z) \mathcal{L}_4(z) \\
&= \frac{1}{gL^3} \int_0^L \mathrm{d}z [z^2 + (L-z)^2] = \frac{2}{3} g^{-1},
\end{aligned}
\tag{4.6}
$$

where double brackets denote cumulants normalized to the average. The correlation is proportional to $1/g$, the inverse power of $g$ counts the number of four-point vertices. It can be looked upon as the chance that two intensities interfere. Later we see more calculations supporting this interpretation.

We discuss some other simplified cases. Consider the case of angular correlations and no absorption but taking the boundary effects into account:

$$
C_2(Q) = \frac{3\pi L}{k^2 l A} F_2(QL),
\tag{4.7}
$$

in which we define the dimensionless function $F_2$

$$
\begin{aligned}
F_2(QL) &= [\sinh 2QL - 2QL + Qz_0 \left( 6 \sinh^2 QL - 2Q^2 L^2 \right) \\
&\quad + 4Q^2 z_0^2 \left( \sinh 2QL - QL \right) + 6Q^3 z_0^3 \sinh^2 QL + Q^4 z_0^4 \sinh 2QL] \times \\
&\quad [2QL \left\{ (1 + Q^2 z_0^2) \sinh QL + 2Qz_0 \cosh QL \right\}^2]^{-1}.
\end{aligned}
\tag{4.8}
$$

which decays like $1/Q$ for large Q (*i.e.* large angles). However, note that it still should hold that $Q\ell \ll 1$. Neglecting boundary effects we find the result known from diffusion approximation [101]:

$$
F_2(QL) = \frac{\sinh(2QL) - 2QL}{2QL \sinh^2(QL)}.
\tag{4.9}
$$

In figure 4.5 we have plotted the $F_2$ function for various situations.

Above, we have studied the correlation as function the two-dimensional momentum $Q$. One also can study the real space correlations, which relate the momentum space correlation by a Fourier transform. The functional form in real space is of roughly similar shape as Fig. 4.5, with typical decay length $L$ [96].

We can also obtain frequency correlations. Without absorption, neglecting boundary effects, one finds [72]

$$
F_2(\Omega) = \frac{2}{\sqrt{2\Omega}\, L} \left( \frac{\sinh(\sqrt{2\Omega}\, L) - \sin(\sqrt{2\Omega}\, L)}{\cosh(\sqrt{2\Omega}\, L) - \cos(\sqrt{2\Omega}\, L)} \right).
\tag{4.10}
$$

As one can see in the appendix, the functional form changes when including boundary effects.

Without internal reflections ($m = 1$) at large $M$, $C_2$ correlation function decays as $1/M$. Lisyansky and Livdan found that this behavior changes at high internal reflectivity [53]. We do not confirm such a behavior, as at large $z_0$ the decay remains of order $1/M$.

### 4.3.1  Focus

Let us now study the influence of the beam profile on the correlation. If the spot of the incoming beam is finite, amplitudes with different angles, i.e. different transverse momenta are present. They can combine into incoming diffusons with perpendicular momentum. We suppose that the incoming beam has a Gaussian profile. We decompose it into plane waves defined in Eq. (2.48) (for convenience we assume the average incidence perpendicular)

$$\psi_{\mathrm{in}} = \frac{2\pi}{W} \sum_a \phi(Q_a)\psi_{\mathrm{in}}^a, \qquad \phi(Q) = \frac{\rho_0}{\sqrt{2\pi}}\mathrm{e}^{-Q^2\rho_0^2/4}, \qquad (4.11)$$

where $\rho_0$ is the beam diameter. In order to have two diffusons with a momentum $Q$ and $-Q$, we find that the four incoming amplitudes combine to a weight function

$$\int \mathrm{d}^2P_1\,\mathrm{d}^2P_3 \; \phi(P_1)\phi^*(P_1+Q) \; \phi(P_3)\phi^*(P_3-Q) = \mathrm{e}^{-\rho_0^2Q^2/4}. \qquad (4.12)$$

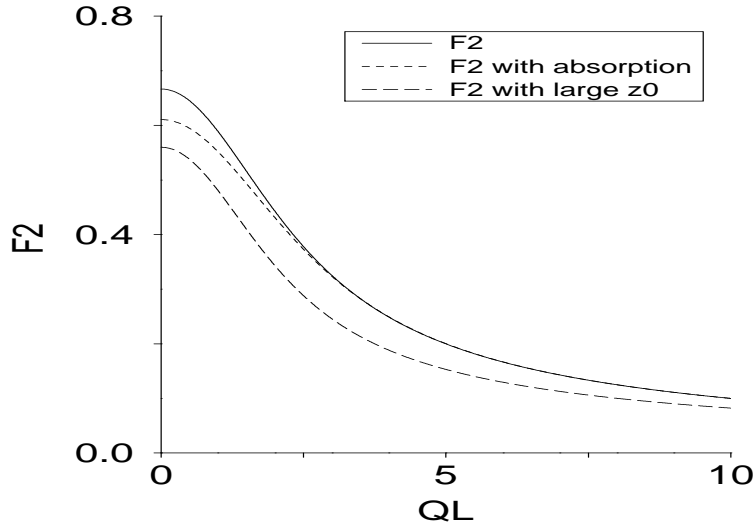

Figure 4.5: $F_2$ angular correlation function plotted against the scaled perpendicular momentum difference. Solid line: small $z_0$, no absorption. Short dashed line: absorption ($\kappa = 2/L$), small $z_0$. Long dashed line: large skin-layers ($z_0 = L/3$), no absorption.

We get the correlation function by integrating the $Q$-dependent correlation function over the momentum with the corresponding weight.

We again neglect boundary layers. For a Gaussian beam profile we find for the top of the correlation [72]

$$\langle\langle T_a^2 \rangle\rangle = \frac{\rho_0^2}{4\pi g} \int \mathrm{d}^2Q \; \mathrm{e}^{-\rho_0^2Q^2/4} F_2(QL), \qquad (4.13)$$

with $F_2(x) = [\sinh(2x)-2x]/[2x\sinh^2 x]$. If the incoming beam is again very broad, $\rho_0 \gg L$, only the term $F_2(Q=0) = 2/3$ contributes and one recovers the plane wave behavior $\langle\langle T_a^2 \rangle\rangle = 2/3g$. Note that this agreement is found by identifying the area of a Gaussian profile with $A = \pi\rho_0^2$. The second cumulant decreases as $1/\rho_0^2$ at large $\rho_0$. In a real space picture evidently the correlation increases if the two incoming channels are closer to each other, i.e. if the beam diameter is smaller. At smaller beam diameters the top of the correlation is proportional to $1/\rho_0$, and would diverge in our current approximation, in which $Q\ell \sim \ell/\rho_0 \ll 1$. In the experiments of De Boer, Van Albada and Lagendijk [71, 72, 105], they kept the focus small in order to minimize the dimensionless conductance $g$ and therefore to maximize the fluctuations.

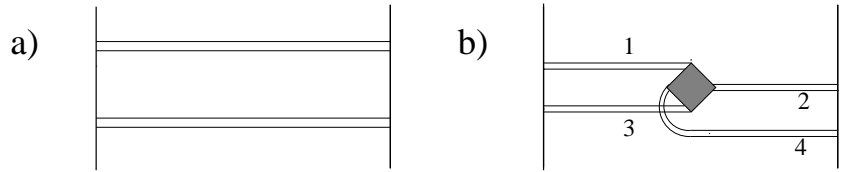

Figure 4.6: The two contributions to the second moment of the total transmission. In diagram (a) the transmission channels are independent; this process is of order unity and is almost completely reducible to the mean value squared. Diagram (b) corresponds to two interfering channels; this is the second cumulant or $C_2$ diagram, it is of order $1/g$. The close parallel lines are diffusons.

### 4.3.2 Contributions from disconnected diagrams to $C_2$

So far we calculated the leading contributions to the long range correlation. They are the connected diagram in Fig. 4.6(b). Yet there are also contributions to the $C_2$ from the disconnected $C_1$ diagrams of Fig. 4.6(a). The disconnected diagram corresponds to cumulant contributions that are not due to interference inside the slab. It describes effects that have little to do with the interference effects we are after. Here we calculate the contribution and show that it is small.

We discuss the contribution of these diagrams only for the top of the $C_2$ correlation function; due to the much slower decay of $C_2$ as compared to the $C_1$ the correction should be the largest at the top of the $C_2$.

As a start we use the model of a waveguide. We assume that the disorder couples one incoming mode $a$ to all outgoing modes. A waveguide has discrete modes, and for the moment we assume that different outgoing modes are uncorrelated. The second moment $\langle T_a^2 \rangle$ splits into a connected part $\langle T_a^2 \rangle_{\mathrm{con}}$, Fig. 4.6(b) and a disconnected part $\langle T_a^2 \rangle_{\mathrm{dis}}$, Fig. 4.6(a). The total transmission is the summation over all outgoing modes $T_a = \sum_b T_{ab}$. The disconnected part of the second moment is

$$
\begin{aligned}
\langle T_a^2 \rangle_{\mathrm{dis}} &= \sum_{b_1 \neq b_2}^{N,N} \langle T_{ab_1} T_{ab_2} \rangle + \sum_{b_1 = b_2}^{N} \langle T_{ab_1} T_{ab_2} \rangle \\
&= N(N-1)\langle T_{ab} \rangle^2 + N\langle T_{ab}^2 \rangle \\
&= \langle T_a \rangle^2 + N\langle T_{ab} \rangle^2,
\end{aligned} \tag{4.14}
$$

where $N$ is the number of modes supported by the waveguide. For the last equality sign we used that the averaged second moment of the intensity speckle is given by the speckle distribution function Eq. (7.1), $\langle T_{ab}^2 \rangle = 2\langle T_{ab} \rangle^2$. From Eq.(4.14) we see that the disconnected diagram $\langle T_a^2 \rangle_{\mathrm{dis}}$ does not completely factorize into the average squared $\langle T_a \rangle^2$. Therefore it contributes to the second cumulant. As shown above, the connected part of the second moment of the total transmission $\langle T_a^2 \rangle_{\mathrm{con}}$ is proportional to $L/N\ell$. For the sum of the disconnected and the connected contribution to the second cumulant

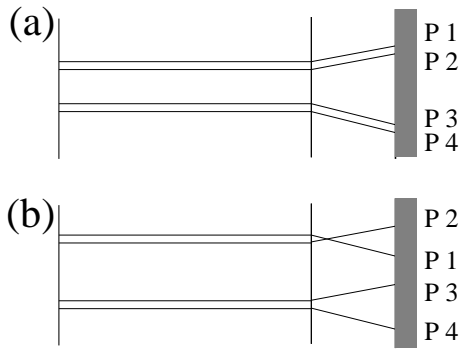

Figure 4.7: The disconnected contribution to the second moment. In (a) the two transmissions factorize into the average value squared, and does not contribute to the second cumulant. Diagram (b) is much smaller, but gives a contribution to the second cumulant. The amplitudes making up the two diffusons propagate out in different directions.

one thus finds

$$\langle\langle T_a^2 \rangle\rangle = \langle\langle T_a^2 \rangle\rangle_{\text{con}} + \langle\langle T_a^2 \rangle\rangle_{\text{dis}} = \frac{L}{2N\ell} + \frac{1}{N}, \qquad (4.15)$$

which also holds for plane wave case.

Now we turn to the more realistic situation of a diffusely scattering slab, with a finite focus of the incoming beam. The incoming beam broadens in the transverse direction by diffusion, changing the above result. To calculate the different contributions to the cumulants, we need the intensity distribution at the exit interface at transverse coordinates $R_1$ and $R_2$.

The amplitudes, making up each diffuson, can propagate from the outgoing surface in different directions, respectively $P_1$, $P_2$, $P_3$ and $P_4$.

$$\Psi(R_1)\mathrm{e}^{iP_1R_1}\Psi^*(R_1)\mathrm{e}^{-iP_2R_1}\Psi(R_2)\mathrm{e}^{iP_3R_2}\Psi^*(R_2)\mathrm{e}^{-iP_4R_2}. \qquad (4.16)$$

To get the contribution to the second moment of the total transmission, one first integrates over the transversal coordinates $R_1$ and $R_2$ to get the contribution of the whole exit interface to the intensity in a certain direction, then one integrates over all directions to get the total transmission. To obtain intensities the amplitudes need to be paired giving the following possibilities. The first possible pairing of the amplitudes is $P_1 = P_2$, $P_3 = P_4$, see Fig. 4.7(a), and brings

$$\int \mathrm{d}R_1 \int \mathrm{d}R_2 \, \langle\Psi(R_1)\Psi^*(R_1)\Psi(R_2)\Psi^*(R_2)\rangle = I^2(0), \qquad (4.17)$$

where we have defined $I(Q)$ as the transmission by a diffuson with transverse momentum $Q$. Including the incoming Gaussian beam profile, it is proportional to, see Eq.(2.42),(4.11)

$$I(Q) \propto \frac{|Q|\mathrm{e}^{-Q^2\rho_0^2/8}}{\sinh |Q|L}. \qquad (4.18)$$

Integrating over all outgoing directions results in

$$\frac{\langle T_a^2 \rangle}{\langle T_a \rangle^2} = \frac{\pi^2 k^4 I^2(0)}{\pi^2 k^4 I^2(0)}. \qquad (4.19)$$

But this is just the factorizing contribution $\langle T_a \rangle \langle T_a \rangle$, and therefore does not contribute to the second cumulant.

The second possible pairing of the amplitudes, $P_1 = P_4$, $P_2 = P_3$, does give a contribution to the second cumulant. It is the diagram in Fig. 4.7(b)

$$\int dR_1 dR_2 \ \langle \Psi(R_1) \Psi^*(R_1) \Psi(R_2) \Psi^*(R_2) \rangle e^{iR_1(P_1 - P_2)} e^{-iR_2(P_1 - P_2)} I(P_1 - P_2) I(P_2 - P_1). \tag{4.20}$$

Subsequent integration over the outgoing directions $P_1$ and $P_2$ yields

$$\langle T_a^2 \rangle = \pi k^2 \int_{|Q| < k} d^2Q \ I^2(Q). \tag{4.21}$$

The integral can be extended to infinity since $I(Q)$ is an exponentially decaying function. The contribution of the disconnected diagram Fig. 4.6(b) to the second cumulant is thus

$$\langle\langle T_a^2 \rangle\rangle_{\text{dis}} = \frac{\pi k^2 \int_{-\infty}^{\infty} d^2Q \, I^2(Q)}{\pi^2 k^4 I^2(0)} \equiv \frac{1}{N'} \tag{4.22}$$

In analogy to the waveguide, this result describes irreducible contributions from disconnected diagrams. It can be interpreted as the inverse of the number of independent speckle spots in transmission at the exit interface [106, 105].

We did not take the vector character of the light into account yet. The two independent polarizations of each outgoing direction effectively double the number of independent speckle spots with respect to Eq. (4.22), thus reducing Eq. (4.22) by a factor *2*.

## Reflection correlations

Similar to transmission correlations, also reflection correlations can be studied. We have seen before that the transmission corresponds to the diffusons, but in the reflection there are two intensity diagrams: The diffusons bring a constant background, the most crossed diagrams yield the enhanced backscatter cone, which is only of importance if incoming and outgoing angle are the same. This leads to more diagrams and to more peaks in the reflection correlation than in the transmission correlation.

For the $C_1$ one can both pair diffusons and most crossed diagrams, see Berkovits *et al.* for some theory [107]. Indeed in optical experiments two peaks were seen by Freund and Rosenbluh [108].

The $C_2$ in reflection was treated by Berkovits [109]. Measurement of the $C_2$ in reflection by collecting all reflected light is tricky as one does not want to interfere with the incoming beam. Nevertheless, the measurement is may be possible as a long range component in the angular resolved reflection (this however requires quite small values of $g$).

For both $C_1$ and $C_2$ in reflection we don't expect a very good agreement with the theory. The problem is that the precise behavior of the diffuson near the surface is important. The diffusion approximation $q\ell \ll 1$ is not valid anymore, in particular if the indices of refraction match (then the intensity gradient near the surface is the largest). The calculation of the $C_1$ is probably easily extended, yet for the $C_2$ one also would need the Hikami box beyond first order in $q\ell$.

## 4.4   Comparison with literature

When comparing our findings to results in literature, care has to be taken of the definition of the beam profile as this changes the prefactor of $C_2$. Also the definition of the transmission $T$ is different is most places, and not all results are normalized by the transmission as done here in formula (4.1). A diagrammatic analysis has previously been carried out by Stephen and Cwilich [96]. Their result exceeds our equations (4.7) and (4.9), by a factor 9. Pnini and Shapiro used the Langevin approach to calculate $C_2$ in the absence of internal reflections [98]. Their result agrees with ours in the limit $z_0 \ll L$. It can also be seen that equation (4.10) coincides with the Langevin approach of de Boer *et al.* [72]. In conclusion, we have shown that the diagrammatic and the Langevin approach lead to the same result for the calculation of $C_2$. We calculated the short and long range, angular and frequency correlations. Previous approaches are extended by the presence of absorption, skin layer effects and scattering beyond second order Born approximation.

# Appendix

Here we present the general expression for the long range correlation function, given by formula (4.4) with the diffusons of formula (2.64). The resulting expression is a fourth order polynomial in $z_0$. It is convenient to split $M_2$ and $M_4$ in their real and imaginary part, $M_{2,4} = \sqrt{Q^2 + \kappa^2 \pm i\Omega} \equiv a \pm ib$.

$$a = \sqrt[4]{(Q^2 + \kappa^2)^2 + \Omega^2} \, \cos(\varphi/2), \qquad b = \sqrt[4]{(Q^2 + \kappa^2)^2 + \Omega^2} \, \sin(\varphi/2)$$
$$\tan \varphi = \frac{\Omega}{Q^2 + \kappa^2} \, . \tag{4.23}$$

We find

$$C_2 = \frac{3\pi L}{k^2 l A} \frac{N_a + N_b + N_\kappa}{BL} \, , \tag{4.24}$$

where we have defined:

$$
\begin{aligned}
B &= [1 + 2z_0^2(a^2 - b^2) + z_0^4(a^2 + b^2)](\cosh 2aL - \cos 2bL) \\
&\quad + 2z_0[1 + z_0^2(a^2 + b^2)](a \sinh 2aL + b \sin 2bL) + 4z_0^2(a^2 + b^2)(\cosh 2aL + \cos 2bL) \\
N_a &= \frac{1}{2a(a^2 - \kappa^2)} \times \{[(2a^2 - \kappa^2) + z_0^2(8a^4 + 2a^2b^2 - b^2\kappa^2 - 3a^2\kappa^2 + \kappa^4) + \\
&\quad z_0^4(a^2 + b^2)(2a^4 - 2a^2\kappa^2 + \kappa^4)] \sinh 2aL + \\
&\quad [4z_0 a(3a^2 - \kappa^2) + 4z_0^3 a(3a^4 + a^2b^2 - 2a^2\kappa^2 + \kappa^4)] \sinh^2 aL\} \\
N_b &= \frac{1}{2b(b^2 + \kappa^2)} \times \{[-(2b^2 + \kappa^2) + z_0^2(2a^2b^2 + 8b^4 + a^2\kappa^2 + 3b^2\kappa^2 + \kappa^4) - \\
&\quad z_0^4(a^2 + b^2)(2b^4 + 2b^2\kappa^2 + \kappa^4)] \sin 2bL + \\
&\quad [4z_0 b(3b^2 + \kappa^2) - 4z_0^3 b(3b^4 + a^2b^2 + 2b^2\kappa^2 + \kappa^4)] \sin^2 bL\} \\
N_\kappa &= \frac{a^2 + b^2}{2\kappa(a^2 - \kappa^2)(b^2 + \kappa^2)} \times \{[-\kappa^2 - z_0^2(6a^2b^2 + 5a^2\kappa^2 - 5b^2\kappa^2 + \kappa^4) + 
\end{aligned}
$$

$$z_0^4 \kappa^2 (-2a^2b^2 - a^2\kappa^2 + b^2\kappa^2)] \sinh 2\kappa L +$$
$$[4z_0(-a^2b^2 - a^2\kappa^2 + b^2\kappa^2 - \kappa^4) + 4z_0^3(-3a^2b^2 - 2a^2\kappa^2 + 2b^2\kappa^2)] \sinh^2 \kappa L\} \quad (4.25)$$

Without internal reflection we find

$$
\begin{aligned}
C_2 \;=\; & \frac{3\pi}{2k^2 lA} \frac{1}{(\cos(2bL) - \cosh(2aL))} \times \{\frac{(\kappa^2 - 2a^2)}{a(a^2 - \kappa^2)} \sinh(2aL) \\
& + \frac{(\kappa^2 + 2b^2)}{b(\kappa^2 + b^2)} \sin(2bL) + \frac{\kappa(a^2 + b^2)}{(a^2 - \kappa^2)(\kappa^2 + b^2)} \sinh(2\kappa L)\}
\end{aligned}
\qquad (4.26)
$$

# 5

# Conductance Fluctuations: $C_3$

In this chapter we study the optical analog of the universal conductance fluctuations in metals. The optical conductance of a multiple scattering medium is the total transmitted light of a diffuse incoming beam. This quantity, analogous to the electronic conductance, exhibits conductance fluctuations. In the electronic case these fluctuations are known as *universal conductance fluctuations* (UCF). The optical conductance and its fluctuations are difficult to measure and measurements were not performed untill now.[1] Yet measurements of this quantity would form a cornerstone in the analogy between optical and electronic mesoscopic systems. We again use the Landauer approach, which we used also for $C_1$ and $C_2$, see chapter 4. However, when calculating the $C_3$ in the Landauer approach, unphysical divergencies show up and showing their cancellation turns out to be elaborate.

First, we review the electronic UCF. In section 5.2 we present the long distance diagrams and analyze the divergencies arising from these diagrams in the diffusion approximation in 5.3. In section 5.4 we show that the divergencies cancel if we analyze *all* diagrams in detail. We calculate the general form of the fluctuations in section 5.5 and applied to optical systems, where absorption and internal reflections may be present.

## 5.1 Introduction

Let us briefly review the situation for mesoscopic electron systems. The electronic conductance of mesoscopic samples shows reproducible sample-to-sample fluctuations. This was seen in experiments on mesoscopic electronic samples by Umbach *et al.* in 1984 [103]. The theoretical explanation was given by Altshuler [104] and by Lee and Stone [102]. For reviews on the subject see Lee, Stone and Fukuyama [110] and the book edited by Altshuler *et al.* [5]. The discovery of the fluctuations boosted the field of mesoscopics, as it showed that interference effects, are important in electronic systems, even far from the localization transition !

The fluctuations are a consequence of scattering from static impurities, and hence they are static. Their magnitude is independent of the sample parameters such as the mean free path, the sample thickness and, most remarkable, the average conductance. Therefore, they are called universal conductance fluctuations. The mean conductance in the considered regime comes from multiple scattered diffuse electrons. The UCF are a consequence of interference of multiple scattered waves, causing correlations between two diffuse paths. The conductance fluctuates when the phases of the waves in the dominant paths change. This happens, of course, if one changes the position of the scatterers, e.g. by taking another sample. One also may keep the scatterers fixed but apply a magnetic field or vary the Fermi energy. (In optical system one can vary the frequency of the light.) In all these cases one modifies the phases of scattered waves, so that different propagation paths become dominant. The fluctuations are much larger then one would obtain classically by modeling the system by a random resistor array, in which interference effects are neglected. That approach is valid only on a length scale exceeding the phase coherence length, where the fluctuations reduce to their classical value.

---

[1]Although attempts are now made in the group of George Maret in Strasbourg.

As both the electronic and optical systems are in the mesoscopic regime governed by the same equation, the scalar wave equation, one expects the fluctuations to be present also in our optical systems. Unfortunately, despite all advantages of optical systems, the optical analog of the UCF has not yet been observed in optical systems. Such experiments turn out to be difficult. Although the magnitude of the fluctuations is universal, they occur on a background of order $g$, where $g$ is the dimensionless conductance (in optical experiments one typically has $g \sim 10^3$). The relative value of the fluctuations to the background is thus $1/g$, so that the $C_3$ correlation function is of order $1/g^2$, typically of order $10^{-6}$. For electrons this problem is absent as moderate values of $g$ are achievable. This is also the reason that electrons are easier brought near Anderson localization, for which $g$ has to take a critical value of order unity. In the electronic case the moderate values of $g$, combined with very sensitive techniques for current measurements, have led to many observations of the universal conductance fluctuations. Recent optical experiments suggest, however, that the optical analog of UCF should just be experimentally accessible: In chapter 6 we discuss the third cumulant present in the total transmission. This quantity is of the same order as the optical UCF, namely $1/g^2$, and was recently measured [105]. We expect that similar techniques can be applied to measure the optical UCF. Microwave scattering is also interesting as it combines lower values of $g$ with many advantages of optical systems.

We are not only interested in just the size of the fluctuations, but also in the somewhat more general frequency correlation. A change in the frequency alters the interference pattern, just as occurs by changing the magnetic field or the Fermi energy in the electronic case. One knows from experiments that using the frequency as tunable parameter provides a good way for measuring the fluctuations. In contrast to electronic systems, in optical systems both angular resolved and angular integrated measurements of the transmission are possible. This led us to introduce three correlation functions, $C_1$, $C_2$, and $C_3$, see chapter 4. We recall that the $C_1$ is the leading correlation in angular resolved measurements, the $C_2$ is the main contribution in the total transmission correlation functions. And the $C_3$, finally, is the correlation function in the conductance. The $C_1$ and $C_2$ correlation were successfully calculated using a diagrammatic technique based on the Landauer approach, see chapter 4 and, for instance, Stephen [85]. One might hope that the calculation of the $C_3$ or UCF in this approach is also straightforward. It is well known, however, that the calculation in the Landauer approach is quite cumbersome, since divergencies show up on scales of one mean free path when one treats the problem on a macroscopic level using diffusons.

To avoid these difficulties, one might be tempted to use the Kubo approach, often used in electronic systems to calculate the UCF [104, 102, 110]. Furthermore, the results for the conductance obtained by Kubo or Landauer formalism should be identical, see Fisher and Lee [111] and Janßen [112]. Yet the Kubo approach cannot be applied directly to optical systems, since it is not clear how external lines should replace current vertices, and how absorption and internal reflections are to be included. Therefore, we use the Landauer approach. Technically, the vertices for partner exchange of two diffusons, the Hikami boxes, cause the difficulties in the Landauer approach. Each Hikami box brings the square of the internal momentum, whereas the current vertices are momentum-independent in the Kubo formula. As a result, the integral over the internal momentum of the closed loop is convergent in the Kubo formula, while naively divergent in Landauer approach.

We know of two studies of the $C_3$ in the Landauer approach. In the first, Kane, Serota, and Lee [86] consider electronic systems and consider *local* conductance fluctuations. In contrast to the global conductance where one averages over the whole sample, divergencies are present. Kane *et al.* run in to similar problems as we do, they then make elegantly use of current conservation to derive an expression for the correlation function. Although in optical systems the conserved quantity is not the intensity but the energy, their prediction applies to optical systems as well, since it amounts to a result for the same sums of scattering diagrams, involving different parameters only. This result has not been confirmed by a direct derivation, however. Moreover, since it relies on a conservation law, it is not clear what happens when absorption is present. The second study was done by Berkovits and Feng [90]. After giving a clear discussion of the problem, they calculated one of the macroscopic diagrams (presented earlier by Feng, Kane, Lee and Stone [91]) and subtracted the divergent parts by hand. In this way the correct order of magnitude and the qualitative frequency dependence was obtained.

It is our goal to clarify the situation by calculating the optical $C_3$ diagrammatically. We need a complete analysis of all leading diagrams, in which finally the divergent parts should cancel. We specialize to the case where one measures the conductance. In this setup the amplitudes of the incoming and outgoing diffusons are exactly in phase. If the $C_3$ correlation is measured as a (small) part of the correlation in the angular transmission or total transmission, this phase condition need not be full filled. Other contributions of order $1/g^2$ that are angular dependent are then present. These contributions are both the diagrams presented below, with different decay rates for the incoming and outgoing diffusons, but also new diagrams contribute, see Fig. 4.2. In conductance measurements these complications do not occur. Our approach immediately allows for inclusion of effects due to boundary layers and absorption. Our calculations, although specialized to optical systems, are valid for any mesoscopic system.

The $C_3$ correlation function, defined in (4.1), involves incoming diffusons $\mathcal{L}_{\text{in}}^{a,c}$, and outgoing ones $\mathcal{L}_{\text{out}}^{b,d}$. Due to the factorization of external direction dependence, see (2.68), (2.77), and (2.78), it cancels from $C_3$. We can write

$$C_3^{abcd}(\kappa, \Omega) = C_3(\kappa, \Omega) = \frac{1}{\langle T \rangle^2} \sum_Q F(Q, \kappa, \Omega). \tag{5.1}$$

where $Q$ is the $(d-1)$-dimensional transversal momentum. The function $F$ is the main object to be determined in this chapter. We thus calculate it at fixed $Q$ and with external diffusons being total-flux diffusons. One finds from (4.1) and (5.1) the conductance fluctuations

$$
\begin{aligned}
C_T(\kappa, \Omega) &= \langle T(\omega) T(\omega + \Delta\omega) \rangle - \langle T(\omega) \rangle \langle T(\omega + \Delta\omega) \rangle = \sum_{abcd} \langle T \rangle_{ab} \langle T \rangle_{cd} C_3 \\
&= \sum_Q F(Q, \kappa, \Omega) \\
&= F(0, \kappa, \Omega) && \text{quasi 1-d,} && (5.2) \\
&= W \int \frac{dQ}{2\pi} F(Q, \kappa, \Omega) && \text{quasi 2-d,} && (5.3) \\
&= W^2 \int \frac{d^2 Q}{(2\pi)^2} F(Q, \kappa, \Omega) && \text{3-d.} && (5.4)
\end{aligned}
$$

In contrast to the $C_1$ and the $C_2$, the $C_3$ depends on the (quasi) dimensionality of the system. For electronic systems one finds

$$
\langle G(k) G(k + \frac{1}{3}\ell\Omega) \rangle - \langle G(k) \rangle \langle G(k + \frac{1}{3}\ell\Omega) \rangle = \left( \frac{2e^2}{h} \right)^2 C_T(0, \Omega). \qquad (5.5)
$$

These results can be extended for other geometries. If the width is comparable to the thickness of the slab, the momentum integral discretizes into a sum over transversal eigenmodes [110]. The result can be generalized further to arbitrary geometries by taking $x, y$ dependence into account, and calculating the diffusons using appropriate boundary conditions.

## 5.2   Long range diagrams

At this point we recall the structure of the leading diagrams for the correlation functions defined in (4.1), see again Fig. 4.1. For all diagrams there are two incoming advanced fields, which we momentarily term $i$ and $j$, and two retarded ones, $i^*$ and $j^*$. The first term on the r.h.s. in Eq. (4.1) follows from the diagram where $i$ and $i^*$ pair into an incoming diffuson, and the same for $j$ and $j^*$. These diffusons have no common scatterers, so for this contribution the expression factorizes into a product of averages. For the $C_1$ term in (4.1) such a factorization also takes place [91]. However, in this term the pairings are $ij^*$ and $ji^*$. In the $C_2$ correlation function there are two terms. In the first the incoming diffusons have pairings $ij^*$ and $ji^*$. These diffusons interfere in some point in space, where they exchange partners through a Hikami-box. The outgoing pairings are then $ii^*$, $jj^*$. It is this term that contributes in measurements of the total transmission.

In electronic experiments where one measures the conductivity, and in optical experiments where one uses an integrating sphere on the incoming side for creating a diffuse beam, the two amplitudes in the diffuson have to be exactly in phase in order to be leading. Therefore, the incoming diffusons cannot have a momentum or frequency difference, and the pairing must be $ii^*$ and $jj^*$. In Fig. 5.1(a) the incoming diffusons interfere somewhere in the slab. In a diagrammatic language the diffusons interchange a propagator so that the pairing changes into $ij^*$ and $ji^*$. Propagation continues with these diffusons, which, due to the different pairing can have non-zero frequency difference and non-zero momentum. But to be dominant, the outgoing diffusons can also not have a momentum or frequency difference. Therefore, somewhere else in the slab a second interference occurs. Again exchanging an amplitude, the original pairing, $ii^*$ and $jj^*$, is restored and the two diffusons propagate out, see Fig. 5.1(a)i. Some other contributions occur as well, yet as the incoming and outgoing pairings are always $ii^*$ and $jj^*$: In Fig. 5.1(a)ii the first incoming diffuson meets an outgoing diffuson and amplitudes exchange. These internal diffusion lines meet at a second point where the original pairings are restored. Clearly in this process the intermediate paths are traversed in time reversed order. Due to time-reversal symmetry they give a similar contribution as previous Fig. 5.1(a)i. In Fig. 5.1(b)i a diffuson breaks up such that one of its amplitudes makes a large detour, returns to the breaking point and recombines into an outgoing diffuson. The second incoming diffuson crosses this excess path of the amplitude, and one of its amplitudes follows the same contour as the first diffuson. The fourth amplitude resides and finally recombines with its original partner amplitude to form an outgoing diffuson. Finally, Fig. 5.1(c) depicts the situation where only one internal diffuson occurs. Its endpoints must lie within a distance of a few mean free paths. Because of its local character this class does not show up in the final result; we need it, however, in the regularization process, since it contains terms that cancel divergencies from the other two classes.

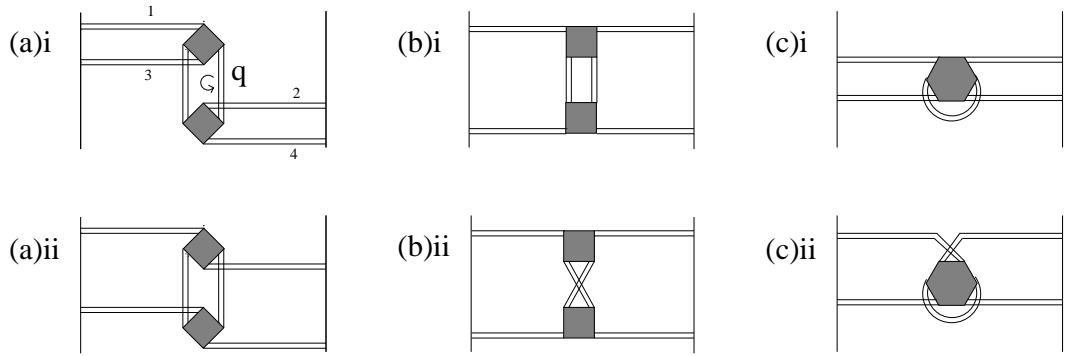

Figure 5.1:   The leading contributions to the conductance fluctuations, apart from some special short distance processes dealt with in section 5.4. The incoming diffusons from the left interfere twice before they go out on the right. The close parallel lines correspond to diffusons; the shaded boxes are Hikami vertices; $\mathbf{q}$ denotes the free momentum which is to be integrated over.

Due to the diffusive behavior of the internal propagators we call all these the "long range diagrams". Their internal lines are diffusons (ladder diagrams), that is to say, in these lines there can be an arbitrary number of scatterers. Note that these long range diagrams also include terms with only a few scatterers, e.g. one or two. The latter contributions are of course not really long range; however they contribute to the geometric series that represents the ladder diagram. We state this explicitly, since below we will discuss some unexpected problems of these short ranged contributions to the long range diagrams. There are also some special short ranged contributions. Due to various subtleties, they resist a general treatment; we postpone their calculation to section 5.4.

## 5.3   Divergencies in the diffusion approximation

The diagrams for the conductance fluctuations contain a loop; the two internal diffusons have a free momentum, over which one has to integrate. In Fig. 5.1(a) we denote this momentum $\mathbf{q}$. Physically, one expects important contributions to the conductance fluctuations if the distance between the two interferences vertices ranges from the mean free path to the sample size. Yet in this section we will show that the $\mathbf{q}$−integral for the long range diagrams diverges for large momentum, i.e. when the two interference processes are close to each other. The standard picture of diffuse transport with diffusons and interference described by Hikami vertices, that works so well for loop-less diagrams such as the $C_2$ correlation function (chapter 4), and the third cumulant of the total transmission (chapter 6), now becomes spoiled by these divergencies. In section 5.4 we solve this problem by going back to mesoscopic scales, and considering all scattering events.

The problem becomes clear if we calculate the diagrams of Fig. 5.1. First, we need the expressions for the Hikami vertices. To derive the vertices, we expand the momenta to leading order in $(\mathbf{q}\ell)$. For the large $\mathbf{q}\ell$ this is in principle not allowed, but in practice it could still work. The four point vertex and the six point vertex are found by summation of the bare vertex and its dressings. We recall from chapter 3

$$H_4 = h_4[-\mathbf{q}_1 \cdot \mathbf{q}_3 - \mathbf{q}_2 \cdot \mathbf{q}_4 + \frac{1}{2}\sum_{i=1}^{4}(\mathbf{q}_i^2 + \kappa_i^2 + i\Omega_i)], \tag{5.6}$$

$$H_6 = -h_6[\mathbf{q}_1 \cdot \mathbf{q}_2 + \mathbf{q}_2 \cdot \mathbf{q}_3 + \mathbf{q}_3 \cdot \mathbf{q}_4 + \mathbf{q}_4 \cdot \mathbf{q}_5 + \mathbf{q}_5 \cdot \mathbf{q}_6 + \mathbf{q}_6 \cdot \mathbf{q}_1$$
$$+ \sum_{i=1}^{6}(\mathbf{q}_i^2 + \frac{1}{2}\kappa_i^2 + \frac{i}{2}\Omega_i)]. \tag{5.7}$$

We defined the pre-factors as

$$h_4 = \frac{\ell^5}{48\pi k^2}, \qquad h_6 = \frac{\ell^7}{96\pi k^4}. \tag{5.8}$$

We denote the momenta of the diffusons attached to these vertices by $\mathbf{q}_i$, where the diffusons are numbered clockwise on the vertex and their momenta are directed towards the vertex. In the actual calculations we use the Fourier transforms in the $z$-direction of the vertices. The additional frequency and absorption terms, according to the diffusion equation (2.39), together with the $\mathbf{q}^2$ terms lead to a source $\delta(z - z')$. For external diffusons, we neglect such terms as they bring contributions of the order $\ell/L$. This approximation simplified the calculation of for instance the long range correlation function, see chapter 4. For the internal diffusons, however, the source terms are of leading order and cause divergencies. They correspond to the situation where the two interferences take place within a distance of a few mean free paths.

As an example we calculate the diagram presented in Fig. 5.1(a)i. This diagram was first depicted by Feng, Kane, Lee and Stone [91] and considered in detail by Berkovits and Feng [90]. These authors pointed out that a short distance divergency appears. For the case of external momenta approximately zero, the Hikami-box (5.6) yields $H_4(\mathbf{q}, 0, -\mathbf{q}, 0) = 2h_4q^2$, while the internal diffuson has the form $\mathcal{L}_{\text{int}}(q) = 12\pi/(\ell^3 q^2)$.

Omitting the external lines, the diagram Fig. 5.1(a)i then simply leads to

$$\int \frac{\mathrm{d}^3 q}{(2\pi)^3} H_4^2(\mathbf{q}, 0, -\mathbf{q}, 0)\mathcal{L}_{\mathrm{int}}^2(q) = \frac{\ell^4}{4k^4}\int\frac{\mathrm{d}^3 q}{(2\pi)^3}q^0 = \frac{\ell^4}{4k^4}\delta^{(3)}(\mathbf{r} = 0). \tag{5.9}$$

This is indeed a cubic divergence in three dimensions, and in general, a $d$-dimensional divergency in $d$ dimensions. As it is arising from the physically innocent situation where the two interference vertices are close to each other, we expect that the divergency has to disappear finally.

We now calculate the diagram for the slab geometry. For simplicity we first consider a quasi one-dimensional system in which frequency differences and absorption are absent, therefore the decay rate or "mass", Eq. (2.7), vanishes, i.e. $M = 0$ for all diffusons. (Beyond quasi 1d one would have to take non-zero $M = Q$ and sum over the allowed $Q$.) From Fig. 5.1(a)i one directly reads off its corresponding expression $F_{a.i}$

$$F_{a.i} = \int\int \mathrm{d}z\mathrm{d}z' \mathcal{L}_{\mathrm{in}}^2(z)H_4(z)\mathcal{L}_{\mathrm{int}}^2(z, z')H_4(z')\mathcal{L}_{\mathrm{out}}^2(z). \tag{5.10}$$

We label the two incoming diffusons 1 and 3, the outgoing ones 2 and 4, and the internal ones $\mathcal{L}(z_5, z_7)$ and $\mathcal{L}(z_6, z_8)$, (with $5, 7$ at $z$ and $6, 8$ at $z'$). The real space expressions for the Hikami boxes become

$$H_4(z) = h_4[\partial_{z_1}\partial_{z_3} + \frac{1}{2}(\partial_{z_5} + \partial_{z_7})^2 - \partial_{z_5}^2 - \partial_{z_7}^2], \tag{5.11}$$

$$H_4(z') = h_4[\partial_{z_2}\partial_{z_4} + \frac{1}{2}(\partial_{z_6} + \partial_{z_8})^2 - \partial_{z_6}^2 - \partial_{z_8}^2], \tag{5.12}$$

in which $\partial_{z_i}$ is the derivative of the corresponding diffuson; after performing the differentiation $z_{1,3,5,7}$ should be put equal to $z$, while $z_{2,4,6,8}$ should be put equal $z'$. Keeping $z_{2,4,6,8}$ fixed, we obtain for the $z$-integral after some partial integrations

$$\int_0^L \mathrm{d}z H_4(z)\mathcal{L}_{\mathrm{in}}(z_1)\mathcal{L}_{\mathrm{in}}(z_3)\mathcal{L}_{\mathrm{int}}(z_5, z_6)\mathcal{L}_{\mathrm{int}}(z_7, z_8)$$

$$= \frac{12\pi h_4}{\ell^3}\int \mathrm{d}z[\mathcal{L}_{\mathrm{int}}(z, z_6)\delta(z - z_8) + \mathcal{L}_{\mathrm{int}}(z, z_8)\delta(z - z_6)]\mathcal{L}_{\mathrm{in}}^2(z)$$

$$+ 2h_4\int \mathrm{d}z\mathcal{L}_{\mathrm{int}}(z, z_6)\mathcal{L}_{\mathrm{int}}(z, z_8)\mathcal{L}_{\mathrm{in}}'^2(z)$$

$$= \frac{12\pi h_4}{\ell^3}\mathcal{L}_{\mathrm{int}}(z_8, z_6)\mathcal{L}_{\mathrm{in}}^2(z_8) + \frac{12\pi h_4}{\ell^3}\mathcal{L}_{\mathrm{int}}(z_6, z_8)\mathcal{L}_{\mathrm{in}}^2(z_6)$$

$$+ 2h_4\int \mathrm{d}z\mathcal{L}_{\mathrm{int}}(z, z_6)\mathcal{L}_{\mathrm{int}}(z, z_8)\mathcal{L}_{\mathrm{in}}'^2(z).$$

Here we also used the diffuson equation, which in this simplified case reads $\partial_z^2\mathcal{L}_{\mathrm{in}} = 0$ and $\partial_z^2\mathcal{L}_{\mathrm{int}}(z, z') = 12\pi\delta(z - z')/\ell^3$. Also carrying out the $z'$-integral we find after performing again some partial integrations

$$F_{a.i} = \int \mathrm{d}z' H_4(z')\mathcal{L}_{\mathrm{out}}(z_2)\mathcal{L}_{\mathrm{out}}(z_4)\left[\frac{12\pi h_4}{\ell^3}\mathcal{L}_{\mathrm{int}}(z_8, z_6)\mathcal{L}_{\mathrm{in}}^2(z_8) + \right.$$

$$\left. \frac{12\pi h_4}{\ell^3}\mathcal{L}_{\mathrm{int}}(z_6, z_8)\mathcal{L}_{\mathrm{in}}^2(z_6) + 2h_4\int \mathrm{d}z\mathcal{L}_{\mathrm{int}}(z, z_6)\mathcal{L}_{\mathrm{int}}(z, z_8)\mathcal{L}_{\mathrm{in}}'(z)\right]$$

$$
\begin{aligned}
=\ &\frac{\ell^4}{4k^4}\delta(0)\int \mathrm{d}z\, \mathcal{L}_{\mathrm{in}}^2(z)\mathcal{L}_{\mathrm{out}}^2(z)\\
&+\frac{h_4\ell^2}{k^2}\int \mathrm{d}z\, \mathcal{L}_{\mathrm{int}}(z,z)\left[\mathcal{L}_{\mathrm{in}}'^2(z)\mathcal{L}_{\mathrm{out}}^2(z)+\mathcal{L}_{\mathrm{in}}^2(z)\mathcal{L}_{\mathrm{out}}'^2(z)\right.\\
&\qquad\left.+2\mathcal{L}_{\mathrm{in}}'(z)\mathcal{L}_{\mathrm{out}}'(z)\mathcal{L}_{\mathrm{in}}(z)\mathcal{L}_{\mathrm{out}}(z)\right]\\
&+4h_4^2\int \mathrm{d}z'\int \mathrm{d}z\, \mathcal{L}_{\mathrm{int}}^2(z,z')\mathcal{L}_{\mathrm{in}}'^2(z)\mathcal{L}_{\mathrm{out}}'^2(z').
\end{aligned}
\tag{5.13}
$$

We denote the spatial derivative of $\mathcal{L}(z)$ by $\mathcal{L}'$. All diffusons are simple linear functions in this case, yielding

$$
F_{a,i}=\frac{2}{15}\delta(0)L+\frac{8}{45}.
\tag{5.14}
$$

Note that the pre-factors of the diffusons and the Hikami boxes have canceled precisely. This is closely related to the universal character of conductance fluctuations in electronic systems, see (5.5).

The term $\delta(0)$ is a linear divergency, which is the cause of all troubles. In the three-dimensional case one has to take $Q\neq 0$. The $\delta(0)L$ term will occur also for transversal momentum $Q\neq 0$, so that the $Q$-sum yields the cubic divergency

$$
W^2\int \frac{\mathrm{d}^2Q}{(2\pi)^2}\delta(0)L=W^2L\delta^{(3)}(0),
\tag{5.15}
$$

as expected from the above bulk consideration.

We now give the results of all diagrams of Fig. 5.1. We no longer restrict to the $M=0$ case. We label the expressions according to the diagrams in figure, $F_a$, $F_b$ and $F_c$. In the diagrams of Fig. 5.1(a), the decay rates of Eq.(2.7) for the internal diffusons are each other complex conjugate, $M$ and $M^*$. In the diagrams of Fig. 5.1(b) both internal diffusons have the same decay rate. Using the definition of the Hikami vertices and the diffusion equation, we obtain

$$
\begin{aligned}
F_a(M)\ =\ &\frac{\ell^4}{2k^4}\delta(0)\int \mathrm{d}z\, \mathcal{L}_{\mathrm{in}}^2\mathcal{L}_{\mathrm{out}}^2\\
&+\frac{h_4\ell^2}{2k^2}\mathrm{Re}\int \mathrm{d}z\, \mathcal{L}_{\mathrm{int}}(z,z;M)\left[3\mathcal{L}_{\mathrm{in}}'^2\mathcal{L}_{\mathrm{out}}^2+3\mathcal{L}_{\mathrm{in}}^2\mathcal{L}_{\mathrm{out}}'^2\right.\\
&\qquad\left.+10\mathcal{L}_{\mathrm{in}}'\mathcal{L}_{\mathrm{out}}'\mathcal{L}_{\mathrm{in}}\mathcal{L}_{\mathrm{out}}-4i\Omega\mathcal{L}_{\mathrm{in}}^2\mathcal{L}_{\mathrm{out}}^2\right]\\
&+4h_4^2\int\int \mathrm{d}z\,\mathrm{d}z'\, \mathcal{L}_{\mathrm{int}}(z,z';M)\mathcal{L}_{\mathrm{int}}(z,z';M^*)\times\\
&\qquad\left[\mathcal{L}_{\mathrm{in}}'^2(z)\mathcal{L}_{\mathrm{out}}'^2(z')+\mathcal{L}_{\mathrm{in}}'(z)\mathcal{L}_{\mathrm{out}}'(z)\mathcal{L}_{\mathrm{in}}'(z')\mathcal{L}_{\mathrm{out}}'(z')\right],\\
F_b(M)\ =\ &\frac{\ell^4}{4k^4}\delta(0)\int \mathrm{d}z\mathcal{L}_{\mathrm{in}}^2\mathcal{L}_{\mathrm{out}}^2\\
&+\frac{h_4\ell^2}{2k^2}\mathrm{Re}\int \mathrm{d}z\, \mathcal{L}_{\mathrm{int}}(z,z;M)\left[-2\mathcal{L}_{\mathrm{in}}'\mathcal{L}_{\mathrm{out}}'\mathcal{L}_{\mathrm{in}}\mathcal{L}_{\mathrm{out}}+\mathcal{L}_{\mathrm{in}}'^2\mathcal{L}_{\mathrm{out}}^2+\mathcal{L}_{\mathrm{in}}^2\mathcal{L}_{\mathrm{out}}'^2\right.\\
&\qquad\left.-4\kappa^2\mathcal{L}_{\mathrm{in}}^2\mathcal{L}_{\mathrm{out}}^2\right]\\
&+\frac{1}{2}h_4^2\int\int \mathrm{d}z\,\mathrm{d}z'\left[\mathcal{L}_{\mathrm{int}}^2(z,z';M)+\mathcal{L}_{\mathrm{int}}^2(z,z';M^*)\right]\times
\end{aligned}
$$

$$\frac{d^2}{dz^2}\left[\mathcal{L}_{\text{in}}(z)\mathcal{L}_{\text{out}}(z)\right]\frac{d^2}{dz'^2}\left[\mathcal{L}_{\text{in}}(z')\mathcal{L}_{\text{out}}(z')\right],$$

$$
\begin{aligned}
F_c(M) = & -\frac{\ell^4}{k^4}\delta(0)\int \mathrm{d}z\,\mathcal{L}_{\text{in}}^2\mathcal{L}_{\text{out}}^2 \\
& -\frac{2h_4\ell^2}{k^2}\text{Re}\int \mathrm{d}z\,\mathcal{L}_{\text{int}}(z,z;M)\times \\
& \left[2\mathcal{L}_{\text{in}}'\mathcal{L}_{\text{out}}'\mathcal{L}_{\text{in}}\mathcal{L}_{\text{out}}+\mathcal{L}_{\text{in}}'^2\mathcal{L}_{\text{out}}^2+\mathcal{L}_{\text{in}}^2\mathcal{L}_{\text{out}}'^2-(\kappa^2+i\Omega)\mathcal{L}_{\text{in}}^2\mathcal{L}_{\text{out}}^2\right]
\end{aligned}
\tag{5.16}
$$

where we used the short hand notation that in the single integrals all incoming and outgoing diffusons have argument $z$. To obtain the variance of the fluctuations in quasi-one dimension, $F$'s is evaluated at transverse momentum $Q = 0$. The internal momentum $Q$ enters the equations via the decay rate of the internal diffusons, defined by $M^2 = Q^2 + \kappa^2 + i\Omega$. In two and three dimensions a sum or integral over the transversal momentum has to be done.

In $F_a$ and $F_b$ we can distinguish three contributions. In the first term, both Hikami boxes operate on the internal diffusons, yielding in the diffusion approximation a delta function evaluated in zero. The resulting term is independent of all the momenta of the external diffusons. One can see from the diffusion equation that a diffuson decays rapidly if its momentum becomes large. Terms of the diffusons with few scatterers are dominant at large momentum; they cause the divergence. Our present description of these processes is incomplete. To find the cancellation of this divergency, calculation of the long range diagrams is not sufficient, so that the short distance processes have to be examined in detail. This will be done in the next section. The second term of (5.16a,b) is a single integral, it comes about when the boxes act on one internal and on one external diffuson. This corresponds to the case where one internal diffuson is almost empty, while the other diffuson contains an arbitrary number of scatterers. In two and three dimensions the momentum integral diverges, since for large $Q$ it behaves as $\int \mathrm{d}z\int \mathrm{d}^{d-1}Q\,\mathcal{L}_{\text{int}}(Q;z,z)\sim \int \mathrm{d}^{d-1}Q\,Q^{-1}$. Yet when summing the $a$, $b$ and $c$ contribution this term cancels. The sum gives

$$
\begin{aligned}
& F_a(M) + F_b(M) + F_c(M) \\
= & -\frac{\ell^4}{4k^4}\delta(0)\int \mathrm{d}z\mathcal{L}_{\text{in}}^2\mathcal{L}_{\text{out}}^2 \\
& +4h_4^2\int\int \mathrm{d}z\,\mathrm{d}z'\,\mathcal{L}_{\text{int}}(z,z';M)\mathcal{L}_{\text{int}}(z,z';M^*) \\
& \qquad\times\left[\mathcal{L}_{\text{in}}'^2(z)\mathcal{L}_{\text{out}}'^2(z')+\mathcal{L}_{\text{in}}'(z)\mathcal{L}_{\text{out}}'(z)\mathcal{L}_{\text{in}}'(z')\mathcal{L}_{\text{out}}'(z')\right] \\
& +\frac{1}{2}h_4^2\int\int \mathrm{d}z\,\mathrm{d}z'\,\left[\mathcal{L}_{\text{int}}^2(z,z';M)+\mathcal{L}_{\text{int}}^2(z,z';M^*)\right] \\
& \qquad\times\frac{d^2}{dz^2}\left[\mathcal{L}_{\text{in}}(z)\mathcal{L}_{\text{out}}(z)\right]\frac{d^2}{dz'^2}\left[\mathcal{L}_{\text{in}}(z')\mathcal{L}_{\text{out}}(z')\right].
\end{aligned}
\tag{5.17}
$$

The last two terms involve a double integral describing interference vertices at different points in space; technically it arises when both boxes act on external diffusons, or from terms where they do so after partial integrations. This term is absent in the expression $F_c$, which contains only one $z$−dependence as one sees from Fig. 5.1(c). When performing the integral of $F$ over the transverse momentum $Q$, the double integral term behaves at large $Q$ as $\int dz \int dz' \int d^{d-1} Q \mathcal{L}_{\text{int}}^2(Q; z, z') \sim \int d^{d-1} Q \, Q^{-3}$. It is thus convergent. One expects that finally only this term will survive. It is also the only contribution depending solely on derivatives of the external diffusons.

In the quasi one-dimensional case where absorption and frequency terms are absent, the expressions reduce to

$$F_a(0) + F_b(0) + F_c(0) = -\frac{2}{15}\delta(0)L + \frac{2}{15}. \tag{5.18}$$

The second part is a well known result for the UCF in one dimension, but a singular part is annoyingly present. Before we can obtain the UCF and correlation functions, we have to show its cancellation. We expect the unphysical divergency to disappear by summing all diagrams, thus we set out finding *all* leading diagrams.

## Generating diagrams by computer

Finding the correct and complete set of diagrams by educated guess proves very difficult, especially for the low order diagrams. The set is quite large and not very systematic. In the article closely related to this chapter [113], we pursue the technique founded by Kadanoff and Baym [114, 115] to generate all diagrams. This method provides a systematic way to construct the diagrams in a particular approximation. The approximation is made on the level of a generating functional. The method was very successful for our purpose and was almost fully done by dr. Ruud Vlaming. The interested reader is referred to Ref. [113]. We verified the set of diagrams generated by the Kadanoff-Baym approach using a computer program. We developed this program to generate all diagrams with six or fewer scatterers automatically. In the appendix we describe this method in detail, in short it does the following: After generating the set, the program checks it against double counting. Next, momenta are assigned to all propagators, such that momentum conservation is obeyed. It then decides whether a particular diagram is of leading order in $(k\ell)^{-1}$; this is the case if all propagators can have a momentum approximately equal to $k$. The leading diagrams are expressed in terms of the standard integrals defined in section 5.4. We did the subsequent analytical calculation and summation of diagrams by hand. The program turns out to be especially useful in finding the precise set of leading short-distance diagrams, but also the long distance diagrams of section 5.2 were reproduced. It should be mentioned that apart from the diagrams from the Kadanoff-Baym approach we also found extra diagrams; in section 5.4.3 we discuss them and we show that they cancel. Both the set of diagrams from the Kadanoff-Baym approach as the computer generated set were calculated in second order Born approximation only. Going beyond second order Born would increase the number of diagrams dramatically, without bringing different physics. The cancelation present in second order Born, will in the end also be present beyond second order Born.

## 5.4   Cancellation of short distance contributions

In this section we show that the strong divergence, that occurs if both Hikami boxes act on internal diffusons, cancels. Since a diffuson decays rapidly if its momentum becomes large, terms of the diffusons with few scatterers are dominant when the momentum is large; they cause the divergence. We thus show the cancellation by considering in great detail the short distance processes.

It turned out that complications arise for diagrams with less than four scatterers. The external diffusons may still contain an arbitrary number of scatterers. For simplicity we perform the calculation in this section in the bulk and the external diffusons are not attached to the diagrams. Since the divergence is independent of the external momenta, the cancellation is generally proven at zero external momentum. For simplicity we may then consider an infinite system. As a result all factors can be expressed in the $d$-dimensional momentum $\mathbf{q} = (Q, q_z)$. We first look at fixed value of the $d$-dimensional internal momentum $\mathbf{q}$ and postpone the integration. We will show that under present conditions the integrand is zero for all $\mathbf{q}$, so that there is no divergency after integration.

In the calculation of the diffuson and the Hikami vertices in the previous sections we expanded the integrals in $q\ell$. Since we are after contributions for $q \sim 1/\ell$, this expansion is not allowed, instead the diagrams can be factorized in products of the $I_{k,l}^{m,n}(q)$ integrals, defined in section 3.4.1. We still assumed that $q \ll k$, as we do not need the physics on length scales comparable to the wavelength, but comparable to one mean free path only.

## 5.4.1 Short distance behavior of long range diagrams

First, look at the long range diagrams with few scatterings. By writing out the dressing of the vertices one obtains the detailed structure of the diagrams, given in Fig. 5.2. In the diagrams Fig. 5.1(a)i-(c)i, the incoming external diffusons are connected to 1 and 3. The diagrams in Fig. 5.1(a)ii-(c)ii are obtained by connecting external diffusons 1 and 2 to the incoming side. If such a diagram contains only a few scatterers, it can be grouped into one of the classes of the long range diagrams. A simple example is the upper left diagram of Fig. 5.2, drawn there with six scatterers. Interchanging the external diffusons numbered 2 and 3 clearly leads to another topology. However, with only two scatterers present, the topology does not change under this operation, and one must be careful not to overcount this term. Bearing this important observation in mind, we sum all long range diagrams for an arbitrary number of scatterers. The expression for each diagram and its combinatorial factor is given in Table 5.1. It is convenient to collect diagrams with an equal number $m$ of scatterers that connect different propagators. Scatterers on which one given amplitude scatters twice are thus, momentarily, not counted as new scatterers. Looking at Fig. 5.2, $m$ equals the total number of scatterers minus the number of scatterers indicated by curved lines. After this re-summation we find the important result that each class of "long range" terms vanishes for $m$ not equal to 2. Denoting the terms of Fig. 5.2 respectively by $R_a$, $R_b$, and $R_c$ we find $R_a = R_b = R_c = 0$, if $m > 2$. If $m = 2$, the classes cancel against each other, since we then obtain $R_a = R_b = -R_c/2 = \frac{\ell^4}{4k^4}A_1^2$. This implies

$$R_a + R_b + R_c = 0. \tag{5.19}$$

We have thus shown that the long range diagrams for an infinite system cancel and thus cause no divergency after integration over $\mathbf{q}$. It is essential that we count the degeneracy of the low order contributions properly.

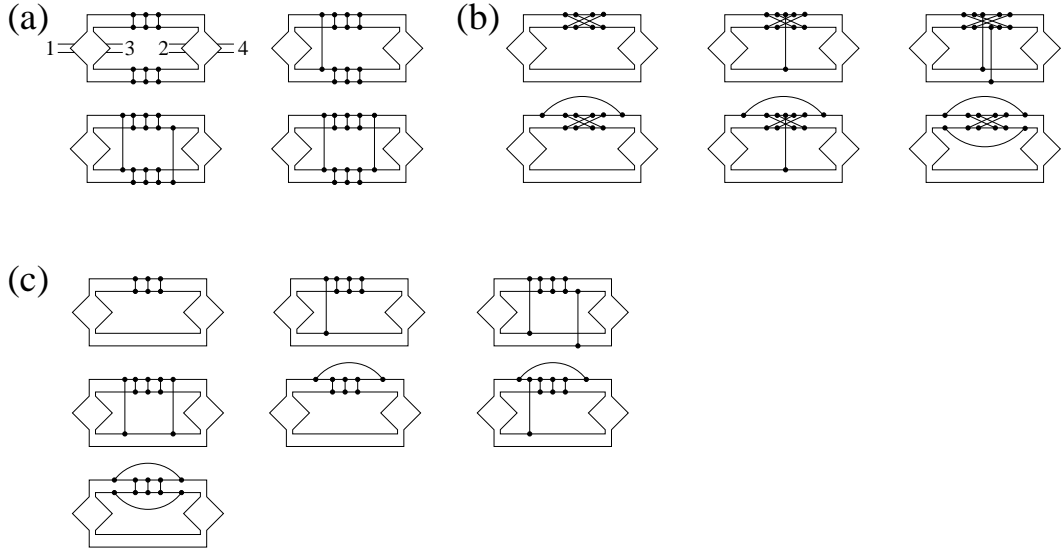

Figure 5.2: The dressings of the shaded vertices of Fig. 5.1 are written out explicitly, this shows the detailed structure of the long range contributions to the conductance fluctuations. The close parallel lines, only explicitly shown in the first diagram, are the external diffusons. Fig. 5.1(a)i corresponds to the case were 1 and 3 are incoming diffusons; Fig. 5.1(a)ii corresponds to the case were 1 and 2 are incoming diffusons. The vertical, diagonal or curved lines linking two dots represent a common scattering of two amplitudes. From top to bottom the horizontal lines in a diagram are advanced, retarded, advanced and, retarded propagators, respectively. The number of scatterers in the internal and external diffusons is arbitrary. Topologically equivalent diagrams are not shown.

| diagr. | expression | n=1 | $n=2$ | $n=3$ | $n \geq 4$ | $n-m$ |
|---|---|---|---|---|---|---|
| a1 | $\left[I_{1,1}^{1,1}\right]^2\left[I_{0,1}^{1,0}\right]^{n-2}$ | 0 | 1=2-1 | 4 | $2n-2$ | 0 |
| a2 | $I_{1,1}^{1,1}\left[I_{1,1}^{1,0}\right]^2\left[I_{0,1}^{1,0}\right]^{n-3}$ | 0 | 0 | 8 | $8n-16$ | 0 |
| a3+a4 | $\left[I_{1,1}^{0,1}\right]^4\left[I_{0,1}^{1,0}\right]^{n-4}$ | 0 | 0 | 0 | $8n-24$ | 0 |
| b1 | $I_{2,1}^{0,1}I_{1,2}^{1,0}\left[I_{0,1}^{1,0}\right]^{n-2}$ | 0 | 4 | 8 | $4n-4$ | 0 |
| b2 | $I_{2,1}^{0,1}\left[I_{1,1}^{1,0}\right]^2\left[I_{0,1}^{1,0}\right]^{n-3}$ | 0 | 0 | 8 | $8n-16$ | 0 |
| b3 | $\left[I_{1,1}^{0,1}\right]^2\left[I_{1,1}^{1,0}\right]^2\left[I_{0,1}^{1,0}\right]^{n-4}$ | 0 | 0 | 0 | $4n-12$ | 0 |
| b4 | $I_{2,1}^{0,0}I_{1,2}^{1,0}I_{2,0}^{0,1}\left[I_{0,1}^{1,0}\right]^{n-3}$ | 0 | 0 | 8 | $8n-16$ | 1 |
| b5 | $I_{2,1}^{0,0}I_{2,0}^{0,1}I_{1,1}^{1,0}\left[I_{0,1}^{1,0}\right]^{n-4}$ | 0 | 0 | 0 | $8n-24$ | 1 |
| b6 | $I_{2,1}^{0,0}I_{2,0}^{0,2}I_{1,2}^{0,0}I_{0,2}^{1,0}\left[I_{0,1}^{1,0}\right]^{n-4}$ | 0 | 0 | 0 | $4n-12$ | 2 |
| c1 | $I_{1,2}^{2,1}\left[I_{0,1}^{1,0}\right]^{n-1}$ | 0=4-4 | 4 | 4 | 4 | 0 |
| c2 | $I_{1,2}^{1,1}I_{1,1}^{0,1}\left[I_{0,1}^{1,0}\right]^{n-2}$ | 0 | 8=16-8 | 16 | 16 | 0 |
| c3 | $I_{1,1}^{1,0}I_{1,1}^{0,1}I_{1,1}^{1,1}\left[I_{0,1}^{1,0}\right]^{n-3}$ | 0 | 0 | 8 | 8 | 0 |
| c4 | $\left[I_{1,1}^{1,0}\right]^2I_{2,1}^{1,0}\left[I_{0,1}^{1,0}\right]^{n-3}$ | 0 | 0 | 8 | 8 | 0 |
| c5 | $I_{2,1}^{0,0}I_{1,2}^{2,0}\left[I_{0,1}^{1,0}\right]^{n-2}$ | 0 | 0=8-8 | 8 | 8 | 1 |
| c6 | $I_{1,1}^{0,0}I_{1,1}^{1,0}I_{2,0}^{1,1}\left[I_{0,1}^{1,0}\right]^{n-3}$ | 0 | 0 | 8=16-8 | 16 | 1 |
| c7 | $I_{1,2}^{0,0}I_{2,1}^{0,0}I_{2,0}^{0,2}\left[I_{0,1}^{1,0}\right]^{n-3}$ | 0 | 0 | 0=4-4 | 4 | 2 |

Table 5.1: Table used for the calculation of the diagrams presented in Fig. 5.2 (for zero external momenta). The expression for each diagram is given in the second column. Its combinatorial prefactor in the other columns for given number of scatterers. The last column counts the number of scatterers that are not included as such in the re-summation. There are six diagrams with degeneracy lower than expected: c1 for $n=1$, also a1,c2,c5 for $n=2$, and c6,c7 for $n=3$.

Another, more standard way to verify the important cancellation is the following. Recall that we look at a bulk situation with fixed internal momentum $\mathbf{q}$, while we put the external momenta zero. Beyond the diffusion approximation an internal diffuson is given by Eq.(3.24). We also need the Hikami-boxes beyond the diffusion approximation. It holds that

$$H_4(\mathbf{q}, 0, -\mathbf{q}, 0) = I_{11}^{11}(q) + \frac{4\pi}{\ell} \left\{ I_{11}^{10}(q) \right\}^2 + \frac{4\pi}{\ell} \left\{ I_{11}^{01}(q) \right\}^2. \quad (5.20)$$

Inserting the values from the table and doing similar but longer calculations for two other vertices we find

$$H_4(0, \mathbf{q}, 0, -\mathbf{q}) = 2H_4(\mathbf{q}, -\mathbf{q}, 0, 0) = \frac{\ell^3}{8\pi k^2} A_1(1 - A_1), \quad (5.21)$$

$$H_6(0, 0, \mathbf{q}, 0, 0, -\mathbf{q}) = \frac{\ell^5}{16\pi k^4} A_1(1 - A_1)(1 - 3A_1). \quad (5.22)$$

Now the long range diagrams of Fig. 5.1 can be evaluated and we insert for the diffuson Eq.(3.24, that is $\mathcal{L}_{\text{int}} = \frac{4\pi}{\ell}(1 - A_1)^{-1}$.

For the internal part one has at fixed $\mathbf{q}$

$$2H_4(\mathbf{q}, 0, -\mathbf{q}, 0)^2 \mathcal{L}_{\text{int}}^2(q) + 4H_4(\mathbf{q}, -\mathbf{q}, 0, 0)^2 \mathcal{L}_{\text{int}}^2(q)$$
$$+2H_6(\mathbf{q}, 0, 0, -\mathbf{q}, 0, 0)\mathcal{L}_{\text{int}}(q)$$
$$= \frac{\ell^4}{4k^4} \{2A_1^2 + A_1^2 + 2A_1(1 - 3A_1)\} = \frac{\ell^4}{4k^4}(2A_1 - 3A_1^2). \quad (5.23)$$

It is essential that the denominators $1 - A_1$ have disappeared from this expression. As mentioned, it means that all high order terms cancel, allowing a cancellation of the remainder by low order contributions. From Table 5.1 one sees that we have over-counted six types of low order terms. The correction to be subtracted is

$$\frac{4\pi}{\ell} \left\{ 4 \frac{\ell^5}{32\pi k^4}(A_1 + A_2 + A_3) \right\}$$
$$+ \left(\frac{4\pi}{\ell}\right)^2 \left\{ \left(\frac{\ell^3 A_1}{8\pi k^2}\right)^2 + 8 \frac{-i\ell^4(2A_1 + A_2)}{32\pi k^3} \frac{-i\ell^2 A_1}{8\pi k} + 8 \frac{i\ell^2}{8\pi k} \frac{i\ell^4}{32\pi k^3}(A_2 + 2A_3) \right\}$$
$$+ \left(\frac{4\pi}{\ell}\right)^3 \left\{ 8 \frac{i\ell^2}{8\pi k} \frac{i\ell^2}{8\pi k} A_1 \frac{-\ell^3}{16\pi k^2} A_2 + 4 \frac{-i\ell^2}{8\pi k} \frac{i\ell^2}{8\pi k} \frac{\ell^3}{8\pi k^2} A_3 \right\}$$
$$= \frac{\ell^4}{4k^4}(2A_1 - 3A_1^2). \quad (5.24)$$

It indeed exactly cancels expression (5.23), which arose as a remainder of the long range terms.

From the results of present section one may be tempted to conclude that the long range diagrams do not lead to divergencies if the correct degeneracies of the low order terms is properly taken into account. Though this conclusion is correct, it is too early to draw it, since we show in next parts that some further complications arise.

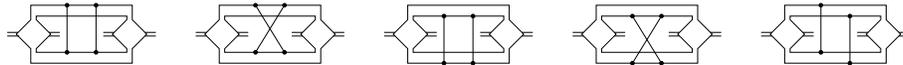

Figure 5.3: Leading diagrams with topologies not yet contained in the long range diagrams. They all contain two scatterers.

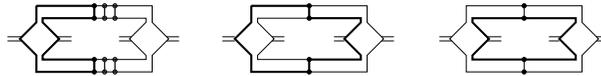

Figure 5.4: Different choices for the loop momenta, all yielding leading contributions; thick and thin lines depict different large momenta, both with length $k$. The first diagram is leading for any number of scatterers. When only two scatterers occur also the other two diagrams are leading.

### 5.4.2 Extra short distance contributions

Unfortunately, we have not yet finished with the calculation as also other terms are of leading order at short distances. First, diagrams with a different topology also occur. There are five new diagrams, all with two scatterers, see Fig. 5.3. They can be looked upon as diagrams from the set of Fig. 5.2(a), but without any scatterers in the ladders.

Second, diagrams can have more than one configuration for a resonant arrangement of the momenta, i.e. several configurations can be leading. The most important contribution to the integral (3.21) arises if the *amplitude* loop momentum $\mathbf{p}$ is close to the wave-vector, i.e. $|\mathbf{p}| = k + O(\mathbf{q})$. With more loops present it is sometimes possible to find more than one choice for the momenta of the amplitude propagators. This we illustrated in Fig. 5.4, where the diagram of Fig. 5.2(a)i is repeated. The expression for the shown diagram has contribution from three different poles. The first arrangement of the momenta is dominant for any number of scatterers, this one is part of the set of long range diagrams for any number of scatterers. The two configurations on the right give an extra leading contribution. This only occurs with two scatterers present; with more scatterers the two diagrams on the right become sub-leading, and only the left diagram remains. Apart from this diagram, the same thing occurs for the diagrams of Fig. 5.2(c)i and Fig. 5.2(c)ii also with two scatterers, and for the diagram Fig. 5.2(c)vi with three scatterers.

Also the diagrams of Fig. 5.3 have more than one pole: The four left diagrams of Fig. 5.3 have two resonant momenta configurations, the right one has three. We sum all the contributions thus found (41 in total) and denote them as $S_n$ for $n$ scatterers. They are only non-zero for diagrams with two or three scatterers, and read

$$
\begin{aligned}
S_2 &= \left(\frac{4\pi}{\ell}\right)^2 \left[5(I_{1,1}^{1,1})^2 + 8I_{1,1}^{1,1}I_{1,2}^{0,1} + 12(I_{1,2}^{0,1})^2 + 8I_{1,1}^{0,1}I_{1,2}^{1,1}\right] \\
&= \frac{-\ell^4}{2k^4}A_1A_2, \tag{5.25}
\end{aligned}
$$

$$
S_3 = 8\left(\frac{4\pi}{\ell}\right)^3 I_{1,1}^{0,1}I_{1,1}^{0,2}I_{1,2}^{0,0} = \frac{\ell^4}{2k^4}A_1A_2. \tag{5.26}
$$

One sees that these special contributions cancel.

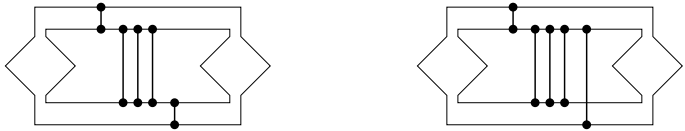

Figure 5.5: Leading diagrams without amplitude exchange with one internal diffuson. The two drawn diagrams cancel against each other. All diagrams of this class cancel.

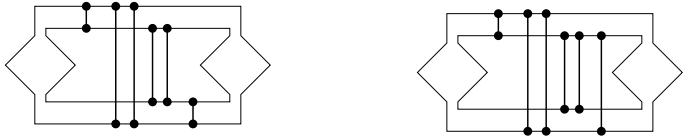

Figure 5.6: Leading diagrams without amplitude exchange with two internal diffusons. The two drawn diagrams cancel against each other. All diagrams of this class cancel.

### 5.4.3   Interference vertices without partner exchange

Apart from the diagrams generated by the Kadanoff-Baym approach, we found another class of leading long range diagrams when we generated the diagrams by computer. In these diagrams the two diffusons have scatterers in common, but no amplitude exchanges. They either have one internal diffuson, see Fig. 5.5 or two, see Fig. 5.6. These diagrams are actually of similar type as Fig. 5.1. The diagram with one internal diffuson is analogous to Fig. 5.1(c)i, while the ones with two internal diffusons are analogous to Figs. 5.1(a)i,(b)i. The equivalents of Figs. 5.1(a)ii,(b)ii and (c)ii also occur. The important difference with the previous diagrams, though, is that at the interference vertices no amplitudes exchange. In this respect they differ qualitatively from Hikami-boxes, where partner exchange does occur. Explicit calculation shows that these terms cancel both at zero and non-zero external momentum, this was also already noted by Kane, Serota and Lee [86], see their Fig. 5(c). These classes of diagrams can thus be fully neglected. This cancellation is due to time reversal invariance. Apart from these two classes, the Kadanoff-Baym approach generates all presented diagrams as we checked with the computer program.

In conclusion, the analysis of this section 5.4 shows that when the external momenta vanish, all leading terms cancel at fixed value of the internal momentum. Upon integration over the loop momentum one still has zero and, in particular, not a divergent contribution. For an infinite system the theory is thus well-behaved at short distances. The divergence has canceled in a careful study of the short-distance process; renormalization as known from field theories was not needed.

## 5.5   Application to optical systems

We have now seen that in infinite systems all divergent terms cancel. In realistic systems, such as a slab, the same diagrams describe the relevant physics. They have to be evaluated with appropriate diffuson propagators. Generally they can be written as the bulk term with additional mirror terms. For a slab we gave them in chapter 2. Knowing that the short distance behavior is regular, we can self-consistently consider all scattering diagrams in the diffusion approximation. For the long range diagrams this was done already in section 5.3, where we evaluated all long range contributions. The discussion of previous section has shown that the only new effect comes from the subtraction of Eq. (5.24). In the diffusion limit for a quasi one-dimensional system at fixed transversal momentum $Q$ it results in a contact term, labeled $F_d$.

$$F_d(M) = \int \mathrm{d}z \mathcal{L}_{in}^2(z) \int_{-\infty}^{\infty} \frac{dq_z}{2\pi} \frac{\ell^4}{4k^4} \mathcal{L}_{out}^2(z) = \frac{\ell^4}{4k^4} \delta(0) \int dz \mathcal{L}_{in}^2(z) \mathcal{L}_{out}^2(z). \quad (5.27)$$

This indeed cancels exactly the leading divergency that remained in Eq.(5.17) for the sum of $F_a$, $F_b$ and $F_c$. The milder divergency present in the individual terms $F_a$, $F_b$, $F_c$ in two and three dimensions canceled already by summing them. Diagrammatically this one sees as follows: The diagrams responsible for the latter divergence contain one internal diffuson (such in Fig. 5.1(c)). For these diagrams one does not have the complications of the previous section; double counting corrections, different resonant momentum configurations, and extra diagrams are absent. This was confirmed by our computer generated diagrams. The only contributions that remain in (5.17) are thus the double $z-$integrals. This finite remainder of our theory yields the sought conductance fluctuations. We thus obtain, with again $h_4 = \ell^5/(48\pi k^2)$,

$$\begin{aligned}
F(Q, \kappa, \Omega) = \ & F_a(M) + F_b(M) + F_c(M) + F_d(M) \\
= \ & 4h_4^2 \iint \mathrm{d}z \, \mathrm{d}z' \, \mathcal{L}_{int}(z, z'; M) \mathcal{L}_{int}(z, z'; M^*) \\
& \times \left[ \mathcal{L}_{in}'^2(z) \mathcal{L}_{out}'^2(z') + \mathcal{L}_{in}'(z) \mathcal{L}_{out}'(z) \mathcal{L}_{in}'(z') \mathcal{L}_{out}'(z') \right] \\
+ \ & \frac{h_4^2}{2} \iint \mathrm{d}z \, \mathrm{d}z' \left[ \mathcal{L}_{int}^2(z, z'; M) + \mathcal{L}_{int}^2(z, z'; M^*) \right] \frac{d^2}{dz^2} \left[ \mathcal{L}_{in}(z) \mathcal{L}_{out}(z) \right] \\
& \times \frac{d^2}{dz'^2} \left[ \mathcal{L}_{in}(z') \mathcal{L}_{out}(z') \right],
\end{aligned} \quad (5.28)$$

in which again $M^2 = Q^2 + \kappa^2 + i\Omega$. This equation is the central result of the present chapter. The upper line of (5.28) corresponds to the diagrams of Fig. 5.1(a), whereas the lower line corresponds to the diagrams of Fig. 5.1(b). Note that only derivatives of external diffusons are present. As compared to the first line, extra terms are present in the second one. According to (2.39) we have

$$\frac{d^2}{dz^2} \left[ \mathcal{L}_{in}(z) \mathcal{L}_{out}(z) \right] = 2 \mathcal{L}_{in}'(z) \mathcal{L}_{out}'(z) + 2\kappa^2 \mathcal{L}_{in}(z) \mathcal{L}_{out}(z) \quad (5.29)$$

The $\kappa-$terms are extra terms arising when absorption is present. Finally, with Eq.(5.4) the value at vanishing transversal momentum gives the variance of the conductance in one dimension, integration over the transversal momentum yields the correlation in two and three dimensions.

Using the general result of (5.28) and (5.4) we consider various cases by inserting the diffusons derived in chapter 2. First consider the case of fully transmitting surfaces; if we neglect absorption and frequency differences this gives

$$F(Q) = \frac{3}{2} \frac{2 + 2Q^2 L^2 - 2\cosh 2QL + QL \sinh 2QL}{Q^4 L^4 \sinh^2 QL}, \quad (5.30)$$

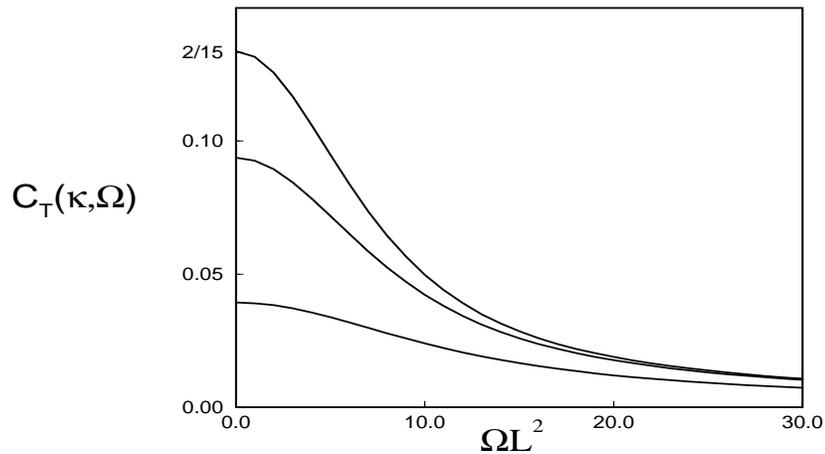

Figure 5.7: The correlation of the conductance as a function of the frequency for various absorption strengths in one dimension without internal reflection. From upper to lower curve: no absorption($\kappa = 0$), $\kappa = 1/L$ and, $\kappa = 2/L$.

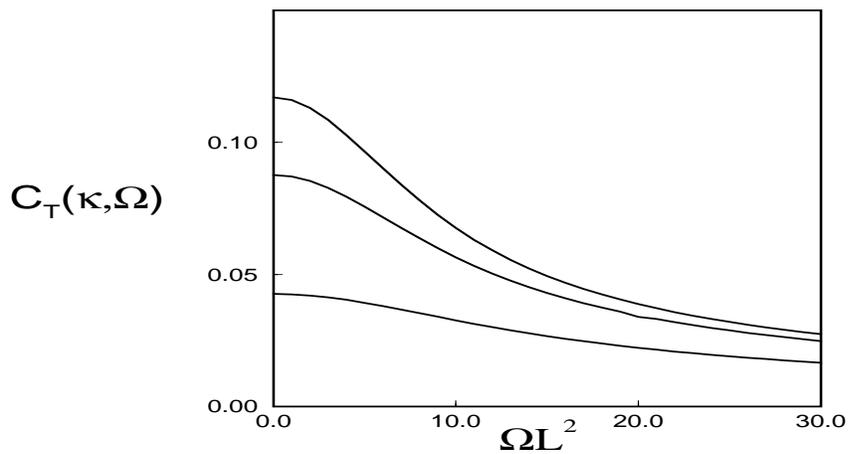

Figure 5.8: The correlation of the conductance as a function of the frequency for various absorption strengths in a two dimensional slab without internal reflection. From upper to lower curve: $\kappa = 0$, $\kappa = 1/L$ and, $\kappa = 2/L$.

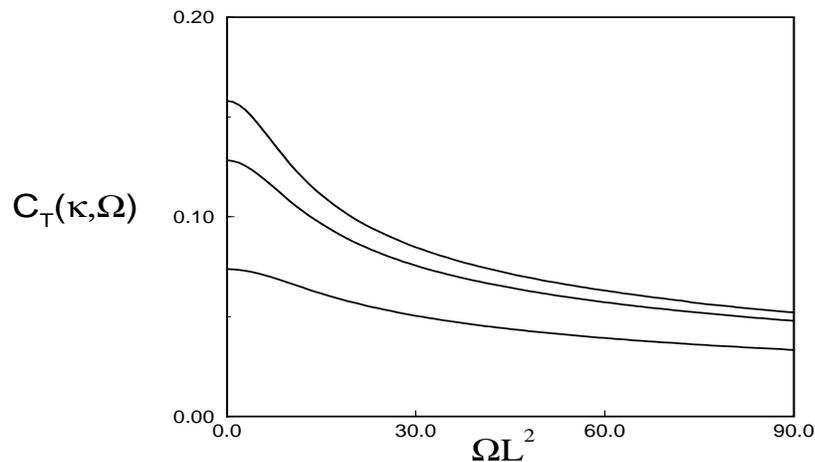

Figure 5.9: The conductance correlation function versus the frequency difference for various absorption strengths in a three dimensional slab without internal reflection. From upper to lower curve: $\kappa = 0$, $\kappa = 1/L$ and, $\kappa = 2/L$.

which decays for large $Q$ as $Q^{-3}$. Here we recover

$$\langle T^2 \rangle_c = \frac{2}{15} \approx 0.133, \qquad \text{quasi 1d} \tag{5.31}$$

$$= \frac{3}{\pi^3} \zeta(3) \frac{W}{L} \approx 0.116 \frac{W}{L}, \qquad \text{quasi 2d} \tag{5.32}$$

$$= \frac{1}{2\pi} \frac{W^2}{L^2} \approx 0.159 \frac{W^2}{L^2} \qquad \text{3d}, \tag{5.33}$$

in which $\zeta$ is Riemann's zeta function. For *cubic* samples Lee *et al.* [110] find in 2D 0.186 and in 3D a value of 0.296 (the value in 1D is of course unaltered).

We determine also the frequency dependency of the correlation; this is important as it determines the frequency range of the light needed to see the fluctuations. Taking the frequency dependency into account we obtain

$$F(Q,\omega) = \frac{4\left(M^{*2} - M^2 + M^2 M^* L \coth M^* L - M^{*2} ML \coth ML\right)}{L^4 M^2 M^{*2}(M^2 - M^{*2})}$$
$$+ \text{Re}\, \frac{2 + 2M^2 L^2 - 2\cosh 2ML + ML\sinh 2ML}{2M^4 L^4 \sinh^2 ML}. \tag{5.34}$$

The correlation decays for large frequency differences as $\Omega^{2-d/2}$, as was stated by Lee, Stone and Fukuyama [110].

We were unable to perform the double integral over the position analytically in presence of absorption. In Figs. 5.7, 5.8, and 5.9 we show the one, two, and three dimensional correlation functions for various values of the absorption. One sees that especially the top of the correlation reduces due to absorption.

Next we applied our theory to the case of partial reflection at the surfaces of the sample. We assume an index of refraction $m = \sqrt{\epsilon_0/\epsilon_1} \neq 1$. For our purpose the internal reflections code in only one parameter, the injection depth $z_0$, see (2.70), (2.75), (2.41). For optically thick samples the correlations are now determined by $z_0/L$. It is physically clear that the internal reflections lead to a less steep diffuse intensity in the sample as function of the depth. The fluctuations are proportional to the space derivatives and thus reduce. We present the results in the Figs. 5.10, 5.11 and 5.12 where we plotted the correlation function for various values of the ratio between extrapolation length and sample thickness. One sees that already for small ratios of $z_0/L$ the correlation is significantly lower than without surface internal reflection. One sees that neither the variance (the value at vanishing $\Omega$), nor the form of the correlations are universal, they are sensitive to absorption and internal reflections.

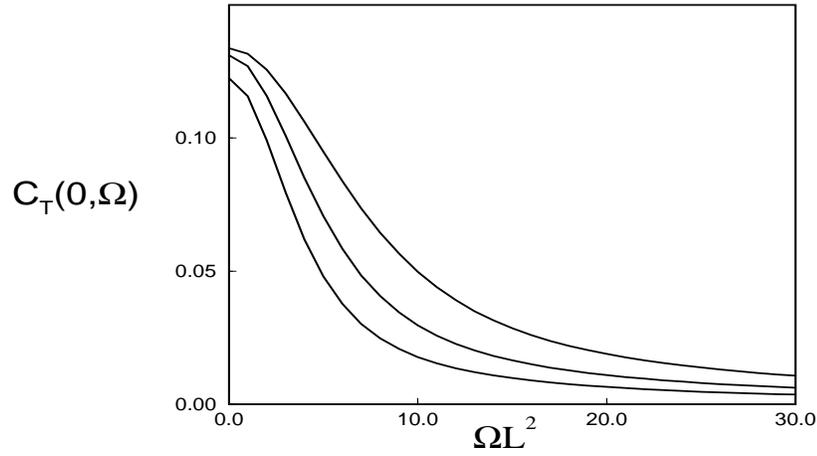

Figure 5.10: The influence of internal reflection on the one dimensional frequency correlation. With $z_0 = 0, L/10, L/5$ (upper, middle, lower curve); no absorption. Also here the fluctuations reduce.

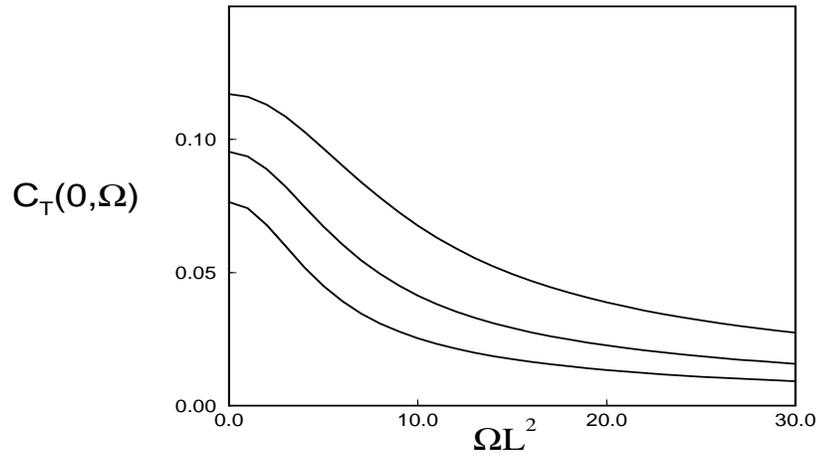

Figure 5.11: Influence of internal reflection on the frequency correlation in 2D; $z_0 = 0, L/10, L/5$ (upper, middle, lower curve); no absorption.

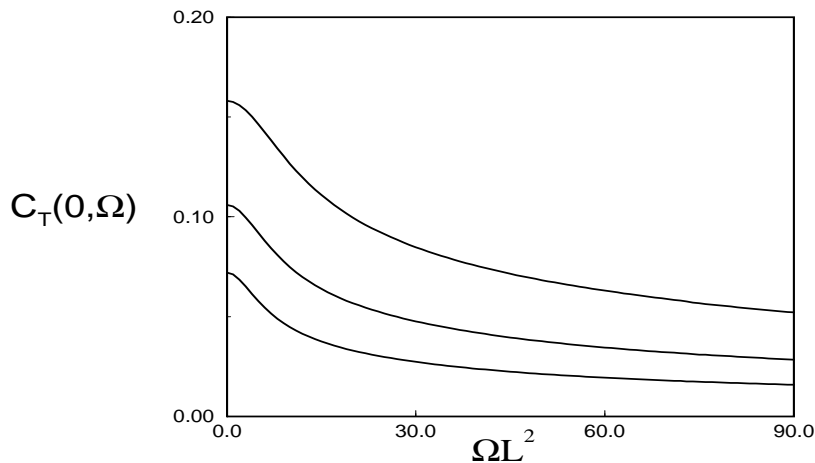

Figure 5.12: Influence of internal reflection on the frequency correlation in 3D; $z_0 = 0, L/10, L/5$ (upper, middle, lower curve); no absorption.

## 5.6   Discussion and outlook

The universal conductance fluctuations (UCF) of mesoscopic electronic systems have a direct counterpart in other mesoscopic systems with multiply scattered classical waves. The corresponding average normalized correlation function is the $C_3$. It is known that a naive calculation of this object in the Landauer approach is plagued with short distance divergencies. In this chapter we have presented a detailed diagrammatic approach to the calculation of the $C_3$ correlation function. We first evaluated the leading long range diagrams. As expected, each diagram, but also their sum, contains a short distance divergency. Also a sub-leading divergency occurs, but for a slab geometry this cancels automatically. The study of the cancellation of the leading divergency then was the main theme of the chapter. Consistency of the approach requires finding extra contributions that exactly cancel the already determined divergency, and therefore serves as proof that the set of long range diagrams exhaust all of them.

We developed a diagrammatic method that systematically generates all leading scattering diagrams. This set was checked by computer. Many diagrams had to be consid-ered in detail. We summed these diagrams for an infinite system at fixed value of the loop momentum. It was seen that also beyond the diffusion approximation the action of Hikami boxes on the long range propagators is to eliminate the long range terms, and leave only some low order contributions. Moreover, some low order terms of diagrams with diffusons have a lower degeneracy than their higher order equivalents. Taking this into account led exactly to a cancellation of all terms in an infinite system. Next we discussed that some extra classes of leading diagrams occur, but they all add up to zero. Thus in an infinite system all diagrams cancel, so that, in particular, no short range divergency occurs. All short range contributions could be coded in a contact term (5.27). It would be interesting to investigate how one derives it in a non-linear sigma-model formulation of the theory. Probably this would go along the lines of Serota, Esposito and Ma [116], who introduce so called higher gradient vertices (see section 3.4.1) for these terms.

Subsequently we applied the theory to systems of finite size. Here the short distance divergencies cancel as well, because the large scale geometry of the system does not have influence on short range effects. The final result is non-zero as it describes the correlation function in terms of derivatives of external diffusion propagators of the geometry considered; such terms have no meaning in an infinite system.

Our central result for the correlation function of the conductance is Eq. (5.28). It is obtained by adding the result of the long range diagrams (5.17) and the contact term (5.27). When there is no absorption, we find agreement with the result of Kane, Serota and Lee [86]. All external diffusons are differentiated once. When absorption is present, however, their approach is no longer valid. We found that extra terms appear where some external diffusons are differentiated twice or are proportional to $\kappa^2$, see (5.29).

We have applied the results to realistic optical systems. The frequency dependent $C_3$ correlation function was calculated for the case where a diffuse incoming beam is used and all outgoing intensity is collected. It was seen that both absorption and internal reflections decrease the correlations by a considerable amount. This is important for a quantitative analysis of experimental data.

Electromagnetic measurements that involve the $C_3$ correlation have been reported by Genack *et al.* [117], who described the data of their infrared experiments by adding the $C_1$, $C_2$ and $C_3$ contributions, but they incorrectly assumed that the $C_3$ is frequency independent. The experimental investigation of optical universal conductance fluctuations is known to be very difficult. One problem is that if the incoming beam has to be diffusive, it has a low intensity.

We propose here a different way to measure the same interference effect. Consider a laser beam coming in a given direction $a$ and measure the frequency dependent total transmission. Such can be done using an integrating sphere [72]. Then repeat the measurement for a very different incoming direction $c$. Each of these two signals will exhibit the large $C_2$ correlation function [72]. However, when the directions $a$ and $c$ are not close to each other, the $C_2$ will not contribute to the *cross correlate*. The cross correlation of the total transmission is much smaller than the auto-correlates; it just represents the typical term of the UCF in (5.5). As this cross correlate is of relative order $1/g^2$, it is of the same order of magnitude as the third cumulant of the total transmission. A very precise measurement of that quantity, see chapter 6, was reported

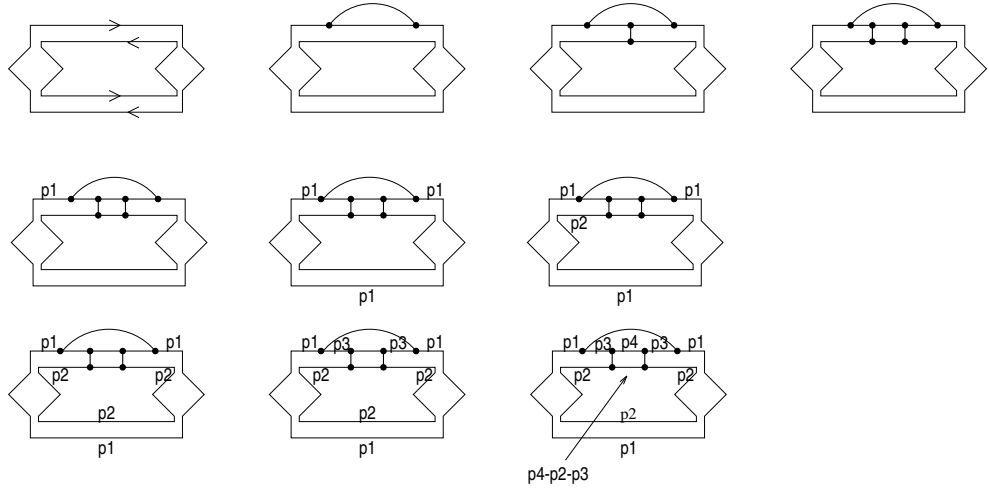

Figure 5.13: Example of the generation and calculation of a UCF diagram: diagram c5 in Fig. 5.2. We follow the placing of the scatterers (top line) and the momenta (lower two lines). For two choices of the momenta this diagram yields leading contributions if $\mathbf{p}_3 = -\mathbf{p}_2 + \mathbf{q}$ or $\mathbf{p}_4 = \mathbf{p}_2 + \mathbf{q}$ (where $\mathbf{q} \sim \ell \ll \mathbf{p}_i$).

[105]. It may therefore be expected that it is just possible to measure $C_3$, and thus essentially the UCF, with visible light.

# Appendix: Generating diagrams automatically

In this part we discuss the program to generate diagrams in more detail. We used this program to generate to short-distance UCF contributions. We did this calculation with zero external momenta. Also absorption and frequency differences were neglected. It proved very difficult to produce all diagrams by hand for the short distances. With only a few scatterers present, the systematic treatment with Hikami boxes breaks down. On one hand some diagrams are over-counted, on the other hand new diagrams appear. The goal of the program was to produce all diagrams without double counting.

## Placing scatterers

The first step in the calculation is to place the scatterers. Starting from the upper left we choose the two points on the propagators where we place the scatterer. We allow every placing in this step. Recursively we construct the tree of diagrams, until the desired number of scatterers is reached.

We thus have constructed a diagram with $n$ scatterers. First thing we do next, is to check the topology. First, it is possible that a certain topology is not allowed. For instance the case where the same propagator scatters twice on a scatterer directly after one another is already included in the self-energy. Yet also scatterers that can be 'pulled' into the external diffuson are not allowed, as they are also already considered. Second, diagrams also can be of lower order. An example is the class of diagrams where one amplitude scatters on scatterer a), then on b), now on c), then again on a) and finally on b). The sub-leading diagrams were thrown away.

Next we checked whether the topology was seen before. The recursion tree is very degenerate. By its construction, we also get permutations of the placing of scatterers, but they should only be counted once. If the topology was seen before, we skipped the diagram. After these steps we are left with distinct topologies that can not be excluded solely on their topology. This set is quite large and is a bottleneck in computation: some 500,000 diagrams for 6 scatterers.

## Placing momenta

The other important reduction of the number of leading terms asks for the explicit evaluation of the diagrams. We place momenta in the diagram and calculate the diagram. First we fix the momentum in the left part of the uppermost propagator. Because of momentum conservation and the fact that the outgoing diffusons carry no momentum, the momentum in the left part of the lowest propagator is the same. Now we repeatedly use momentum conservation at the external legs of the diagram and at the scatterers. If after the use of momentum conservation there are still undetermined momenta, we introduce a new independent momentum. We repeat this procedure until no momenta are left undetermined, see Fig. 5.13. Although at first sight this procedure looks straightforward, it is possible that, if the new momenta are put randomly, momentum conservation is broken in the end. By carefully choosing the order in which we introduce new momenta, it is possible to conserve momentum everywhere.

Now we know all momenta. To have a resonant, i.e. leading, diagram the momenta of all propagators need to be around $|\mathbf{p}| \approx k$. It is possible that this is already the case after the last step. But otherwise one momentum has to be made small as compared to $|k|$; this is the *intensity* loop momentum $\mathbf{q}$. It turned that there are maximally three possibilities for the choice. In Fig. 5.13 the two possible choices are $\mathbf{p}_3 = -\mathbf{p}_2 + \mathbf{q}$ and $\mathbf{p}_4 = \mathbf{p}_2 + \mathbf{q}$. For each momentum configuration thus found, we counted the number of independent of momenta left. This yielded the expression for the diagram in terms of the $I$ integrals of section 5.4 and 3.4.1. If either the number of retarded of advanced propagators is zero, the integral has no poles and becomes sub-leading.

On an IBM RS6000 work station the maximal number of scatterers in the diagrams was limited to six. This proved large enough, as already with four or more scatterers the special features of the short distance diagrams disappeared, as we saw in section 5.4. Thus we were able to check this up to two orders higher.

# 6

# Third cumulant of the total transmission

Uptill now we have considered correlations between intensities defined as

$$\frac{\langle T_{ab}T_{cd}\rangle}{\langle T_{ab}\rangle\langle T_{cd}\rangle} = 1 + C_1 + C_2 + C_3. \tag{6.1}$$

The $C_1(Q,\Omega)$, $C_2(Q,\Omega)$, and $C_3(\Omega)$ correlation functions thus describe correlations between two intensities only. By their definition the value at zero frequency and zero momentum-difference, i.e. the peak value, presents the second cumulant of the distribution functions of the transmission quantities. This interpretation of the correlation functions and the interest in higher moments is rather recent. It is especially of interest near the Anderson transition, where interference gets more and more important (more Hikami boxes "appear") changing the distribution function. In chapter 7 we calculate the full distribution function for the angular resolved transmission and the total transmission in the regime of moderate $g$. Yet first, in this chapter we calculate the third cumulant of the total transmission. Using the good old diagrammatic approach we relate the third cumulant normalized to the average, $\langle\langle T_a^3\rangle\rangle$ to the normalized second cumulant $\langle\langle T_a^2\rangle\rangle$. For a broad Gaussian beam profile we find $\langle\langle T_a^3\rangle\rangle = \frac{16}{5}\langle\langle T_a^2\rangle\rangle^2$, which is in good agreement with experimental data.

## 6.1   Introduction

The interest in the full distribution functions in mesoscopic systems is rather recent. Examples of such distributions are the intensity distribution in speckle patterns for classical waves [95, 118] and the conductance distribution for electronic systems [23]. The size of the fluctuations and the shape of the distribution relates to the 'distance' from the localization transition. Far from localization, diffusion channels are almost uncorrelated and fluctuations are small (except the optical speckle pattern in the angular resolved transmission). The correlation between the channels increases if the localization is approached. The relevant parameter is the inverse dimensionless conductance $1/g$, which can be interpreted as the chance that two channels interfere. We express the dimensionless conductance in the thickness of the sample $L$, the mean free path $\ell$, and the number of channels $N$,

$$g = \frac{4N\ell}{3L} \ . \tag{6.2}$$

We calculate the number of channels in analogy with a waveguide, where it is unambiguously defined. In the diffuse mesoscopic regime $g^{-1}$ is a small parameter of our perturbation theory; experimentally this proves fully justified, as there one has $g \sim 10^3$, [105]. Close to Anderson localization $g$ approaches unity, and fluctuations increase. The central question is how the distributions change as we approach the strong localization regime [23, 119].

The total transmission is a constant superposed with fluctuations. In first order of $g^{-1}$ the fluctuations have a Gaussian distribution [72, 118]. The relative variance of this distribution, the top of the $C_2$, is proportional to $g^{-1}$, it is thus a factor $g$ larger than for the conductance fluctuations. This sensitivity of the total transmission to interference processes and its simple limiting behavior (as compared to the angular resolved transmission) make it an ideal quantity to study mesoscopic transmission. Recently the third cumulant of the distribution was found experimentally by De Boer *et al.* [105]. In this chapter we present the theoretical details of that work. We focus on the Gaussian distribution and the deviation from the Gaussian due to the presence of the third cumulant. The structure of this chapter is as follows. In section 6.3 we discuss the character of the probability distribution. In section 6.3.2 and 6.4.1 we calculate the third cumulant of the probability distribution. Next we calculate experimental corrections to our result in section 6.4.2 and 4.3.2, after which we compare our results with the experimental data in section 6.5.

## 6.2 Intermezzo: Cumulants and moments

In the following we need some generalities from probability theory, see e.g. Van Kampen [120]. To a general probability distribution $P(x)$ one can assign the characteristic function $\phi(k)$, defined as

$$\phi(k) \equiv \langle e^{ikx} \rangle = \sum_{n=0}^{\infty} \frac{(ik)^n}{n!} \langle x^n \rangle \tag{6.3}$$

where $\langle x^n \rangle$ is the $n^{th}$ moment of the distribution: $\int \mathrm{d}x\, P(x) x^n$. From this characteristic function one can again reconstruct $P(x)$. In statistical problems one often encounters the addition of many independent chances. The central limit theorem tells us that the sum will tend to a Gaussian distribution (no matter the distribution of the individual variables). Moments are a bit unapt for Gaussian distributions, as each moment is nonzero. Cumulants are more appropriate here. The cumulants $K_n$ are implicitly defined as

$$\phi(k) = \exp\left[\sum_{n=0}^{\infty} \frac{(ik)^n}{n!} K_n\right] \tag{6.4}$$

Expansion with respect to $k$ yields the relation between cumulants and moments

$$K_{n+1} = \langle x^{n+1} \rangle - \sum_{j=1}^{n} \binom{n}{j} \langle x^j \rangle K_{n+1-j} \tag{6.5}$$

Thus $K_1 = \langle x \rangle$, $K_2 = \langle x^2 \rangle - \langle x \rangle^2$, and $K_3 = \langle x^3 \rangle - 3\langle x^2 \rangle \langle x \rangle + 2\langle x \rangle^3$. For a Gaussian distribution

$$P(x) = \frac{1}{\sigma\sqrt{2\pi}} e^{-\frac{(x - \langle x \rangle)^2}{2\sigma^2}} \tag{6.6}$$

only the first two cumulants are non-zero: $K_1 = \langle x \rangle$ and $K_2 = \sigma^2$, where $\sigma^2$ is the variance.

## 6.3 Cumulants of the probability distribution

In this section we introduce the probability distribution of the total transmission of scalar waves and we discuss some of its properties. The corrections for vector waves will be made in section 4.3.2. We link the moments of the distribution to diagrams. The moments of the probability distribution $P(T_a)$ can be extracted as

$$\langle T_a^k \rangle = \int dT_a \, P(T_a) T_a^k. \tag{6.7}$$

In a diagrammatic approach the $k$-th moment can be represented by a diagram with $k$ diffusons on both incoming and outgoing side. The $k = 1$ term is the average total transmission $\langle T_a \rangle$, as given by the Schwarzschild-Milne equation in Eq.(2.80). This quantity is a single diffuson and is thus independent of channel-to-channel correlations. The second moment can be decomposed in the first two cumulants:

$$\frac{\langle T_a^2 \rangle}{\langle T_a \rangle^2} = \frac{\langle T_a \rangle^2 + \langle T_a^2 \rangle_{\text{cum}}}{\langle T_a \rangle^2} = 1 + \langle\langle T_a^2 \rangle\rangle, \tag{6.8}$$

the double brackets denote cumulants normalized to the average. Diagrammatically we depict the second moment in Fig. 4.6. The decomposition in cumulants proves useful as each cumulant corresponds to a different number of interactions between the diffusons. In the first term, Fig. 4.6(a) there is no interference; it factorizes in the average transmission squared (apart from a small correction discussed in section 4.3.2). The second term, Fig. 4.6(b), is the second cumulant $\langle\langle T_a^2 \rangle\rangle$. It gives the variance of the fluctuations. Interactions between two diffusons are responsible for the presence of this second cumulant. This is just a special case of the $C_2$ correlation function studied in chapter 4. The $C_2$ frequency correlation function was defined: $C_2(\Delta\omega) = \langle\langle T_a(\omega) T_a(\omega + \Delta\omega) \rangle\rangle$. For our case find that $\langle\langle T_a^2 \rangle\rangle = C_2(0)$ and thus corresponds to the peak value of this correlation function.

Similar to the second moment, one can distinguish three different contributions to the third moment,

$$\frac{\langle T_a^3 \rangle}{\langle T_a \rangle^3} = 1 + 3\langle\langle T_a^2 \rangle\rangle + \langle\langle T_a^3 \rangle\rangle. \tag{6.9}$$

We have drawn the corresponding leading diagrams in Fig. 6.1. The first term, Fig. 6.1(a), again corresponds to the transmission without interference. The second term, Fig. 6.1(b), is reducible in a single diffuson and a second cumulant diagram. From the figure it is clear that this decomposition can be done in three ways which reflects in the prefactor of $\langle\langle T_a^2 \rangle\rangle$ in Eq. (6.9). The third contribution stands for the third cumulant in the distribution and expresses the leading deviation from the Gaussian distribution. This is the term we are mainly interested in. It consists of two related diagrams: Fig. 6.1(c+d). The three intensities can interfere twice two by two, or the intensities can interact all three together, with a so called Hikami six-point vertex. Both contributions will prove to be of the same order of magnitude. The strength of the effect can be easily estimated using the interpretation of $1/g$ as an interaction probability. By looking at the diagram, the third cumulant is proportional to the chance of two diffusons meeting twice, thus of the order $1/g^2$. We find the basic result

$$\langle\langle T_a^3 \rangle\rangle \propto \langle\langle T_a^2 \rangle\rangle^2. \tag{6.10}$$

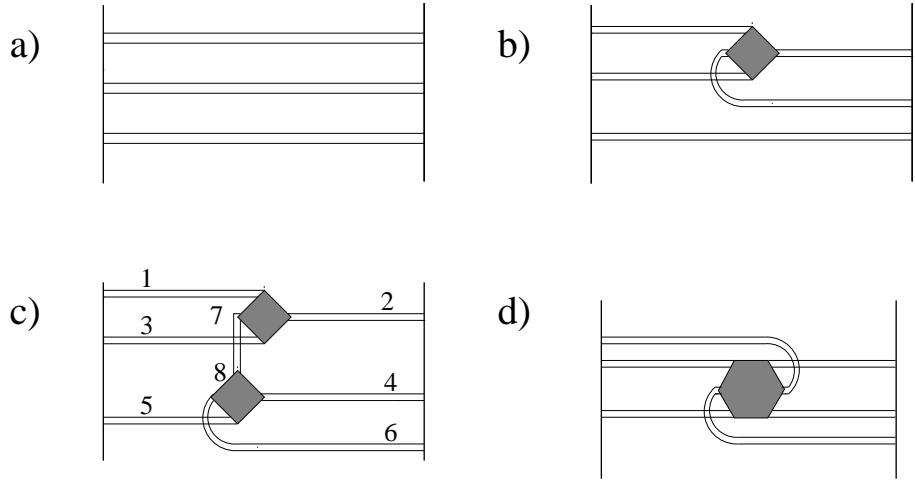

Figure 6.1: The three contributions to the third moment of the total transmission. Diagram a) corresponds to independent transmission channels; it is of order 1. In diagram b) there is correlation but it can be decomposed into the second cumulant; this is of order $g^{-1}$. The diagrams c) and d) are the contributions to the third cumulant, $O(g^{-2})$.

The rest of the chapter essentially consist of proving this relation and determination of the prefactor. Finally we will compare this relation to experimental data of Ref. [105].

Because one samples only a finite number of channels in an experiment, the law of large numbers predicts a distribution with some non-zero width even if we only would consider disconnected diagrams. However, we have shown in section 4.3.2 that this effect brings only a negligible contribution to the measured second cumulant; this is also true for the third cumulant as we will show below. The large fluctuations are almost only due to the interference.

Note that in our calculations we again neglect diagrams with loops. An example of a loop diagram is the $C_3$ contribution to the second cumulant, see chapter 5. The $C_3$ diagram contains two Hikami boxes, therefore it gives a contribution of order $g^{-2}$ to the second cumulant. In general, one easily sees that to create a loop, one needs more interference vertices. Therefore, these diagrams are of higher order in $1/g$, and we did not calculate them. The leading contributions to the cumulants are by far sufficient for the description of the experiment of Ref. [105].

### 6.3.1 The second cumulant

First we recover the results for the second cumulant as calculated in chapter 4. In the plane wave limit of the incoming beam, all transverse momenta are absent and the diffusons are simple linear functions. One obtains for the second cumulant

$$
\begin{aligned}
\langle\langle T_a^2\rangle\rangle &= \langle T_a\rangle^{-2} \int\int \mathrm{d}x\mathrm{d}y \int_0^L \mathrm{d}z\, H_4 \mathcal{L}_1(z)\mathcal{L}_2(z)\mathcal{L}_3(z)\mathcal{L}_4(z) \\
&= \frac{1}{gL^3}\int_0^L \mathrm{d}z[z^2 + (1-z)^2] = \frac{2}{3}g^{-1}.
\end{aligned}
\tag{6.11}
$$

Taking only this second cumulant we find for the moment a Gaussian probability distribution in the plane wave limit [118, 72]

$$
P(T_a) = \sqrt{\frac{3g}{4\pi}} \exp\left[-\frac{3g}{4}\left(T_a - \langle T_a\rangle\right)^2\right].
\tag{6.12}
$$

We also recall the influence of the beam profile on the correlation. There the $Q$-dependence of the incoming diffusons should be taken into account. One finds the total cumulant by integrating over the momentum with the corresponding weight. For a Gaussian beam profile we found

$$
\langle\langle T_a^2\rangle\rangle = \frac{\rho_0^2}{4\pi g}\int \mathrm{d}^2 Q\; e^{-\rho_0^2 Q^2/4} F_2(QL),
\tag{6.13}
$$

with $F_2(x) = [\sinh(2x) - 2x]/[2x\sinh^2 x]$. If the incoming beam is again very broad, $\rho_0 \gg L$, one recovers the plane wave behavior $\langle\langle T_a^2\rangle\rangle = 2/3g$. Note that this agreement is found by identifying the area of a Gaussian profile with $A = \pi\rho_0^2$. This definition is somewhat arbitrary and other choices are also possible. After fixing this definition, no freedom remains and we will see that for the third cumulant a Gaussian profile leads to other results than a plane wave.

### 6.3.2 The third cumulant

We now discuss the calculation the third cumulant. As mentioned, there are two processes contributing. One with two four point vertices, that we term $\langle\langle T_a^3\rangle\rangle_c$ and one with a six point vertex $\langle\langle T_a^3\rangle\rangle_d$, where we have chosen the subscript according to Fig. 6.1. These processes were, to our knowledge, not studied before in detail. Our calculation follows the lines of the second cumulant calculation.

#### Interference via two four point vertices

First consider the diagram in Fig. 6.1(c). We have labeled to incoming diffusons with odd numbers, the outgoing ones with even numbers. Two incoming diffusons, $\mathcal{L}_1$ and $\mathcal{L}_3$ meet at a position $z$. In a Hikami box the diffusons $\mathcal{L}_1$ and $\mathcal{L}_3$ interfere into $\mathcal{L}_2$ and an internal diffuson $\mathcal{L}_{78}^{\mathrm{int}}$. The $\mathcal{L}_2$ propagates out, whereas $\mathcal{L}_{78}^{\mathrm{int}}$ interferes again at $z'$ with incoming diffuson $\mathcal{L}_5$ into two outgoing ones $\mathcal{L}_4$ and $\mathcal{L}_6$. Apart from this process, also three other sequences of interference are possible. This means that the diffusons can also be permuted as: $(\mathcal{L}_1, \mathcal{L}_3, \mathcal{L}_5, \mathcal{L}_2, \mathcal{L}_4, \mathcal{L}_6) \rightarrow (\mathcal{L}_3, \mathcal{L}_5, \mathcal{L}_1, \mathcal{L}_4, \mathcal{L}_6, \mathcal{L}_2) \rightarrow (\mathcal{L}_5, \mathcal{L}_1, \mathcal{L}_3, \mathcal{L}_6, \mathcal{L}_2, \mathcal{L}_4)$. We denote the sum over these permutations as $\sum_{\text{per}}$. As the diagrams also can be complex conjugated, there is also a combinatorial factor 2 for all diagrams. (Note that complex conjugation for the *second* cumulant diagram, yields no different diagram and thus it should not be counted.) The expression for the diagram of Fig. 6.1(c) is now

$$
\begin{aligned}
\langle\langle T_a^3 \rangle\rangle_c &= \langle T_a \rangle^{-3} \, 2 \sum_{\text{per}} A \int_0^L \mathrm{d}z \int_0^L \mathrm{d}z' \, H_4(z) H_4(z') \mathcal{L}_1(z) \mathcal{L}_2(z) \mathcal{L}_3(z) \times \\
&\quad \mathcal{L}_4(z') \mathcal{L}_5(z') \mathcal{L}_6(z') \mathcal{L}_{78}^{\text{int}}(z, z').
\end{aligned}
\tag{6.14}
$$

In turns out that it is useful to rewrite the form of the Hikami boxes as introduced in Eq.(3.5) into an equivalent expression, using momentum conservation $\mathbf{q}_a + \mathbf{q}_{a'} + \mathbf{q}_b + \mathbf{q}_{b'} = 0$. In real space the use of momentum conservation corresponds to partial integration. The Hikami box is again simplified using the fact that there are no transversal momentum terms, or $Q$ terms, for the outgoing diffusons. Also we neglect the source terms $q^2$ of incoming and outgoing diffusons. Using the numbering in Fig. 6.1, we obtain

$$
H_4(z) = \frac{-\ell^5}{48\pi k^2}[2\partial_{z_1}\partial_{z_2} + 2\partial_{z_2}\partial_{z_3}], \qquad H_4(z') = \frac{\ell^5}{48\pi k^2}[2\partial_{z_4}\partial_{z_6} - \partial_{z_8}^2 + Q_8^2].
\tag{6.15}
$$

Source terms, i.e. $q_i^2$−terms of the incoming and outgoing diffusons were again neglected, but the source term of the diffuson between the vertices is important. As one sees with the diffusion equation (2.39), it brings

$$
-\partial_{z_8}^2 \mathcal{L}_{78}^{\text{int}}(z, z') + Q_8^2 \mathcal{L}_{78}^{\text{int}}(z, z') = \frac{\ell^3}{12\pi}\delta(z - z').
\tag{6.16}
$$

The contribution from the source term, i.e. $H_4(z') \propto -\partial_{z_8}^2 + Q_8^2$, $H_4(z) \propto \partial_{z_1}\partial_{z_2} + \partial_{z_2}\partial_{z_3}$, is

$$
\begin{aligned}
&-4A\left(\frac{\ell^5}{48\pi k^2}\right)^2 \sum_{\text{per}} \int \mathrm{d}z (\partial_{z_1}\partial_{z_2} + \partial_{z_2}\partial_{z_3}) \mathcal{L}_1 \mathcal{L}_2 \mathcal{L}_3 \int \mathrm{d}z' (-\partial_{z_8}^2 + Q_8^2) \mathcal{L}^{\text{int}} \mathcal{L}_4 \mathcal{L}_5 \mathcal{L}_6 \\
&= \frac{-\ell^7 A}{48\pi k^4} \int_0^L \mathrm{d}z [\partial_{z_1}\partial_{z_2} + \partial_{z_2}\partial_{z_3} + \ldots + \partial_{z_6}\partial_{z_1}] \mathcal{L}_1 \ldots \mathcal{L}_6.
\end{aligned}
\tag{6.17}
$$

Although this corresponds to a local process , just one $z$−coordinate is involved, it is of leading order. Together with the expression coming from $H_4(z')$ proportional to $\partial_{z_4}\partial_{z_6}$, we find for the total contribution of the process in Fig.6.1(c)

$$
\begin{aligned}
\langle\langle T_a^3 \rangle\rangle_c &= \frac{1}{\langle T_a \rangle^3}\left(\frac{l^5}{48\pi k^2}\right)^2 8 \sum_{\text{per}} A \int_0^L \mathrm{d}z \, \mathcal{L}_1(z) \mathcal{L}_2'(z) \mathcal{L}_3(z) \times \\
&\qquad \int_0^L \mathrm{d}z' \, \mathcal{L}_4'(z') \mathcal{L}_5(z') \mathcal{L}_6'(z') \partial_z \mathcal{L}^{\text{int}}(z, z') \\
&\quad -\frac{\ell^7 A}{48\pi k^4 \langle T_a \rangle^3} \int_0^L \mathrm{d}z [\partial_{z_1}\partial_{z_2} + \partial_{z_2}\partial_{z_3} + \partial_{z_3}\partial_{z_4} + \partial_{z_4}\partial_{z_5} + \partial_{z_5}\partial_{z_6} + \partial_{z_6}\partial_{z_1}] \\
&\quad \times \mathcal{L}_1(z) \mathcal{L}_2(z) \mathcal{L}_3(z) \mathcal{L}_4(z) \mathcal{L}_5(z) \mathcal{L}_6(z),
\end{aligned}
\tag{6.18}
$$

where $\mathcal{L}'(z)$ denotes the derivative towards $z$ of $\mathcal{L}(z)$. Calculated for a plane wave it gives

$$\langle\langle T_a^3\rangle\rangle_c = \frac{22}{15g^2}, \tag{6.19}$$

which is indeed proportional to $g^{-2}$, as predicted.

## Contribution of the six point vertex

There is another diagram contributing to the third cumulant which is of the same order of magnitude as the process calculated above; it is depicted in Fig. 6.1(d). The six sided polygon is again the Hikami six point vertex $H_6$, see chapter 3. It can be thought of the following way: The use of Hikami box in the previous section assumes that the outgoing legs scatter at least once before they propagate out or interfere again. This is a reasonable assumption for the outgoing diffusons, but for the internal diffuson $\mathcal{L}_{78}^{\text{int}}$ it also possible that coming from $z$ it directly, i.e. without scattering, interferes again at $z'$. This process is not included in the calculation of the previous section but has to be studied separately. The unscattered intensity decays exponentially over one mean free path, therefore this process is only important if $z$ and $z'$ are within one mean free path. to this diagram. It is not allowed to dress the bare six-point vertex (leftmost r.h.s. diagram in Fig. 3.2) with a scatterer that connects two opposite propagators. This dressing gives also a leading contribution even if the dressing done with an arbitrary number of scatterers, but the resulting diagram is the same as the composed diagram with two four point vertices and should thus not be counted. Yet this observation is useful to check the combinatorial ratio between the six point vertex and the composed diagram: the forbidden dressing can be performed in three ways. As the diagrams can also be complex conjugated, there is also a factor 2 for all diagrams.

In the lowest order of $(q\ell)$ we found for the six point vertex

$$H_6 = \frac{-\ell^7}{96\pi k^4}[\mathbf{q}_1\cdot\mathbf{q}_2 + \mathbf{q}_2\cdot\mathbf{q}_3 + \mathbf{q}_3\cdot\mathbf{q}_4 + \mathbf{q}_4\cdot\mathbf{q}_5 + \mathbf{q}_5\cdot\mathbf{q}_6 + \mathbf{q}_6\cdot\mathbf{q}_1 + \sum_i q_i^2]. \tag{6.20}$$

After a Fourier-transformation in the $z-$direction the six-point vertex yields a contribution to the third cumulant

$$\langle\langle T_a^3\rangle\rangle_d = \langle T_a\rangle^{-3}\frac{\ell^7 A}{48\pi k^4}\int_0^L \mathrm{d}z[\partial_{z_1}\partial_{z_2} + \partial_{z_2}\partial_{z_3} + \partial_{z_3}\partial_{z_4} + \partial_{z_4}\partial_{z_5} + \partial_{z_5}\partial_{z_6} + \partial_{z_6}\partial_{z_1}]$$
$$\times\mathcal{L}_1(z)\mathcal{L}_2(z)\mathcal{L}_3(z)\mathcal{L}_4(z)\mathcal{L}_5(z)\mathcal{L}_6(z). \tag{6.21}$$

Here we used the fact that all outgoing diffusons have zero transversal momentum. Therefore all $Q_iQ_j$-terms are absent. In the limit of an incoming plane wave we find a contribution to the third cumulant

$$\langle\langle T_a^3\rangle\rangle_d = -\frac{2}{5g^2}. \tag{6.22}$$

The contribution from the source term, i.e. Eq.(6.17) of the previous section, exactly cancels the contribution from the six-point vertex. The cancellation seems plausible as one does not expect short distances properties to be important in the total process. Nevertheless, this cancellation depends on the precise form of the Hikami four-point vertex in Eq.(6.15). If we use other equivalent forms of the Hikami-box the contributions of the single and double integral in Eq.(6.18) shift with respect to each other and a full cancellation is not present. Of course, neither the result for Eq.(6.18) nor the final result for $\langle\langle T_a^3 \rangle\rangle$ relies on this choice. The precise mechanism behind this is not clear to us. However, using the cancellation we only need to consider the term in Eq.(6.18), which comes from $H_4(z) \propto \partial_{z_1}\partial_{z_2} + \partial_{z_2}\partial_{z_3}$, $H_4(z') \propto \partial_{z_4}\partial_{z_6}$ and the permutations. We thus obtain for the third cumulant

$$
\begin{aligned}
\langle\langle T_a^3 \rangle\rangle &= \langle\langle T_a^3 \rangle\rangle_c + \langle\langle T_a^3 \rangle\rangle_d \\
&= \langle T_a \rangle^{-3} \frac{\ell^{10}A}{288\pi^2 k^2} \sum_{\text{per}} \int_0^L \mathrm{d}z \; \mathcal{L}_1(z)\mathcal{L}_2'(z)\mathcal{L}_3(z) \times \\
&\qquad \int_0^L \mathrm{d}z' \; \mathcal{L}_4'(z')\mathcal{L}_5(z')\mathcal{L}_6'(z')\partial_z \mathcal{L}^{\text{int}}(z,z').
\end{aligned}
\tag{6.23}
$$

In the next section we calculate this expression for several incoming beam profiles.

## 6.4  Corrections for the experimental situation

### 6.4.1  Influence of incoming beam profile

Now that we know the leading interference processes, inserting the diffusons gives the final value of the third cumulant. We first consider the simple case of incoming plane waves. As there can be no transversal momentum difference in the incoming amplitudes, all $Q_i$ vanish. As a result all diffusons are simple linear functions of $z$. And we find from Eq. (6.23)

$$
\langle\langle T_a^3 \rangle\rangle = \frac{16}{15g^2} = \frac{12}{5}\langle\langle T_a^2 \rangle\rangle^2 \qquad \text{plane wave; } \rho_0 \gg L.
\tag{6.24}
$$

In practice, however, we deal with a Gaussian beam with limited spot size, influencing the cumulants in two ways. First, if the spot size decreases to values comparable to the sample thickness we have to convolute over a range of incoming momenta, just like we did when calculating the second cumulant. Second, the Gaussian profile brings an extra geometrical factor as we will show below.

We need the expression when diffusons with arbitrary momentum are connected to the Hikami boxes. Because of momentum conservation and phase condition on the outgoing diffusons, the transversal momentum $Q_7$ of the diffuson connecting the two four-boxes must equal $Q_5$. The integration over the possible momenta results again in a Gaussian weight function. From the definition (4.11) we derive

$$
\begin{aligned}
&\int \mathrm{d}^2 P_1 \mathrm{d}^2 P_3 \mathrm{d}^2 P_5 \; \phi(P_1)\phi^*(P_1 + Q_1) \; \phi(P_3)\phi^*(P_3 + Q_3) \; \phi(P_5)\phi^*(P_5 + Q_5) \\
&= \exp\left[-\rho_0^2(Q_1^2 + Q_3^2 + Q_5^2)/8\right]
\end{aligned}
\tag{6.25}
$$

We use momentum conservation to eliminate also $Q_5$ and reduce the integration to two transversal momenta. The final result for the third cumulant is obtained by inserting the momentum dependent diffusons into Eq.(6.23). This gives

$$\langle\langle T_a^3\rangle\rangle = \frac{\rho_0^4}{16\pi^2 g^2}\int d^2Q_1 d^2Q_3\, e^{-\rho_0^2[Q_1^2+Q_3^2+(Q_1+Q_3)^2]/8}F_3(|Q_1|L,|Q_3|L,|Q_1+Q_3|L),$$
(6.26)

with

$$
\begin{aligned}
F_3(x_1,x_3,x_5) &= \sum_{\text{per}}\big[\frac{(x_1+x_3)^2x_5\cosh(x_1+x_3)}{(x_1+x_3+x_5)^2(x_1+x_3-x_5)^2} - \frac{(x_1-x_3)^2x_5\cosh(x_1-x_3)}{(x_1-x_3+x_5)^2(x_1-x_3-x_5)^2} \\
&\quad - \frac{(x_1+x_3)x_5\sinh(x_1+x_3)}{(x_1+x_3+x_5)(x_1+x_3-x_5)} + \frac{(x_1-x_3)x_5\sinh(x_1-x_3)}{(x_1-x_3+x_5)(x_1-x_3-x_5)} \\
&\quad + \frac{(x_1+x_3)\cosh(x_1+x_3+2x_5)}{4(x_1+x_3+x_5)^2} - \frac{(x_1+x_3)\cosh(x_1+x_3-2x_5)}{4(x_1+x_3-x_5)^2} \\
&\quad - \frac{(x_1-x_3)\cosh(x_1-x_3+2x_5)}{4(x_1-x_3+x_5)^2} + \frac{(x_1-x_3)\cosh(x_1-x_3-2x_5)}{4(x_1-x_3-x_5)^2}\big]\times \\
&\quad \big[x_5\sinh(x_1)\sinh(x_3)\sinh^2(x_5)\big]^{-1}.
\end{aligned}
$$
(6.27)

These two equations are the main result in this chapter. We study again the behavior if the beam diameters are wide. In the limit of large beam diameter ($\rho_0 \gg L$) one finds $F_3(0,0,0) = \frac{16}{15}$, this means for the third cumulant $\langle\langle T_a^3\rangle\rangle = 4F_3(0,0,0)/3g^2$, or

$$\langle\langle T_a^3\rangle\rangle = \frac{16}{5}\,\langle\langle T_a^2\rangle\rangle^2, \qquad \text{Gaussian profile; } \rho_0 \gg L,$$
(6.28)

which differs by a factor $\frac{4}{3}$ from the plane wave limit Eq.(6.24). This is purely a geometrical effect, depending on the profile of the incoming beam. In a real space picture one understands this effect best. The correlation depends on the distance: it is strongest if the incoming intensities are close. Therefore it is not surprising to see the influence of the overlap. In next chapter we calculate this geometrical factor for higher orders also (the area of a Gaussian beam is defined differently there). For the experimental relevant case that the beam diameter is roughly equal to the thickness, we calculated Eqs.(6.26) and (6.27) numerically. It then turns out that the behavior of Eq.(6.28) is found for a large range of beam diameters. The increase of the correlation for smaller beams, turns out to be roughly the same for both the third cumulant and the second cumulant squared. All corrections to (6.28) turn out to be relatively small, as we will discuss below. Apart from this advantage, errors in the sample thickness and the mean free path cancel by presenting the results as the ratio between the second cumulant squared and the third cumulant.

## 6.4.2 Influence of internal reflection

In this section we calculate the influence of internal reflection on our results. We saw in chapter 4 that surface reflection decreases the $C_2$ correlation. In Eq.(6.28) corrections from boundary reflection cancel partly. We did not calculate the influence of internal reflections for the general case, but only for the case of very broad beams (i.e. only for $Q$ independent diffusons). One expects that this behavior may be extrapolated to the $Q$−dependent case. At least, for the second cumulant this is a good approximation [68]. The $Q$− independent diffusons in the presence of internal reflections are given in chapter 2, Eqs.(2.60,2.75). If internal reflections are absent $z_0$ equals $0.71\ell$ and the corrections, which are of order $z_0/L$, are often negligible. With internal reflection present $z_0$ increases and should be taken into account. In first order of $z_0/L$ the second and third cumulant behave as

$$\langle\langle T_a^2\rangle\rangle = \frac{2}{3g}\left(1 - 3\frac{z_0}{L}\right), \qquad \langle\langle T_a^3\rangle\rangle = \frac{16}{15g^2}\left(1 - \frac{15}{2}\frac{z_0}{L}\right), \tag{6.29}$$

Therefore, the central relation (6.28) has a correction

$$\frac{\langle\langle T_a^3\rangle\rangle}{\langle\langle T_a^2\rangle\rangle^2} = \frac{16}{5}\left(1 - \frac{3}{2}\frac{z_0}{L}\right). \tag{6.30}$$

The experimental determination of the index of refraction of the sample, which gives $z_0$, is difficult [64]. Fortunately, the correction is rather small for the experimental situation considered.

### 6.4.3  Contributions from disconnected diagrams

So far we calculated the leading contributions to the second and third cumulant. They are the connected diagrams in Fig. 4.6(b) and Figs. 6.1(c+d), respectively. Yet there are also contributions to the second and third cumulant from disconnected diagrams. As we saw in chapter 4, the diagram in Fig. 4.6(a) gives an additional contribution to the second cumulant, and likewise, the diagrams of Figs. 6.1(a) and (b) give a contribution to the third cumulant. These disconnected diagrams correspond to cumulant contributions that are not (fully) due to interference. They describe effects that have little to do with the interference effects we are after. Here we calculate their contribution and show that they are small.[1]

The contribution of the disconnected diagram Fig. 4.6(b) to the second cumulant was calculated in chapter 4

$$\langle\langle T_a^2\rangle\rangle_{\text{dis}} = \frac{\pi k^2 \int_{-\infty}^{\infty} \mathrm{d}^2 Q\, I^2(Q)}{\pi^2 k^4 I^2(0)} \equiv \frac{1}{N'} \tag{6.31}$$

In analogy to a waveguide, this result describes irreducible contributions from disconnected diagrams. It can be interpreted as the inverse of the number of independent speckle spots in transmission at the exit interface [106, 105].

**Extra contributions to the third cumulant**

We apply the same method from chapter 4 for contributions to the third cumulant. Following the waveguide argument as above, we find the contributions to the third cumulant of the diagrams of Figs. 6.1(c+d), 6.1(b), and 6.1(a), respectively,

$$\langle\langle T_a^3\rangle\rangle = \frac{3}{5}\frac{L^2}{N^2\ell^2} + \frac{6}{2}\frac{L}{N^2\ell} + \frac{2}{N^2}. \tag{6.32}$$

---

[1] This calculation was initiated by Drs. Johannes de Boer for our joint articles Refs. [105, 73]

The first r.h.s. term is the connected diagram. Clearly the second term, the diagram Fig. 6.1(b), gives a much larger contribution to the third cumulant than the Fig. 6.1(a), as the diagram Fig. 6.1(b) is already enhanced by some interference. In the following we only consider this diagram.

As one sees from Fig. 6.1(b) there are three possibilities to combine the three diffusons into two connected diffusons. and a single diffuson. Attaching outgoing directions to the amplitudes at the exit interface gives (see Fig. (6.2),

$$\Psi(R_1)e^{iP_1R_1}\Psi^*(R_1)e^{-iP_2R_1}\Psi(R_2)e^{iP_3R_2}\Psi^*(R_2)e^{-iP_4R_2}\Psi(R_3)e^{iP_5R_3}\Psi^*(R_3)e^{-iP_6R_3} \quad (6.33)$$

Obviously there are six possibilities to pair the outgoing directions into intensities. Integrating over the transversal coordinates $R_1$, $R_2$ and $R_3$ gives the following six contributions,

$$I_{\text{con}}(0)I_{\text{con}}(0)I(0)$$
$$I_{\text{con}}(P_1 - P_3)I_{\text{con}}(P_3 - P_1)I(0)$$
$$I_{\text{con}}(0)I_{\text{con}}(P_1 - P_5)I(P_5 - P_1)$$
$$I_{\text{con}}(0)I_{\text{con}}(P_3 - P_5)I(P_5 - P_3)$$
$$I_{\text{con}}(P_1 - P_5)I_{\text{con}}(P_3 - P_1)I(P_5 - P_3)$$
$$I_{\text{con}}(P_1 - P_3)I_{\text{con}}(P_3 - P_5)I(P_5 - P_1), \quad (6.34)$$

where $I_{\text{con}}$ denotes the transmitted intensity coming from the connected part of diagram Fig. 6.1(b). The third cumulant is in the discrete mode model given by

$$\langle\langle T_a^3\rangle\rangle = \frac{1}{\langle T_a\rangle^3}\left[\sum_{b,b',b''}\langle T_{ab}T_{ab'}T_{ab''}\rangle - 3\sum_b\langle T_{ab}\rangle\sum_{b,b'}\langle T_{ab}T_{ab'}\rangle + 2\sum_b\langle T_{ab}\rangle^3\right] \quad (6.35)$$

Similarly to the second cumulant we insert all pairings of Eq. (6.34) and find for the contribution from disconnected diagrams

$$\langle\langle T_a^3\rangle\rangle_{\text{dis}} = \frac{6}{\langle T_a\rangle^3}\int d^2Q_1 d^2Q_2 d^2Q_3\left[I_{\text{con}}(0)I_{\text{con}}(Q_1)I(-Q_1) + I_{\text{con}}(Q_1)I_{\text{con}}(Q_2)I(Q_3)\right]. \quad (6.36)$$

The integrand is dominated by its first term, which is depicted in Fig. 6.2. We explicitly evaluated this term. The calculation follows completely the line of the second cumulant calculation, yet it is slightly more complicated as it has a $Q-$dependent outgoing diffuson.

The outgoing diffusons have, from top to bottom in Fig. 6.2, momenta 0, $Q_1$, and $-Q_1$, respectively. The incoming diffusons therefor have momenta $Q$, $Q_1 - Q$, and $-Q_1$, respectively, where $Q$ is an extra free momentum. The weight factor for the incoming diffusons is

$$\exp[-\frac{\rho_0^2}{8}(Q_1^2 + Q^2 + (Q_1 - Q)^2)] \quad (6.37)$$

For fixed momenta the connected part of the diagram is

$$\frac{1}{\langle T_a\rangle^2}\int_0^L dz \mathcal{L}_1(z;Q)\mathcal{L}_3(z;Q_1 - Q)H_4(z)\mathcal{L}_2(z;0)\mathcal{L}_4(z;Q_1)$$

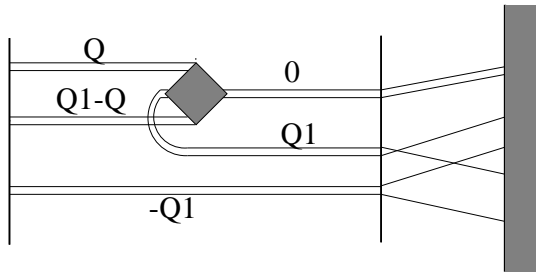

Figure 6.2: The leading disconnected contribution to the third cumulant, the box symbolizes again the Hikami vertex.

$$
\begin{aligned}
= \quad & \frac{1}{g} \big[ 4x_1^2 xy \sinh x_1 + x_1[x^3 + y^3 - x_1^2(x + y) - xx_1(x + y)] \sinh(x + y) \\
& + x_1[-x^3 + y^3 - x_1^2(-x + y) - xx_1(-x + y)] \sinh(x - y)] \times \\
& \big[ (x_1^4 - 2x_1^2(x^2 + y^2) + x^4 + x_2^4 + 2x^2 y^2) \sinh x \sinh x_1 \sinh y \big]^{-1}, \quad (6.38)
\end{aligned}
$$

where we have defined $x = |Q|L$, $x_1 = |Q_1|L$, and $y = |Q_1 - Q|L$. As for the connected diagram there is again a weight factor from the Gaussian incoming beam profile. The integration over the free momenta $Q_2$ and $Q_3$ in Eq. (6.36) gives a factor $(\pi k^2)^2$. Comparing the result thus found for Eq. (6.36) with the last two terms in third cumulant in the waveguide model, Eq. (6.32), we define $N^*$ as

$$
\langle\langle T_a^3 \rangle\rangle_{\text{dis}} = 6\langle\langle T_a^2 \rangle\rangle_{\text{con}} / N^*. \quad (6.39)
$$

The number $N^*$ is inversely proportional to the contribution of disconnected diagrams to the third cumulant for a Gaussian profile.

**Polarization effects**

We did not take the vector character of the light into account yet. The two independent polarizations of each outgoing direction effectively double the number of independent speckle spots $N'$. Thus for an incoming plane wave of unit intensity with fixed polarization the total transmission and the conductance $g$ are twice as large as they would be in the scalar case. As we work with normalized cumulants, this effect reduces the value of the second cumulant ($\propto 1/g$) and the value of the third cumulant ($\propto 1/g^2$). Therefore, one immediately sees that the relation $\langle\langle T_a^3 \rangle\rangle \propto \langle\langle T_a^2 \rangle\rangle^2$ (Eq. 6.28) is not affected. The vector character does reduce the correction Eq. (4.22) by a factor $2$. For the experimental data of Ref. [105] we list the number of modes $N'$ as well as $N^*$ in Table 6.1 (including the doubling). The difference between $N'$ and $N^*$ is small. Yet due to the diffuse broadening of the beam, $N'$ is much larger than the number of modes in the waveguide $N = k^2 A/4\pi$ (typically some 30 times in the experiment).

Summarizing the previous sections we have included three corrections. We first obtained the result for very broad Gaussian beams, in the large $L/\ell$ limit, Eq. (6.28). The first correction was the influence of a finite beam diameter, it changes the diffuse intensity from linear into an exponential decaying, see Eq. (2.42). This correction is contained in Eq. (6.27) and Eq. (6.26). The presence of internal reflections also changes the spatial dependence of the diffuse intensity resulting in a correction Eq. (6.30). The third correction is of another nature, it is the only process that does not come from interference, but from disconnected diagrams. Only this term depends the number of modes, which in the vector case is twice as large as in the scalar case.

## 6.5 Comparison with experiments

We reproduce the data set found experimentally in Ref. [105] in Table 6.1 and Fig. 6.3. The experiments reported there were performed with seven different samples. The experimental setup and measurement technique used is extensively described in Ref. [72]. Samples consisted of 36 vol.% rutile $TiO_2$ pigment on a transparent substrate. The extrapolation length was estimated from the effective index of refraction to be $z_0 \approx 1.1 \mu m$. The absorption length $\ell_a$ was determined to be $\simeq 70 \mu m$. Different values of the conductance $g$ were probed by taking several sample thicknesses and by varying the beam diameter. The fluctuations in the total transmission were measured by varying the wavelength of the light.

For a very broad beam we found the simple relation (6.28) between second and third cumulant. A weighted least square fit to

$$\langle\langle T_a^3 \rangle\rangle = const.\langle\langle T_a^2 \rangle\rangle)^2 \tag{6.40}$$

of the raw experimental data yields a prefactor $2.9 \pm 0.6$. However, as discussed above, there are three corrections to be made. First, if the beam width becomes comparable to sample thickness we have to perform the integrals (6.13) and (6.26): If the beam width reduces, $g$ decreases accordingly and both cumulants increase in absolute size. Yet the precise increase is somewhat different resulting in a somewhat smaller prefactor in Eq. (6.28). We corrected each data point individually for its finite focus, mapping it to the infinite focus case. The third cumulant was multiplied by a factor that ranged from 1.03 to 1.13, as $L/\rho_0$ ranged from 0.41 to 6.3, see Table 6.1. This is the largest correction; it changes the prefactor some 10%. Secondly, we corrected for internal reflections according to Eq.(6.30). The third correction comes from the disconnect diagrams. After all these corrections the data should again obey the law: $\langle\langle T_a^3 \rangle\rangle = 3.2 \langle\langle T_a^2 \rangle\rangle^2$. We have plotted the results in Fig. 6.3, where the points are the corrected data points and the line is the theoretical prediction. A least square fit gives

$$\langle\langle T_a^3 \rangle\rangle = (3.3 \pm 0.6)\langle\langle T_a^2 \rangle\rangle^2. \tag{6.41}$$

Note that there is no adjustable parameter. We find that there is good agreement between experiment and theory. All corrections are minor as compared to the error in the data and the data can even be described rather well disregarding the corrections. Inspecting the figure one might be tempted to use a linear fit, but in Ref. [105] it was shown that this fit is statistically very improbable.

## 6.6 Discussion

We have calculated the third cumulant of the distribution of the total transmission and compared it to the data of Ref. [105]. The second and third cumulant are a conse-quence of interference between diffuse channels. The third cumulant is proportional to the second cumulant squared. We also found a non-trivial dependence on the profile of the incoming beam used. We calculated the cumulants for arbitrary beam diameter, but the influence of a finite focus on the ratio is rather weak. Also boundary reflections were included. Our calculations confirm that the main contributions come from diagrams with interference processes, i.e. connected diagrams, as we have shown that the contributions from disconnected diagrams is small. The experimentally found ratio of the third cumulant versus the second cumulant squared is well described by our theory.

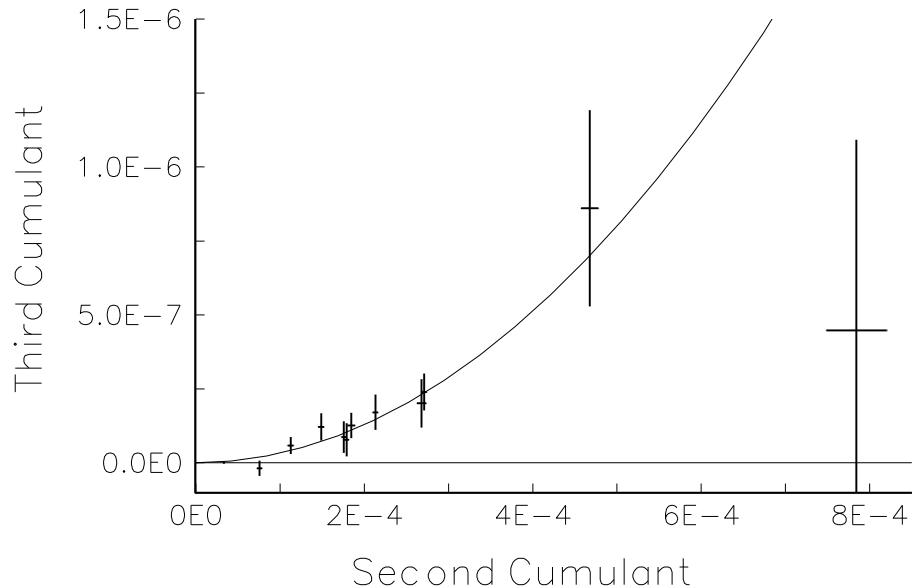

Figure 6.3: The third cumulant plotted against the second cumulant. The points are the corrected experimental data (see text). The line is the theoretical prediction, $\langle\langle T_a^3 \rangle\rangle = 3.2 \langle\langle T_a^2 \rangle\rangle^2$; we introduced no free parameters.

The extension of the calculations to higher cumulants is straightforward. The $n$-th cumulant will contain $(n-1)$ Hikami four-point vertices. So the contribution is $\langle\langle T_a^n \rangle\rangle \propto g^{1-n}$. Also corrections and cancellations from higher order vertices are present, but it is clear that the calculation becomes very laborious at large $n$. We follow another approach to this problem in the next chapter.

Table 6.1: Sample thickness (in microns), beam width (in microns), second cumulant, third cumulant for the different samples as taken from Ref.[105]. Next are the beam diameter correction factor on the third cumulant (see section 6.5), and the number of modes $N'$ and $N^*$. Together they yield the corrected cumulants (plotted, not listed).

| Sample thickn. L | Beam diam. $\rho_0$ | Second cumulant ($\times 10^{-4}$) | Third cumulant ($\times 10^{-7}$) | Beam corr. factor | Number of modes $N'$ | Number of modes $N^*$ |
|---|---|---|---|---|---|---|
| 30 | 77 | 0.36±0.01 | 0.014 ± 0.035 | 1.03 | 388000 | 300000 |
| 12 | 26 | 0.97 ±0.03 | −0.03 ±0.25 | 1.04 | 46800 | 36800 |
| 22 | 32 | 1.24 ±0.04 | 0.68 ±0.28 | 1.06 | 88600 | 72100 |
| 30 | 33 | 1.57 ±0.04 | 1.30 ±0.46 | 1.07 | 119000 | 99300 |
| 53 | 35 | 1.80 ±0.03 | 0.91 ±0.53 | 1.10 | 241000 | 212000 |
| 30 | 26 | 1.90 ±0.03 | 0.92 ±0.56 | 1.09 | 94900 | 81300 |
| 45 | 33 | 1.90 ±0.05 | 1.33 ±0.43 | 1.10 | 187000 | 163000 |
| 53 | 26 | 2.18 ±0.03 | 1.77 ±0.59 | 1.10 | 208000 | 189000 |
| 170 | 27 | 2.69 ±0.06 | 2.02 ±0.82 | 1.13 | 1420000 | 1430000 |
| 78 | 28 | 2.74 ±0.03 | 2.43 ±0.62 | 1.11 | 396000 | 372000 |
| 30 | 17 | 4.82 ±0.10 | 9.1 ±3.3 | 1.11 | 71900 | 64400 |
| 30 | 10 | 8.01 ±0.36 | 5.3 ±6.4 | 1.11 | 60200 | 56400 |

# 7

# Full distribution functions

In this chapter we calculate the full distribution of the total transmission and the angular resolved transmission. Both can be mapped on the eigenvalue distribution of transmission matrix. This distribution is known for moderate values of $g$.

## 7.1 Classics

Already from the autocorrelation functions we know that interference is important at moderate values of the dimensionless conductance $g$. For instance the $C_2$ correlation is proportional to $1/g$ and was clearly seen in experiments. But let us first discuss the 'classical' situation, which is the limit of large $g$. Then correlations between intensities are absent, and we recover well known distribution functions. For the angular resolved transmission, or speckle intensity, there is the Raleigh law [106]

$$P(T_{ab}) = \frac{1}{\langle T_{ab} \rangle} e^{-T_{ab}/\langle T_{ab} \rangle} \tag{7.1}$$

Anyone who ever saw a laser speckle pattern on the wall, will remember the wild pattern with both very bright but also dark spots. This is exactly due to the Rayleigh law.

The Rayleigh law can also be derived in the context of the ladder diagrams [92]. The $n-$th moment of the speckle intensities is made up of $n$ amplitudes and $n$ complex conjugated amplitudes. As stated before, there is no constriction on the pairing of two amplitudes into a diffuson for this type of measurement. Therefore the amplitudes $\{a_1^*, a_2^*, a_3^*, \ldots, a_n^*\}$ have $n!$ possibilities to pair with the amplitudes $\{a_1, a_2, a_3, \ldots, a_n\}$. Thus

$$\langle T_{ab}^n \rangle = n! \langle T_{ab} \rangle^n \tag{7.2}$$

which is easily checked to correspond with the Rayleigh law. Yet, as we know from the example of a laser beam reflecting on the wall, the Rayleigh is more general. We need no *multiple* scattering at all. Indeed, Rayleigh's original derivation goes as follows [106]: The field $\psi$ on a given position on the outgoing side is the sum of many fields, and therefore real and imaginary part have each a independent Gaussian distribution. The intensity $I$ being the amplitude squared has thus an exponential distribution

$$P \sim e^{-[(\mathrm{Re}\psi)^2 + (\mathrm{Im}\psi)^2]/2\sigma^2} \sim e^{-I/\langle I \rangle}. \tag{7.3}$$

Recently Kogan and Kaveh addressed the question of the speckle statistics of (optically) very thin samples [121]. In very thin samples a considerable part of the light does not scatter at all, but is coherent. In the limit of zero thickness, only coherent light, the distribution becomes a delta-function. In the intermediate regime a simple counting argument yields the distribution, which intra-polates between the delta distribution and the Rayleigh law.

In the mesoscopic regime interference modifies the speckle distribution. The leading correction was derived by Shnerb and Kaveh [122]. Genack and Garcia observed a deviation from the Rayleigh law at large intensities [95, 123]. A crossover to stretched exponential behavior was derived by Kogan *et al.* [118].

For the total transmission and the conductance the distributions are similar in the large $g$ limit. For both quantities the outgoing diffusons should be frequency and momentum independent as we explained in chapter 4. As there is no interaction between the diffusons in this limit, the same applies for the incoming diffusons. Thus for both quantities only one type of pairing is possible. This yields in principle a $\delta$-distribution if one probes an infinite number of channels. In practice of course only limited number of channels is probed, and the law of large numbers predicts a narrow Gaussian distribution.

## 7.2  Eigenvalues of the transmission matrix

In mesoscopic systems the observables are random quantities and are therefore not always characterized by the mean values, but their entire distribution functions are of interest (see [124] for a toy model). This is particularly prominent in the distribution of eigenvalues of the transmission matrix. Assuming that all eigenvalues contribute equally to the conductance, one expects a Gaussian distribution. As there are $N$ eigenmodes, the eigenvalue distribution should be peaked $g/N = \frac{\ell}{L}$. But this picture proves wrong. The eigenvalues have a "bimodal" distribution peaked around 0 and 1. This was first put forward by Dorokhov [125] and later also by Imry [126]. The eigenvalues $T_n$ of the transmission matrix $t^\dagger t$ can be expressed as $T_n = 1/\cosh^2(L/\xi_n)$. Not the eigenvalues but the inverse localization lengths $1/\xi_n$ are uniform distributed, see Pendry *et al.* [127] and Stone *et al.* [128]. In all derivations quasi-1D was assumed. Recently, however Nazarov [129] showed that it is not only true in quasi-1D, but under very general conditions. It implies that:

$$\langle T_n^j \rangle = \frac{\ell}{L} \int_0^1 \frac{\mathrm{d}T}{2T\sqrt{1-T}} T^j. \tag{7.4}$$

Or

$$\langle \mathrm{Tr}(tt^\dagger)^j \rangle = \langle \sum_{n=1}^N T_n^j \rangle = g \int_0^1 \frac{\mathrm{d}T}{2T\sqrt{1-T}} T^j. \tag{7.5}$$

This result was first derived using random matrix techniques by Mello *et al.* [130]. We plotted the distribution in Fig.7.1. Note that few eigenvalues have the mean value $\ell/L \ll 1$. Note that, the normalization of the distribution is ill-defined, the distribution should be understood in the sense that all its moments are well defined. This problem can be avoided by taking the proper lower boundary of the integral instead of 0. The eigenvalues are expressed as

$$T_n = \cosh^{-2}(L/\xi_n) \tag{7.6}$$

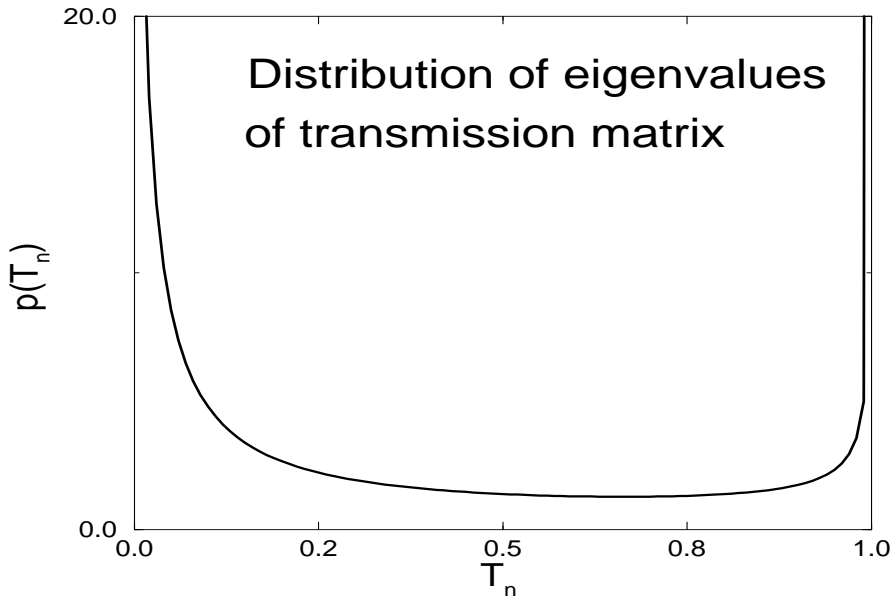

Figure 7.1: The bimodal eigenvalue distribution. The average value has only a small probability, instead the eigenvalues are almost all either 0 or 1.

The channel localization length $\xi_n$ ranges (roughly) from $\ell$ to infinity. The minimal value corresponds to the decay of unscattered intensity (whereas the upper limit corresponds to the case where there is no scattering). Therefore the minimal value for $T_n$ is $T_{\min} = \cosh^{-2}(L/\ell) \approx \exp(-2L/\ell)$ and the distribution Eq. (7.4) can be normalized. Although this cut-off is important for the normalization of the distribution, its influence on the momenta is very small and we refrain from the correction. The moments of the distribution are

$$\langle T_n^k \rangle = \frac{2^{n-1}(n-1)!}{1.3.\ldots.(2n-1)} \tag{7.7}$$

Such that $\langle T_n \rangle = \frac{\ell}{L}$, $\langle T_n^2 \rangle = 2/3\frac{\ell}{L}$, and $\langle T_n^3 \rangle = 8/15\frac{\ell}{L}$. The distribution Eq.(7.5) is valid only in the regime where $g$ is not too small. Loop effects near the Anderson transition, where $g \sim 1$, will change it, as we will discuss below.

The physical meaning of this curious distribution function, is the concept of open and closed channels. An eigenmode of the transmission matrix is, according to this distribution, either essentially blocked or, with a smaller probability, essentially conducting. This picture was confirmed in various computer simulations [127, 131]. Instructive is the simulation of Oakeshott and McKinnon [132], where the conducting channels are made visible in real-space; it can then be seen that the current is carried by only a few channels. Even in the conducting regime most channels are localized. When approaching localization the localization length of more and more channels becomes smaller than the sample size, untill finally the last channel is localized, and the full sample becomes insulating. It also provides us with a very nice picture explaining why the correlations in mesoscopic systems are so large. If all the channels are conducting equally, fluctuations of the channels average out by the law of large numbers. Yet if only a few channels are conducting, fluctuations in one channel will be clearly seen. Yet this reasoning is not always right, consider for example the case of the UCF. In the old picture that all channels contribute equally to the conductance, the variance in the conductance $N\ell/L$ is caused by the finiteness of the number of channels $N$. In the new picture of either closed or open channels the number of open channels $N_{\text{eff}}$ is much smaller

$$N_{\text{eff}} \times 1 = \frac{N\ell}{L} \ll N \tag{7.8}$$

The conductance now fluctuates according the law of large numbers with a variance: $\langle \sum_n T_n^2 \rangle = \frac{2}{3}g$, which is incorrect as we know from chapter 5. Below, we will see why the argument went wrong.

In practice it is very difficult, if not impossible, to measure the eigenmodes and eigenvalues directly. As they are the eigenmodes of the very large random transmission matrix, they have a very complex structure. Nevertheless it has some important observable consequences. First it was shown by Beenakker and Büttiker [133] that the shot-noise of electronic conductors universally reduces by a factor 3 because of this distribution. Shot-noise only occurs in electronic systems and is thus not of much interest for us. Secondly, we will show the total transmission and speckle intensity distribution function are fully determined by the eigenvalue distribution function.

## 7.3   Distribution of total transmission

We first calculate the probability distribution of the total transmission. Already from the distributions without interference effects, see section 7.1, we saw that the total transmission has a simpler distribution (in the sense of cumulants) than the speckle.

We neglect all corrections for absorption, skin layers, and disconnected diagrams. Consider again an incoming plane wave in direction $\mu_a$ ($\mu$ denotes again the cosine with respect to the $z-$axis). The wave is transmitted into outgoing channel $b$ with transmission amplitude $t_{ab} = 2k\sqrt{\mu_a\mu_b}G_{ab}$ and transmission probability $T_{ab} \equiv |t_{ab}|^2$. The average total transmission is obtained by summing all outgoing channels, see chapter 2

$$\langle T_a \rangle = \langle \sum_b T_{ab} \rangle = \frac{\tau_1(\mu_a)\ell}{3L\mu_a}. \tag{7.9}$$

We find the average conductance by

$$g \equiv \sum_a \langle T_a \rangle = \frac{k^2 A\ell}{3\pi L}, \tag{7.10}$$

thus $\langle T_a \rangle = \epsilon_a g$, while one also has $\langle T_{ab} \rangle = \epsilon_a \epsilon_b g$.

We consider the $j^{\text{th}}$ cumulant of $T_a$. In a diagrammatic approach this object has $j$ transmission amplitudes $t_{ab}$ and an equal number of Hermitian conjugates $t_{ba}^\dagger = t_{ab}^*$. The explicit calculation for the second cumulant $C_2$ (chapter 4), and third cumulant (chapter 6) showed that the leading diagrams are connected, yet have no loops.

Let us fix the external diffusons in the term $t_{ab_1}t_{b_1a}^\dagger t_{ab_2}\cdots t_{ab_j}t_{b_ja}^\dagger$. Contributions to the sum over $b_i$ only come from diagrams with outgoing diffusons that have no transversal momentum. These are the diagrams where the lines with equal $b_i$ pair into diffusons. We indicate this pairing of the outgoing diffusons with a underlining bracket

$$\underbrace{t_{ab_1} t_{b_1 a}^\dagger} \underbrace{t_{ab_2} t_{b_2 a}^\dagger} \cdots \underbrace{t_{ab_j} t_{b_j a}^\dagger} \tag{7.11}$$

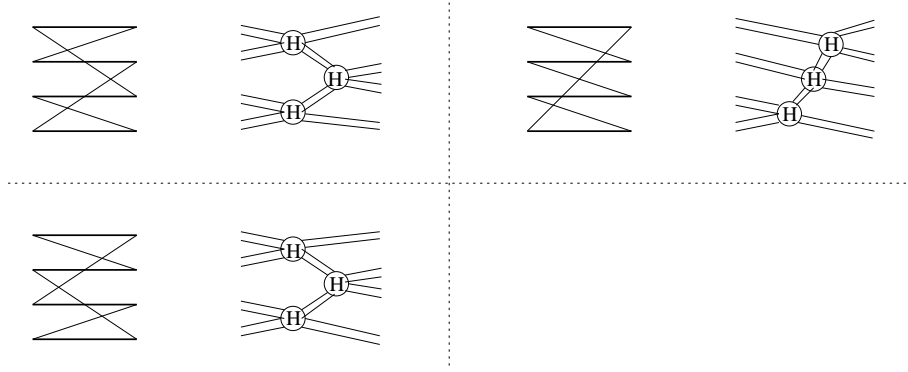

Figure 7.2: Schematic picture of the leading diagrams for the fourth cumulant total transmission. Although there are 6 diagrams, there are only three topologies on intensity level. On the left the thick lines represent $t_{ab}$, the thin lines $t_{ab}^\dagger$; this type of diagram simplifies the counting. On the right the more conventional picture using Hikami boxes.

The outgoing diffusons are fixed now. For the incoming side it is convenient to keep for instance the $t_{ab}$'s fixed, while the $t_{ab}^\dagger$'s are permuted among them, see Fig. 7.2. This leaves us $j!$ possibilities for the incoming side. The number of Hikami boxes in this process "misfits". The number of misfits for diagrams that are connected and loop-less, is $j-1$, giving $(j-1)!$ possibilities. Diagrams with less misfits are disconnected, diagrams with more contain more vertices and are thus sub-leading. (Of course just as happens for the third cumulant there will be local corrections from, for instance the six point $H_6$ vertex, we do not consider them explicitly here). Examples of allowed diagrams are cyclic permutations of $t_{ab}^\dagger$ with respect to the outgoing side.

Next we factor out the incoming and outgoing diffusons and group the remainder of the diagrams into a skeleton $K$. Making use of Eq. (2.68) we obtain:

$$\langle T_a^j \rangle_{\text{con}} = \epsilon_a^j (j-1)! \int d\mathbf{r}_1 d\mathbf{r}_1' \cdots d\mathbf{r}_j d\mathbf{r}_j' \mathcal{L}_{\text{in}}(\mathbf{r}_1) \mathcal{L}_{\text{out}}(\mathbf{r}_1')$$
$$\cdots \mathcal{L}_{\text{in}}(\mathbf{r}_j) \mathcal{L}_{\text{out}}(\mathbf{r}_j') K(\mathbf{r}_1, \mathbf{r}_1', \cdots, \mathbf{r}_j, \mathbf{r}_j'). \tag{7.12}$$

The integral just describes

$$\langle \text{Tr}(tt^\dagger)^j \rangle \equiv \text{Tr} \sum_{b_1, a_2, b_2, \cdots, a_j, b_j} \langle t_{a_1 b_1} t_{b_1 a_2}^\dagger t_{a_2 b_2} \cdots t_{a_j b_j} t_{b_j a_1}^\dagger \rangle$$
$$= \sum_{a_1, b_1, \cdots, a_j, b_j} \langle t_{a_1 b_1} t_{b_1 a_2}^\dagger t_{a_2 b_2} \cdots t_{a_j b_j} t_{b_j a_1}^\dagger \rangle. \tag{7.13}$$

There is only one way to attach incoming and outgoing diffusons to $K$. The sums over the indices lead exactly to the total-flux diffusons in Eq. (7.12). We thus find

$$\langle T_a^j \rangle_{\text{con}} = (j-1)! \epsilon_a^j \langle \text{Tr}(tt^\dagger)^j \rangle, \tag{7.14}$$

which is the crucial step in the derivation. Normalizing with respect to the average, we introduce $s_a = T_a / \langle T_a \rangle$. The generating function of the connected diagrams is easily calculated

$$
\begin{aligned}
\Phi_{\text{con}}(x) & \equiv \sum_{j=1}^{\infty} \frac{(-1)^{j+1} x^j}{j!} \langle s_a^j \rangle_{\text{con}} \\
& = g \log^2 \left( \sqrt{1 + x/g} + \sqrt{x/g} \right).
\end{aligned}
\tag{7.15}
$$

Since the cumulants are solely given by connected diagrams, the distribution of $s_a$ follows as

$$
P(s_a) = \int_{-i\infty}^{i\infty} \frac{\mathrm{d}x}{2\pi i} \exp\left[ x s_a - \Phi_{\text{con}}(x) \right].
\tag{7.16}
$$

In Fig. 7.3 we present the distribution (7.16) for some values of $g$. At moderate $g$ we clearly see the deviation from a Gaussian. Let us examine some properties of the distribution. For $s_a$ near unity and large $g$ we can expand $\Phi$ up to order $x^2$, recovering the Gaussian behavior found by Kogan *et al.* [118]:

$$
P(s_a) \approx \sqrt{\frac{3g}{4\pi}} \exp[-\frac{3g}{4}(s_a - 1)^2]
\tag{7.17}
$$

This corresponds to including only the $C_2$ contribution. The integrand in Eq. (7.16) has a branch cut from $x = -g$ to $x = -\infty$. For $s_a \leq 0$ the contour can be closed to the right half-plane and $P(s_a)$ vanishes. For limiting values of $s_a$ we use a saddle point analysis. The saddle point is found by the condition $\frac{d}{dx}[x s_a - \Phi_{\text{con}}(x)] = 0$. Thus we find

$$
s_a = \frac{\log \left( \sqrt{1 + x/g} + \sqrt{x/g} \right)}{\sqrt{x/g(1 + x/g)}}
\tag{7.18}
$$

The r.h.s. of Eq.(7.18) diverges if $x$ approaches $-g$ (from above) and decreases monotonically for larger $x$. Thus for large $s_a \gg 1$, we find the saddle point near $x = -g$. By inserting the saddle point one finds a simple exponential decay

$$
P(s_a) \approx \exp(-g s_a + g \frac{\pi^2}{4}), \qquad s_a \gg 1.
\tag{7.19}
$$

The saddle point also dominates the shape for small $s_a \ll 1$ (and large $g$). This time $x/g$ has to be large to solve Eq.(7.18), i.e. $s_a \approx 9(\log 4x/g)/2x$. One finds essentially a log-normal growth:

$$
P(s_a) \sim \exp\left[ \frac{g}{4} - \frac{g}{4} \left( \log \frac{2}{s_a} + \log \log \frac{2}{s_a} - 1 \right)^2 \right].
\tag{7.20}
$$

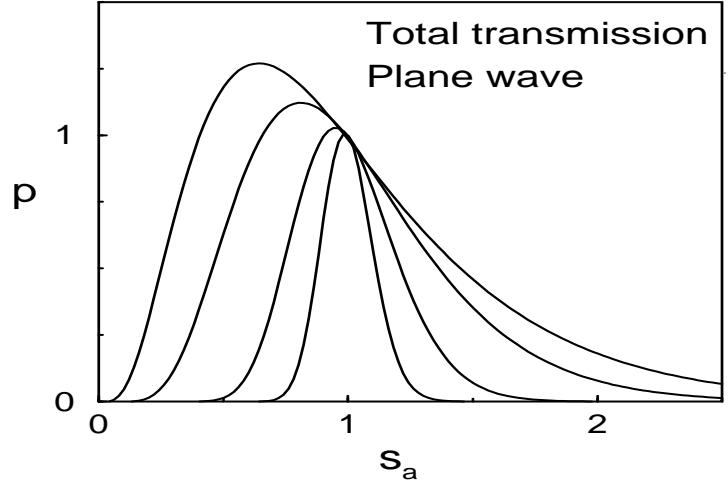

Figure 7.3: Intensity distribution of the total transmission, in units of $\sqrt{3g/4\pi}$, versus the normalized intensity $s_a$ for an incoming plane wave. With $g = 2, 4, 16,$ and $64$ (upper to lower curve).

## 7.3.1 Influence of beam profile

So far we have considered the case of an incoming plane wave. We already mentioned that in optical systems a Gaussian intensity profile is more realistic. For the third cumulant we saw a non-trivial dependence on the incoming beam profile, suggesting that also higher cumulants are sensitive. For perpendicular incidence the incoming amplitude is $\psi_{\text{in}}(\mathbf{r}) = W^{-1} \sum_a \phi(q_a) \psi_{\text{in}}^a(\mathbf{r})$, where $\psi_{\text{in}}^a(\mathbf{r})$ is the plane wave of Eq. (2.48), and where

$$\phi(q_a) = \sqrt{2\pi}\,\rho_0 \exp(-\frac{1}{4}\rho_0^2 q_a^2). \tag{7.21}$$

We consider the limit where the beam is much broader than the sample thickness ($\rho_0 \gg L$) but still much smaller than the transversal size of the slab ($\rho_0 \ll W$). If we would consider a smaller beam diameter, the incoming transverse momenta, which are of order $1/\rho_0$, become of the order of $1/L$. The diffusons will then become the well known cosh-functions, chapter 2. Here the momentum dependence of the diffusons can be neglected; apart from a geometrical factor, the situation is identical with the plane wave case. Due to integration over the center of gravity, each diagram involves a factor $A\delta_{\Sigma q, \Sigma q'}$. In the $j^{\text{th}}$ order term there occurs a factor

$$\begin{aligned}
F_j &= \frac{A}{A^{2j}} \sum_{q_1 q_1' \cdots q_j q_j'} \phi(q_1)\phi^*(q_1') \cdots \phi(q_j)\phi^*(q_j') \delta_{\Sigma q, \Sigma q'} \\
&= \int \mathrm{d}^2\rho \, |\phi(\rho)|^{2j}.
\end{aligned} \tag{7.22}$$

For a plane wave we have $|\phi(\rho)| = \sqrt{A}$, and $F_j = A^{1-j}$. For our Gaussian beam we obtain

$$F_j = \frac{1}{j} \left( \frac{\pi \rho_0^2}{2} \right)^{1-j}. \tag{7.23}$$

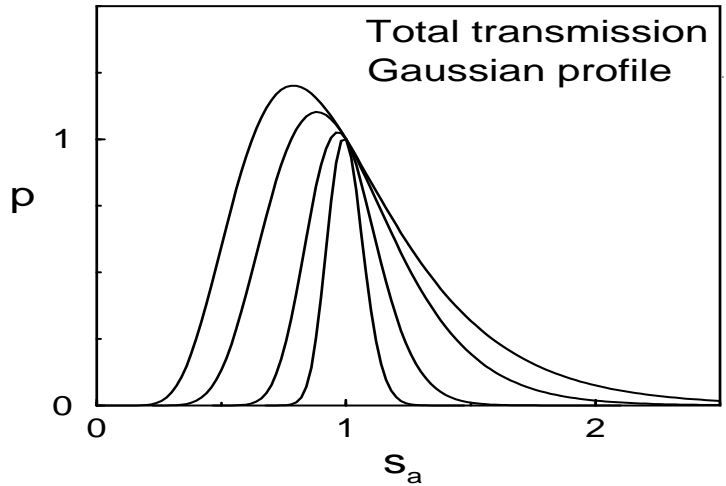

Figure 7.4: Intensity distribution of the total transmission, in units of $\sqrt{3g/2\pi}$, versus the normalized intensity $s_a$ for an incoming wave with Gaussian profile. With $g = 2, 4, 16$, and $64$ (upper to lower curve).

It is thus convenient to identify $A_G = \frac{1}{2}\pi\rho_0^2$ with the effective area of a Gaussian beam. (This definition is different from the one used in previous chapters, were $A_G = \pi\rho_0^2$). As compared to the plane wave case, the $j^{\text{th}}$ order term is smaller by a factor $1/j$ for a Gaussian profile. This implies for the generating function of the connected diagrams

$$\Phi_{\text{con}}(x) = g\int_0^1 \frac{dy}{y}\log^2\left(\sqrt{1 + \frac{xy}{g}} + \sqrt{\frac{xy}{g}}\right). \tag{7.24}$$

For small $s_a$ (and large $g$) there is again a log-normal saddle point. For large $s_a$ the dominant shape of the decay is given by the singularity at $x = -g$ and again yields $P(s_a) \sim \exp(-gs_a)$. In Fig. 7.4 we present the distribution function for different values of $g$. The shape of the distribution is quite similar to the plane case. A difference in the figures is that for a Gaussian beam profile the effects seem smaller. One could say that the Gaussian profile reduces the fluctuations, but it is mainly a question of definition of the number of modes $N$ and the area of a Gaussian intensity profile. In the graphs the distribution for the Gaussian profile at $g$ almost overlaps the plane wave distribution for $2 \times g$. Although the analytic forms are different, the difference is small.

By expansion of the generating function we recover the results for the second and third cumulant, obtained previously

$$\langle s_a^2\rangle_{\text{cum}} = \frac{2}{3g}, \qquad \langle s_a^3\rangle_{\text{cum}} = \frac{12}{5}\langle s_a^2\rangle_{\text{cum}}^2 \tag{7.25}$$

for a plane wave. For a Gaussian beam we recover

$$\langle s_a^2\rangle_{\text{cum}} = \frac{1}{3g}, \qquad \langle s_a^3\rangle_{\text{cum}} = \frac{16}{5}\langle s_a^2\rangle_{\text{cum}}^2. \tag{7.26}$$

The results coincide with the results of chapter 6 and describe the experiments of Ref. [105] well.

## 7.4    Speckle-intensity distribution

We apply the same mapping to obtain the distribution of the angular transmission coefficient. The angular and total transmission can be related to each in fairly simple way, because the interference processes are dominated by the same type of diagrams: the loop-less connected diagrams. This is in contrast to the conductance that is dominated by far more complicated loop diagrams. The different distribution for total transmission and speckle is the consequence of a counting argument only. In the plane wave situation the average angular transmission reads $\langle T_{ab} \rangle = \epsilon_a \epsilon_b g$. Let us count the number of connected loop-less diagrams that contribute to $T_{ab}^j = t_{ab} t_{ba}^\dagger t_{ab} \cdots t_{ab} t_{ba}^\dagger$. Now not only massless outgoing diffusons, but all pairings into outgoing diffusons contribute. This yields an extra combinatorial factor $j!$ in the $j^{\text{th}}$ moment:

$$\langle T_{ab}^j \rangle_{\text{con}} = j!(j-1)! \epsilon_a^j \epsilon_b^j \langle \text{Tr}(t^\dagger t)^j \rangle. \tag{7.27}$$

For the normalized angular transmission coefficient $s_{ab} = T_{ab}/\langle T_{ab} \rangle$, we introduce the following generating function of the connected diagrams:

$$\Psi_{\text{con}}(x) \equiv \sum_{j=1}^\infty \frac{(-1)^{j-1} x^j}{j! j!} \langle s_{ab}^j \rangle_{\text{con}} \tag{7.28}$$

It is easy to see that

$$\Psi_{\text{con}}(x) = \Phi_{\text{con}}(x), \tag{7.29}$$

with $\Phi_{\text{con}}$ given by Eq. (7.15) for plane wave incidence and by Eq. (7.24) for a broad Gaussian beam, respectively. In contrast to the total-transmission distribution, the cumulants are not only given by the connected diagrams. As example one can look again at Fig. 6.1; all the diagrams there contribute to the third moment of both speckle and total transmission, but for the speckle pattern there are 3! diagrams for each topology. Kogan *et al.* [118] showed that the summation of the disconnected diagrams can be done elegantly by performing an additional integral: Consider again the $j$th moment $\langle s_{ab}^j \rangle$. First there is the factor $j!$ from the different pairings of outgoing amplitudes. Consider a disconnected diagram made up of several connected ones. The number of connected parts with $i$ incoming (and $i$ outgoing) diffusons is labeled $m_i$. For instance Fig. 6.2 is characterized with $m_1 = 1$ and $m_2 = 1$, as it contains one non-interacting diffuson and a part connecting 2 diffusons. The total number of diffusons is fixed so $m$ obeys:

$$\sum_{i=1}^j i m_i = j \tag{7.30}$$

The number of different topologies that can be obtained is

$$\frac{n!}{m_1! m_2! \dots m_n!} \tag{7.31}$$

But now we have over-counted as permuting the diffusons of a connected part yields no different contribution. Therefore we have to divide with $(i!)^{m_i}$ for each $i$. Summarizing we thus find for the moments of $P(s_{ab})$

$$\langle s_{ab}^j \rangle = j! \sum_{m_1, m_2, \dots, m_j}' P(m_1, m_2, \dots, m_n) \langle s_{ab} \rangle_{\text{con}}^{m_1} \langle s_{ab}^2 \rangle_{\text{con}}^{m_2} \cdots \langle s_{ab}^j \rangle_{\text{con}}^{m_j} \tag{7.32}$$

Here the primed sum means that Eq.(7.30) should be fulfilled. The $P$ is the combinatorial prefactor

$$P(m_1, m_2, \ldots, m_n) = \frac{j!}{(1!)^{m_1}(2!)^{m_2}\cdots(j!)^{m_j}m_1!m_2!\cdots m_j!} \tag{7.33}$$

This factor is one for the connected diagram ($m_j = 1$). Now we know the moments the distribution follows from them as

$$P(s_{ab}) = \int_{-i\infty}^{i\infty} \frac{dx}{2\pi i} \sum_{j=0}^{n} \frac{(-x)^j}{j!} \langle s_{ab}^j \rangle \exp(x s_{ab}) \tag{7.34}$$

It is useful to write the factorial in $s_{ab}$ in Eq.(7.32) as an integral

$$j! = \int_0^\infty dv\, v^j \exp(-v) \tag{7.35}$$

Yielding

$$P(s_{ab}) = \int_{-i\infty}^{i\infty} \frac{dx}{2\pi i} \exp(s_{ab}x/v) \int_0^\infty \frac{dv}{v} \exp(-v) \sum_{j=0}^{\infty} \frac{(-x)^j}{j!} M_j \tag{7.36}$$

With $M_j = \langle s_{ab}^j \rangle / j!$. Also

$$
\begin{aligned}
\sum_{j=0}^{\infty} \frac{(-x)^j}{j!} M_j &= \sum_{m_1=0}^{\infty} \frac{(-x)^{m_1}}{m_1!} \left( \frac{\langle s_{ab} \rangle_{\mathrm{con}}}{1!} \right)^{m_1} \times \sum_{m_2=0}^{\infty} \frac{(-x)^{m_2}}{m_2!} \left( \frac{\langle s_{ab}^2 \rangle_{\mathrm{con}}}{2!} \right)^{m_2} \cdots \\
&= \exp\left(\sum_{j=0}^{\infty} \frac{(-x)^j}{j!} \langle s_{ab}^j \rangle_{\mathrm{con}}\right)
\end{aligned} \tag{7.37}
$$

Or in words, the sum over the disconnected diagrams replaces the exponent of the sum of connected diagrams. One gets

$$
\begin{aligned}
P(s_{ab}) &= \int_0^\infty \frac{dv}{v} \int_{-i\infty}^{i\infty} \frac{dx}{2\pi i} \exp\left( -\frac{s_{ab}}{v} + xv - \Psi_{\mathrm{con}}(x) \right) \\
&= \int_{-i\infty}^{i\infty} \frac{dx}{\pi i} K_0(2\sqrt{-s_{ab}x}) \exp\left( -\Psi_{\mathrm{con}}(x) \right).
\end{aligned} \tag{7.38}
$$

With $K_0$ the modified Bessel function. We plotted the speckle intensity distribution in Fig. 7.5 for an incoming plane wave. For large $g$ and moderate $s_{ab}$ we have $\Phi_{\mathrm{con}}(x) \approx x$ and we recover the Rayleigh law: $P(s_{ab}) = \exp(-s_{ab})$. The leading correction is found by expanding in $1/g$

$$P(s_{ab}) = e^{-s_{ab}} \left[ 1 + \frac{1}{3g}(s_{ab}^2 - 4s_{ab} + 2) \right] \tag{7.39}$$

This was derived previously by Shnerb and Kaveh [122]; here we have related the prefactor of the correction term to the conductivity. Genack and Garcia fitted their data to this relation and found $g = 14.6$. Our Eq. (7.38) describes these data very well for $g = 14.4$.

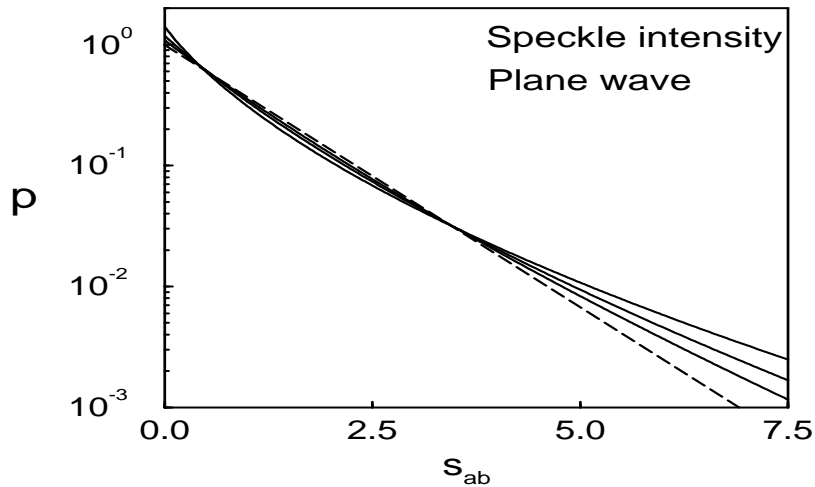

Figure 7.5: Intensity distribution of speckles versus the normalized intensity $s_{ab}$ for an incoming plane wave. With $g = 2, 4$, and 8 (upper to lower curve at $s_{ab} = 5$). The dashed line corresponds to the Rayleigh law ($g = \infty$).

For large $s_{ab}$ one can again apply steepest descent, which yields

$$P(s_{ab}) \sim \exp(-2\sqrt{gs_{ab}}\,). \qquad (7.40)$$

This stretched exponential tail differs from the form $P(s_{ab}) \sim \exp[-\sqrt[3]{(\frac{9}{4}s_{ab})^2 g}]$ asserted by Kogan *et al.* Their findings are based on truncating $\Phi_{\text{con}}(x)$ after order $x^2$. This corresponds to including only the simplest connected diagram: the $C_2$ diagram. With this diagram also all corresponding disconnected diagrams are constructed. Although their theory works quite well, one introduces a extra free parameter for each higher order correction, whereas we find the distribution as a function of a single parameter, i.e. $g$, only. Taking the full generating function into account, we find a qualitatively different saddle point. The stretched exponential was also seen in experiments. Already in 1989 Garcia and Genack observed it in their microwave experiments. But unfortunately their dynamic range was rather small. The fit of stretched exponential is therefore rather unprecise. The maximum intensity was 5 times the average, here one does not expect to see the tail already. Rather one observes a cross-over behavior here. However, more accurate measurements by Genack and Polkosnik are planned. Nevertheless, in the total transmission experiments, which were the inspiration for this research, the higher order corrections are easier to extract from the data.

Using Eq. (7.24) we derive the speckle distribution due to an incoming beam with Gaussian profile. A Gaussian profile leads to a different distribution with the same asymptotic behavior; it is plotted in Fig. 7.6.

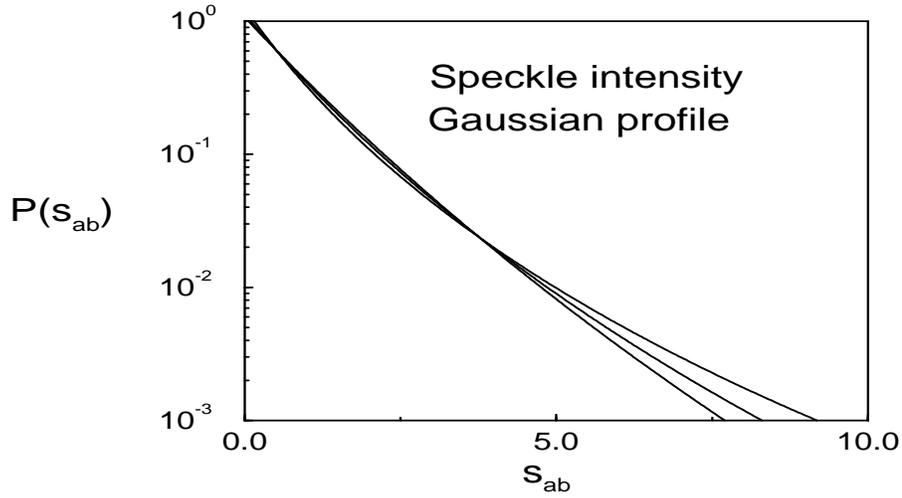

Figure 7.6: Intensity distribution of speckles versus the normalized intensity $s_{ab}$ for an incoming wave with Gaussian profile. With $g = 2, 4$, and $8$ (upper to lower curve at $s_{ab} = 5$).

### 7.4.1  Joint distribution

It turns out that the speckle distribution and the total transmission distribution of a certain incoming direction are related.[1] One can expect this, a large total transmission for a given incoming direction, will also be reflected in the individual speckles. The moments of the joint distribution are

$$\langle s_{ab}^k s_a^l \rangle_{\text{con}} = (k + l - 1)! \, k! \, \langle T_n^{k+l} \rangle \qquad (7.41)$$

The combinatorial factor $k!$ is for the speckle pattern, the factor $(k + l - 1)!$ is the number of possible pairings of incoming beam. Defining $\sigma$ as $s_{ab} = \sigma_{ab}s_a$, one obtains $\langle \sigma_{ab}^k s_a^m \rangle = k! \, (m - 1)! \, \langle T_n^m \rangle$ with $m = k + l$. Thus $P(\sigma_{ab}, s_a) = \exp(-\sigma_{ab})P(s_a)$. Thus one sees that $\sigma_{ab}$ and $s_a$ are independent variables, yet $s_{ab}$ and $s_a$ are dependent. Their joint distribution is

$$P(s_{ab}, s_a) = \frac{\exp(-s_{ab}/s_a)}{s_a} \int \frac{\mathrm{d}x}{2\pi i} \exp(x s_a - \Phi_{con}). \qquad (7.42)$$

We have plotted this distribution in Fig. 7.7. Integration over $s_a$ or $s_{ab}$ yields again the distribution $P(s_{ab})$, or $P(s_a)$, respectively. As the r.h.s in Eq. (7.42) contains only connected diagrams, the same holds thus for the l.h.s. Nevertheless, by analytic continuation for negative $l$ and the subsequent integration over $s_a$ we find the correct expression for $P(s_{ab})$.

## 7.5  Concluding remarks

In the above, we derived the full distribution function of both total transmission and angular transmission coefficient by mapping it to the distribution of eigenvalues of $T_{ab}$.

---

[1]This calculation is an unpublished result of Th. M. Nieuwenhuizen

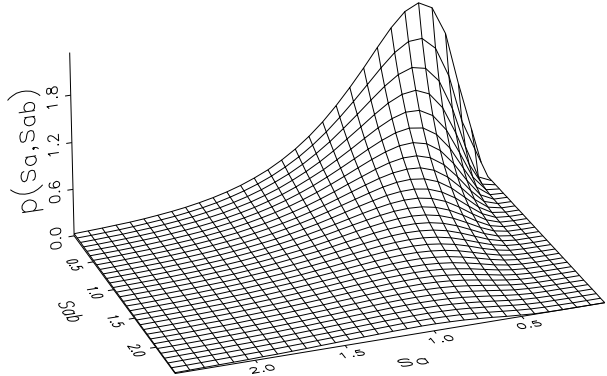

Figure 7.7: The joint distribution function of $s_a$ and $s_{ab}$, for a Gaussian beam profile, for $g = 2$. The maximum as a function of $s_a$ shifts as $s_{ab}$ changes.

But at least for the lower one we can use this mapping the other way around. First, the experiments seem to confirm the validity of the eigenvalue distribution. For a long time the prove for this eigenvalue distribution was done by the sometimes error-ing random matrix theory for quasi 1D systems. Only recently Nazarov showed its validity in arbitrary dimension. Probably the mesoscopic systems using multiple scattered electromagnetic waves, i.e. microwaves or light, are still the cleanest to prove this fundamental distribution.

Secondly, from our detailed evaluation of the low order diagrams, i.e. the second and third cumulant, insight in the eigenvalue distribution can be obtained. It is clear from those calculations that also the eigenvalue distribution is based on loop-less connected diagrams. Therefore the eigenvalue distribution cannot directly be used for the conductance. The explicit evaluation of second and third cumulant made it also clear, that the results are rather insensitive to absorption and surface effects.

Note that mapping from the eigenvalue distribution to the speckle and total transmission distribution is very general. Most interesting would be the inclusion of loops into the eigenvalue distribution, which would clarify the situation for small values of $g$ but also for the tails. In the important paper of Altshuler, Kravtsov and Lerner, [88] it was seen that although loop diagrams may be of lower order, they do become important at high moments (see below for an explanation). Therefore, our results for the tail change maybe. Explicit inclusion of the loops would solve this question.

## 7.6 Higher moments of the conductance

It would be interesting to apply the technique above also to the conductance. The distribution function of the conductance has been studied extensively, mainly with the advanced models mentioned in the introduction, chapter1. For a review see Altshuler, Kravstov, and Lerner in Ref. [23]. Using the non-linear sigma model they find for the lower cumulants ($n < g$) the estimate

$$\langle g^n \rangle_{\mathrm{cum}} \sim g^{2-n}. \tag{7.43}$$

But at large $n$ there is a topological growth of the diagrams: Introducing an extra loop in a diagram does introduce a factor $1/g^2$ making the diagram smaller, but on the same the number of topologically different diagrams increases with roughly a factor $n^2$ for large $n$. It is calculated that the tail of the distribution becomes essentially log-normal [134]. Their calculation seems to violate the one parameter scaling theory [12], as for the higher moments not only the dimensionless conductance $g$, but also extra parameters are necessary (however, see also Shapiro [135]). The question whether this is really the case is not resolved yet. Surprisingly enough, despite the vast number of measurements of the UCF, i.e. $\langle g^2 \rangle$, there are hardly any experimental data on the full distribution of $g$. The question of the precise distribution function is still open, as the pre-factors in Eq. (7.43) are unknown. Recently Macêdo showed in the quasi 1D case, that $\langle g^3 \rangle_{\mathrm{cum}} \propto g^{-2}$, which would mean that some "miraculous" cancelation occurs for the term proportional to $1/g$ [136].

Let us return to our diagrammatic approach. We need to construct connected diagrams with $2n$ amplitudes on the incoming side going to the outgoing side. The paring into diffusons should be the same on incoming and outgoing side. From Eq. (7.43) one observes that the $n$th *relative* cumulant should be proportional to $g^{2-2n}$. As in relative cumulants, each Hikami-box brings a factor $g^{-1}$, the diagrams for the $n$th cumulant should consist of $2n - 2$ Hikami-boxes. Reconsider the UCF diagrams of chapter 5. A possible method for generating these diagrams is: take the diagram for the third cumulant (chapter 6) and join an incoming to an outgoing diffuson, thus creating one extra internal diffuson. The joining of the diffusons can be done using the $K$-operator of section 2.6. Apart from forbidden diagrams with a loop of *amplitudes*, a subset of the $C_3$ diagrams is generated. Yet we do not find diagrams with most crossed diagrams internally instead of diffusons, i.e. the diagrams of lower line in Fig. 5.1, by this method. In two and three dimensions, the extra internal momentum integral has to be inserted. Unfortunately, this research is not concluded yet.

# A

# List of symbols

In general three-dimensional vectors are denoted as lower case boldface letters, and two-dimensional vectors with normal face capitals.
In the diagrams:

- Dots like: • denote a $t$-matrix, except in chapter 5 where they denote a potential.

- Diffusons are often depicted as close parallel lines

- Dashed lines connect identical scatterers

| Symbol | Meaning | Introduction |
|---|---|---|
| $\epsilon_0$ | dielectric constant inside slab | Eq. (2.4) |
| $\epsilon_1$ | dielectric constant outside slab | Eq. (2.4) |
| $\epsilon_a$, $\epsilon_b$ | diffuson prefactor | Eq. (2.67) |
| $\kappa$ | absorption rate | Eq. (2.34) |
| $\mu$ | cosine angle w.r.t $z-$axis | Eq. (2.48) |
| $\rho$ | transverse coordinate | Eq. (2.47) |
| $\rho_0$ | beam diameter | Eq. (4.11) |
| $\tau_1$ | limit intensity | Eq. (2.60) |
| $\omega$ | frequency of light | Eq. (2.4) |
| $\Omega$ | reduced frequency difference in diffuson | Eq. (2.36) |

| Symbol | Meaning | Introduction |
|---:|---|---:|
| $A$ | beam area | Eq. (2.47) |
| $a$ | albedo | Eq. (2.28) |
| $a_0$ | scatterer radius | Eq. (2.6) |
| $C_1$ | short range correlation | Eq. (4.1) |
| $C_2$ | long range correlation | Eq. (4.1) |
| $C_3$ | UCF correlation | Eq. (4.1) |
| $G$ | dressed Green's function | Eq. (2.22) |
| $G_0$ | bare Green's function | Eq. (2.9) |
| $g$ | dimensionless conductance | Eq. (6.2) |
| $H_4$ | Hikami four point vertex | Eq. (3.5) |
| $H_6$ | Hikami six point vertex | Eq. (3.7) |
| $h_4, h_6$ | vertex-prefactor | Eq. (5.8) |
| $I_{k,l}$ | loop integral | Eq. (2.31) |
| $K$ | extra scatterer operator | Eq. (2.82) |
| $k$ | wave vector | Eq. (2.7) |
| $L$ | sample thickness | sec. 2.1 |
| $\mathcal{L}$ | diffuse intensity | sec. 2.4 |
| $\ell$ | mean free path | Eq. (2.26) |
| $M$ | intensity-decay rate | Eq. (2.40) |
| $m$ | index of refraction ratio | Eq. (2.7) |
| $N$ | number of modes | Eq. (2.78) |
| $n$ | scatterer density | Eq. (4.11) |
| $\mathbf{p}$ | 3D microscopic momentum $|\mathbf{p}| \sim k$ | Eq. (2.30) |
| $Q$ | 2D transverse momentum | Eq. (2.40) |
| $\mathbf{q}$ | 3D mesoscopic momentum $\mathbf{q}\ell \simeq 1$ | Eq. (2.30) |
| $R(\mu)$ | boundary reflection coefficient | Eq. (2.44) |
| $s_a$ | normalized $T_a$ | Eq. (7.3) |
| $s_{ab}$ | normalized $T_{ab}$ | Eq. (7.4) |
| $T(\mu)$ | boundary transmission coefficient | Eq. (2.51) |
| $T_{ab}$ | angular resolved transmission | Eq. (2.79) |
| $T_a$ | total transmission | Eq. (2.80) |
| $t$ | $t-$matrix | Eq. (2.19) |
| $z$ | depth in the sample | chapter 2 |
| $z_0$ | extrapolation length | sec. 2.4, (2.41) |
| $\langle\langle O^k \rangle\rangle$ | normalized cumulant | $\langle O^k \rangle_{\mathrm{cum}}/\langle O \rangle^k$ |

# Bibliography

[1] S. Chandrasekhar, *Radiative Transfer* (Dover, New York, 1960).

[2] H. C. van de Hulst, *Multiple Light Scattering, Vols. 1 and 2* (Academic, New York, 1980).

[3] T. R. Kirkpatrick, Phys. Rev. B **31**, 5746 (1985).

[4] D. Sornette, Acoustica **67**, 199 and 251 (1989), and Vol. **68** 199.

[5] *Mesoscopic phenomena in solids*, Vol. 30 of *Modern problems in condensed matter sciences*, edited by B. L. Altshuler, P. A. Lee, and R. A. Webb (North-Holland, Amsterdam, 1991).

[6] *Electron-electron interactions in disordered systems*, Vol. 10 of *Modern problems in condensed matter sciences*, edited by A. L. Efros and M. Pollak (North-Holland, Amsterdam, 1985).

[7] *Analogies in Optics and Micro Electronics*, edited by W. van Haeringen and D. Lenstra (Kluwer Academic, Haarlem, 1990).

[8] *Analogies in Optics and Micro Electronics*, edited by W. van Haeringen and D. Lenstra (Kluwer Academic, Haarlem, 1991).

[9] R. Grosseteste, *Commentaria Roberti Linconiensis in libros postiorum Aristotelis*, Venice (1497), in *Der Weg der Physik*, edited by S. Sambursky (Artemis, Zürich, 1975).

[10] P. W. Anderson, Phys. Rev. **109**, 1492 (1958).

[11] N. F. Mott, *Metal-Insulator transitions* (Taylor and Francis, London, 1990).

[12] E. Abrahams, P. W. Anderson, D. C. Licciardello, and T. V. Ramakrishnan, Phys. Rev. Lett. **42**, 673 (1979).

[13] D. J. Thouless, Phys. Rep. **13**, 93 (1974).

[14] D. J. Thouless, Phys. Rev. Lett. **39**, 1167 (1977).

[15] F. Wegner, Z. Phys. B **25**, 327 (1976).

[16] D. Vollhardt and P. Wölfe, Phys. Rev. Lett. **45**, 842 (1980).

[17] D. Vollhardt and P. Wölfle, Phys. Rev. B **22**, 4666 (1980).

[18] D. Vollhardt and P. Wölfle, in *Electronic phase transitions*, Vol. 32 of *Modern problems in condensed matter sciences*, edited by W. Hanke and Yu. V. Kopaev (North-Holland, Amsterdam, 1992), p. 1.

[19] J. Kroha, Physica A **167**, 231 (1990).

[20] J. Kroha, T. Kopp, and P. Wölfe, Phys. Rev. B **41**, 888 (1990).

[21] F. Wegner, Z. Phys. B **35**, 207 (1979).

[22] K. N. Efetov, Adv. in Phys. **32**, 53 (1983).

[23] B. L. Altshuler, V. E. Kravtsov, and I. V. Lerner, in *Mesoscopic phenomena in solids*, Vol. 30 of *Modern problems in condensed matter sciences*, edited by B. L. Altshuler, P. A. Lee, and R. A. Webb (North-Holland, Amsterdam, 1991), p. 449.

[24] S. Hikami, Phys. Rev. B **24**, 2671 (1981).

[25] S. Hikami, Physica A **167**, 149 (1990).

[26] *Anderson transition and mesoscopic fluctuations*, Vol. 167 of *Physica A*, edited by B. Kramer and G. Schön (North Holland, Amsterdam, 1990), special issue of Physica A.

[27] M. P. van Albada, B. A. van Tiggelen, A. Lagendijk, and A. Tip, Phys. Rev. Lett. **66**, 3132 (1991).

[28] K. M. Leung and Y. F. Li, Phys. Rev. Lett. **65**, 2646 (1990).

[29] *Photonic band gaps and localization*, edited by C. M. Soukoulis (Plenum Press, New York, 1993).

[30] G. Bergmann, Phys. Rev. B **28**, 2914 (1983).

[31] S. Hikami, A. I. Larkin, and Y. Nagaoka, Prog. Theor. Phys. **63**, 707 (1980).

[32] *Anderson Localization*, Vol. 28 of *Springer proceedings in physics*, edited by T. Ando and H. Fukuyama (Springer-Verlag, Berlin, 1988).

[33] *Anderson localization*, Vol. 8 of *Solid state sciences*, edited by Y. Nagaoka and H. Fukuyama (Springer, Berlin, 1982).

[34] *Electronic phase transitions*, Vol. 32 of *Modern problems in condensed matter sciences*, edited by W. Hanke and Yu. V. Kopaev (North-Holland, Amsterdam, 1992).

[35] *Classical Wave Localization*, edited by P. Sheng (World Scientific, Singapore, 1990).

[36] E. Amic, J. M. Luck, and Th. M. Nieuwenhuizen, submitted to Phys. Rev. E (1995).

[37] A. F. Ioffe and A. R. Regel, Prog. Semicond. **4**, 237 (1960).

[38] N. F. Mott, Adv. Phys. **16**, 49 (1967).

[39] M. Kaveh, in *Analogies in Optics and Micro-Electronics*, edited by W. van Haeringen and D. Lenstra (Kluwer Academic, Haarlem, 1991).

[40] E. N. Economou, *Green's functions in quantum physics* (Spinger, Berlin, 1990).

[41] G. D. Mahan, *Many-particle physics* (Plenum Press, New York, 1990).

[42] A. A. Abrikosov, L. P. Gorkov, and I. E. Dzyaloshinski, *Methods of quantum field theory in statistical physics* (Dover, New York, 1963).

[43] H. C. Ohanian, *Classical Electrodynamics* (Allyn and Bacon, Boston, 1988).

[44] Th. M. Nieuwenhuizen, A. L. Burin, Yu. Kagan, and G. V. Shlyapnikov, Phys. Lett. A **184**, 360 (1994).

[45] G. Mie, Ann. Phys. **25**, 377 (1908).

[46] Th. M. Nieuwenhuizen, A. Lagendijk, and B. A. van Tiggelen, Phys. Lett. A **169**, 191 (1992).

[47] B. A. van Tiggelen, A. Lagendijk, and A. Tip, J. Phys. C. M. **2**, 7653 (1990).

[48] I. Polishchuk, A. L. Burin, and L. A. Maksimov, JETP Lett. **51**, 731 (1990).

[49] B. I. Halperin, Phys. Rev. A **139**, 104 (1965).

[50] P. Lloyd, J. Phys. C **2**, 1717 (1969).

[51] M. C. W. van Rossum, Th. M. Nieuwenhuizen, E. Hofstetter, and M. Schreiber, Phys. Rev. B **49**, 13377 (1994).

[52] B. A. van Tiggelen, A. Lagendijk, M. P. van Albada, and A. Tip, Phys. Rev. B **45**, 12233 (1992).

[53] A. A. Lisyansky and D. Livdan, Phys. Lett. A **170**, 53 (1992).

[54] J. X. Zhu, D. J. Pine, and D. A. Weitz, Phys. Rev. A **44**, 3948 (1991).

[55] D. S. Wiersma, M. P. van Albada, and A. Lagendijk (unpublished).

[56] A. Yu. Zyuzin, Europhys. Lett. **26**, 517 (1994).

[57] A. Ishimaru, *Wave propagation and scattering in random media, Vols. 1 and 2* (Academic, New York, 1978).

[58] A. Lagendijk, R. Vreeker, and P. de Vries, Phys. Lett. A **136**, 81 (1989).

[59] Th. M. Nieuwenhuizen and J. M. Luck, Phys. Rev. E **48**, 560 (1993).

[60] H. H. Kagiwada, R. Kalaba, and S. Ueno, *Multiple scattering processes: inverse and direct*, Vol. 8 of *Applied mathematics and computation* (Addison-Wesley, Reading Massachusetts, 1975).

[61] J. C. Maxwell-Garnett, Philos. Trans. Roy. Soc. London **203**, 385 (1904).

[62] D. A. G. Bruggeman, Ann. Phys. **24**, 636 (1935).

[63] P. Sheng, in *Macroscopic properties of disordered media*, Vol. 154 of *Lecture notes in physics*, edited by J. Ehlers *et al.* (Springer, Berlin, 1982), p. 239.

[64] P. N. den Outer and A. Lagendijk, Opt. Comm. **103**, 169 (1993).

[65] P. Molenaar, internal report University of Amsterdam (1991).

[66] P. N. den Outer, private communication.

[67] H. C. van de Hulst and R. Stark, Astron. Astrophys. **235**, 511 (1990).

[68] M. C. W. van Rossum and Th. M. Nieuwenhuizen, Phys. Lett. A **177**, 452 (1993).

[69] R. Landauer, Z. Phys. B **21**, 247 (1975).

[70] M. Büttiker, Phys. Rev. Lett. **57**, 1761 (1986).

[71] M. P. van Albada, J. F. de Boer, and A. Lagendijk, Phys. Rev. Lett. **64**, 2787 (1990).

[72] J. F. de Boer, M. P. van Albada, and A. Lagendijk, Phys. Rev. B **45**, 658 (1992).

[73] M. C. W. van Rossum, J. F. de Boer, and Th. M. Nieuwenhuizen, to be published in Phys. Rev. E (1995), preprint cond-mat/9412122 at babbage.sissa.it.

[74] B. J. Nieuwenhuis, internal report University of Amsterdam (1994).

[75] R. Berkovits and S. Feng, Phys. Rev. Lett. **65**, 3120 (1990).

[76] P. N. den Outer, Th. M. Nieuwenhuizen, and A. Lagendijk, J. Opt. Soc. Am. A **10**, 1209 (1993).

[77] P. N. den Outer, M. C. W. van Rossum, Th. M. Nieuwenhuizen, and A. Lagendijk, *OSA proceedings on Advances in optical imaging and photon migration*, edited by R. R. Alfano (SPIE, Bellingham, 1994), p. 297.

[78] Y. Kuga and A. Ishimaru, J. Opt. Soc. Am. A **1**, 831 (1985).

[79] M. P. van Albada and A. Lagendijk, Phys. Rev. Lett. **55**, 2692 (1985).

[80] P. E. Wolf and G. Maret, Phys. Rev. Lett. **55**, 2696 (1985).

[81] E. Akkermans, P. E. Wolf, and R. Maynard, Phys. Rev. Lett. **56**, 1471 (1986).

[82] Yu. N. Barbaranenkov, Izv. Vysch. Uch. Zav.-Radiofiz. **16**, 88 (1973).

[83] M. J. Stephen and G. Cwilich, Phys. Rev. B **34**, 7564 (1986).

[84] L. P. Gor'kov, A. I. Larkin, and D. E. Khmel'nitskii, JETP Lett. **30**, 228 (1979).

[85] M. J. Stephen, in *Mesoscopic phenomena in solids*, Vol. 30 of *Modern problems in condensed matter sciences*, edited by B. L. Altshuler, P. A. Lee, and R. A. Webb (North-Holland, Amsterdam, 1991), p. 81.

[86] C. L. Kane, R. A. Serota, and P. A. Lee, Phys. Rev. B **37**, 6701 (1988).

[87] I. V. Lerner and Th. M. Nieuwenhuizen, private communication.

[88] B. L. Altshuler, V. E. Kravtsov, and I. V. Lerner, Sov. Phys. JETP **64**, 1352 (1986).

[89] J. R. Gao *et al.* (unpublished).

[90] R. Berkovits and S. Feng, Phys. Rep. **238**, 135 (1994).

[91] S. Feng, C. Kane, P. Lee, and A. D. Stone, Phys. Rev. Lett. **61**, 834 (1988).

[92] B. Shapiro, Phys. Rev. Lett. **57**, 2168 (1986).

[93] I. Freund, M. Rosenbluh, and S. Feng, Phys. Rev. Lett. **61**, 2328 (1988).

[94] I. Freund and R. Berkovits, Phys. Rev. B **41**, 496 (1990).

[95] A. Z. Genack and N. Garcia, Europhys. Lett. **21**, 753 (1993).

[96] M. J. Stephen and G. Cwilich, Phys. Rev. Lett. **59**, 285 (1987).

[97] A. Yu. Zyuzin and B. Z. Spivak, Sov. Phys. JETP **66**, 560 (1987).

[98] R. Pnini and B. Shapiro, Phys. Rev. B **39**, 6986 (1989).

[99] A. Z. Genack, N. Garcia, and W. Polkosnik, Phys. Rev. Lett. **65**, 2129 (1990).

[100] N. Garcia, A. Z. Genack, R. Pnini, and B. Shapiro, Phys. Lett. A **176**, 458 (1993).

[101] R. Pnini and B. Shapiro, Phys. Lett. A **157**, 265 (1991).

[102] P. A. Lee and A. D. Stone, Phys. Rev. Lett. **55**, 1622 (1985).

[103] C. P. Umbach, S. Washburn, R. B. Laibowitz, and R. A. Webb, Phys. Rev. B **30**, 4048 (1984).

[104] B. L. Altshuler, JETP Lett. **41**, 649 (1985).

[105] J. F. de Boer *et al.*, Phys. Rev. Lett. **73**, 2567 (1994).

[106] J. W. Goodman, in *Laser speckle and related phenomena*, edited by J. C. Dainty (Springer, Berlin, 1975), Vol. 9, p. 9.

[107] R. Berkovits and M. Kaveh, Europhys. Lett. **13**, 97 (1990).

[108] I. Freund and M. Rosenbluh, Opt. Commun. **82**, 362 (1991).

[109] R. Berkovits, Phys. Rev. B **42**, 10750 (1990).

[110] P. A. Lee, A. D. Stone, and H. Fukuyama, Phys. Rev. B **35**, 1039 (1987).

[111] D. S. Fisher and P. A. Lee, Phys. Rev. B **23**, 6851 (1981).

[112] M. Janßen, Solid State. Comm. **79**, 1073 (1991).

[113] M. C. W. van Rossum, Th. M. Nieuwenhuizen, and R. Vlaming, to be published in Phys. Rev. E (1995), preprint cond-mat/9412040 at babbage.sissa.it.

[114] G. Baym and L. Kadanoff, Phys. Rev. **124**, 287 (1961).

[115] G. Baym, Phys. Rev. **127**, 1391 (1962).

[116] R. A. Serota, F. P. Esposito, and M. Ma, Phys. Rev. B **39**, 2952 (1989).

[117] N. Garcia and A. Z. Genack, Phys. Rev. Lett. **66**, 1850 (1991).

[118] E. Kogan, M. Kaveh, R. Baumgartner, and R. Berkovits, Phys. Rev. B **48**, 9404 (1993).

[119] Th. M. Nieuwenhuizen and M. C. W. van Rossum, Phys. Rev. Lett. **74**, 2674 (1995).

[120] N. G. van Kampen, *Stochastic processes in physics and chemistry* (North-Holland, Amsterdam, 1992).

[121] E. Kogan and M. Kaveh, preprint cond-mat/9412090, (1994).

[122] N. Shnerb and M. Kaveh, Phys. Rev. B **43**, 1279 (1991).

[123] N. Garcia and A. Z. Genack, Phys. Rev. Lett. **63**, 1678 (1989).

[124] Th. M. Nieuwenhuizen and M. C. W. van Rossum, Phys. Lett. A **160**, 461 (1991).

[125] O. N. Dorokhov, Solid State. Comm. **51**, 381 (1984).

[126] Y. Imry, Europhys. Lett. **1**, 249 (1986).

[127] J. B. Pendry, A. MacKinnon, and A. B. Pretre, Physica A **168**, 400 (1990).

[128] A. D. Stone, P. A. Mello, K. A. Muttalib, and J.-L. Pichard, in *Mesoscopic phenomena in solids*, edited by B. L. Altshuler, P. A. Lee, and R. A. Webb (North-Holland, Amsterdam, 1991), Vol. 30, p. 369.

[129] Yu. V. Nazarov, Phys. Rev. Lett. **73**, 134 (1994).

[130] P. A. Mello, P. Pereyra, and N. Kumar, Ann. Phys. **181**, 290 (1988).

[131] J. B. Pendry, A. MacKinnon, and P. J. Roberts, Proc. R. Soc. London Ser. A **437**, 67 (1992).

[132] R. B. S. Oakeshott and A. MacKinnon, J. Phys.: Condens Matter **6**, 1513 (1994).

[133] C. W. J. Beenakker and M. Büttiker, Phys. Rev. B **46**, 1889 (1992).

[134] B. L. Altshuler, V. E. Kravtsov, and I. V. Lerner, JETP Lett. **43**, 441 (1986).

[135] B. Shapiro, Phys. Rev. Lett. **65**, 1510 (1990).

[136] A. M. S. Macêdo, Phys. Rev. B **49**, 1858 (1994).

# Mesoscopische verschijnselen in veelvoudige lichtverstrooiing

Onze ogen zijn onze belangrijkste zintuigen en het licht dat zij zien kan veel interessante verstrooiingsprocessen ondergaan. Denk bijvoorbeeld aan een zonsondergang of aan de kleurschakeringen van een vlinder. Eén van die verstrooiingsprocessen is uit esthetisch oogpunt niet bijzonder aantrekkelijk, maar voor natuurkundigen een nadere studie waard, namelijk licht dat vele malen verstrooit op een collectie verstrooiers. Het veelvoudig verstrooide licht wordt dan diffuus. Een alledaags voorbeeld hiervan is licht dat zich voortplant door mist of door melk. In ons laboratorium echter worden de experimenten gedaan door kleine plakjes witte verf te beschijnen met een laser. Een laser en wat witte verf, dat klinkt simpel, maar nauwkeurige metingen zijn echt niet eenvoudig. Juist de hoge kwaliteit van de experimenten bleek van groot nut om ook de kleinste effecten te kunnen zien.

Wat is nu zo interessant aan dat veelvuldig verstrooide licht? Alle interessante informatie wordt toch immers door elkaar gehusseld? Laat ik dat proberen uit te leggen: In heldere lucht of glas volgt het licht een rechte weg. In die witte verf wordt door alle verstrooiingen de gevolgde weg kronkelig en de afgelegde weg wordt erg lang. De zogenaamde diffusiebenadering geeft nu een redelijke beschrijving. Maar er is meer aan de hand. Als verstrooiing sterker wordt (de mist dikker wordt) ontstaat er een kans dat lichtbundels elkaar ontmoeten. Omdat licht een golfverschijnsel is, kunnen de bundels dan interfereren. Nu is deze kans op interferentie meestal erg klein en wordt in de diffusiebenadering dan ook verwaarloosd. Maar deze effecten zijn in de plakjes witte verf wel degelijk waarneembaar en voor mij zijn ze het zout in de pap. Men spreekt ook wel over mesoscopische effecten, aanduidende dat men kijkt naar lengteschalen ergens tussen enerzijds de macroscopie (de diffusie, waar alles uitmiddelt) en anderzijds de microscopie (de individuële verstrooiingsprocessen). Overigens komen deze mesoscopische verschijnselen ook voor bij andere diffuse golven zoals geluidsgolven en microgolven. Zelfs het transport van elektronen door een stukje verontreinigd metaal valt onder deze beschrijving. Sinds de introductie van de quantummechanica weten we immers dat ook elektronen een golfkarakter hebben. De theorie in dit werk is op al die systemen binnen zekere grenzen toepasbaar. De optische experimenten hebben echter het voordeel zeer zuiver en nauwkeurig te zijn.

Welke invloed hebben die mesoscopische effecten op het transport? We onderscheiden twee mogelijkheden voor de interferenties. Ten eerste is het mogelijk dat een lichtbundel, na de nodige omzwervingen, zichzelf weer tegenkomt en interfereert. Die terugkeerkans blijkt groter te zijn dan men op grond van de diffusietheorie zou verwachten. Omdat licht vaker terugkeert, is de effectieve voortplantingssnelheid kleiner. Dit effect staat bekend als zwakke localizatie. Voor elektronen is het zelfs mogelijk de verstrooiing zo sterk te maken, dat transport onmogelijk wordt. Op dat moment zijn de elektronen niet meer mobiel en de weerstand van het metaal wordt bijzonder groot. Dit wordt ook wel sterke of Anderson-localizatie genoemd. Voor licht is Anderson-localizatie nog nooit gezien, omdat het niet eenvoudig blijkt licht zo sterk te verstrooien. Optisch wel goed waarneembaar is de tweede mogelijkheid voor interferentie. Dit is het process waarbij de ene bundel die de witte verf doorkruist, interfereert een andere bundel. Vanwege de mogelijke interferentie zullen de bundels zich niet meer onafhankelijk voortplanten, maar verstrengeld raken. Dit komt tot uitdrukking in hun correlatiefunctie, welke de mate van relatie tussen de twee intensiteiten aangeeft. Over dit soort processen gaat deze studie.

Het aardige van optische systemen is dat men niet snel is uitgeëxperimenteerd: In tegenstelling tot electronische systemen, waar men eigenlijk alleen de weerstand kan meten, zijn er voor optische systemen drie transmissiegroootheden. Allereerst kan men hoekopgelost meten. Voor zo'n meting beschijnt men de witte verf met een evenwijdige bundel uit een zekere hoek, en meet de uitgaande intensiteit in een zekere richting. Dit geeft de hoekopgeloste transmissie. Daarnaast is het ook mogelijk al het uitgaande licht te verzamelen om zo de totale transmissie te verkrijgen. Ten slotte is het ook nog mogelijk met vele richtingen in te schijnen en ook weer al het uitgaande licht op te vangen. In dat geval meet men de geleiding. Het blijkt dat de correlatiefuncties van deze drie groootheden zeer verschillend van elkaar zijn, zowel wat betreft orde van grootte, als wat betreft eigenschappen.

Na een inleiding, begin ik dit proefschrift met de diffusievergelijking en de stralings-transportvergelijking. Zij bepalen hoe de diffuse intensiteit zich gedraagt in de plak witte verf. Ik zal bespreken hoe het precieze gedrag nabij de rand van de plak moet worden meegenomen, want dat blijkt van belang te zijn in de experimenten. Ik sta ook stil bij de vraag of het mogelijk is, ondanks de diffusie, objecten in het verstrooiende medium waar te nemen. Dit sluit aan bij onderzoek in de medische fysica naar de mogelijkheid om dergelijke technieken tomografisch te gebruiken voor het opsporen van bloedstolsels of tumoren.

In hoofdstuk 3 introduceer ik een andere belangrijk element voor de berekeningen: de Hikami-box. Deze beschrijft de interferentie tussen twee bundels. De berekeningsmeth-ode die ik gebruik is een diagrammatische techniek. Die komt erop neer dat men alle mogelijke verstrooiingsprocessen van een zekere klasse systematisch tekent, uitrekent en optelt. Het diffuse transport wordt in diverse afbeeldingen aangegeven twee parallelle lijnen, terwijl de Hikami-box wordt aangegeven met een grijs vierkantje. Zo, nu be-grijpt de lezer hopelijk die diagrammetjes, parallelle lijnen: simpel diffuus transport, vierkantjes: een interessant interferentieproces.

Nu ik alle ingrediënten heb, begin ik met de berekening van de mesoscopische ef-fecten. De correlatiefuncties van de drie verschillende transmissiegroootheden worden in de hoofdstukken 4 en 5 uitgerekend. Ik bereken de correlatiefunctie van twee bundels die beide de witte verf doorkruisen, maar en net iets andere invalshoek of een net iets andere kleur hebben. Waar mogelijk vergelijk ik onze uitkomsten met experimentele gegevens. Hoewel de diagrammatische techniek het voordeel heeft recht-door-zee te zijn en uiteindelijk altijd de juiste uitkomst levert, komt in hoofdstuk 5 een nadeel van deze techniek naar voren. Het is namelijk soms moeilijk alle relevante processen uit te vissen en hun aantal kan erg groot worden. De berekening wordt nogal arbeidsintensief. Ik laat zien dat, ondanks scepsis in de literatuur, deze harde noot toch kraakbaar is.

In hoofdstuk 6 beschouw ik de samenhang tussen drie bundels voor het geval men de totale transmissie meet. Voor dit effect moeten dus drie bundels met elkaar interfereren, hetgeen niet zo waarschijnlijk lijkt. Maar hoewel dit inderdaad een erg klein effect blijkt, werd het onlangs experimenteel gemeten. Ik vind goede overeenstemming met de experimenten.

Door deze berekeningen uit te breiden naar nog meer bundels is het mogelijk de volledige verdeling van de uitgaande intensiteiten uit te rekenen voor hoekopgeloste en totale transmissie metingen. Dit gebeurt in het zevende en laatste hoofdstuk. Ook hiermee zijn experimentele gegevens, dit maal van microgolfexperimenten, in overeenstemming. Het blijkt dat de fluctuaties rond het gemiddelde veel groter zijn dan men op grond van de diffusiebenadering zou verwachten. Dit is ook wel te begrijpen: in de diffusiebenadering zijn de lichtbundels onafhankelijk en dus zullen variaties in de ene bundel al snel door een andere bundel uitgemiddelt worden. Maar in werkelijkheid zijn, ten gevolge van al die interferentie processen, de bundels niet onafhankelijk. Door dit gemeenschappelijke gedrag zal er veel minder middeling plaatsvinden en zullen dus de fluctuaties groot zijn.

In het laatste hoofdstuk worden de conclusies van dit werk het duidelijkst: 1) Interferentie processen leiden tot interessante effecten bij veelvoudig verstrooide golven. Er blijken namelijk sterke correlaties en grote fluctuaties in de doorgelaten bundels te ontstaan. 2) Optische systemen zijn bijzonder geschikt om dit waar te nemen.